\documentclass[manuscript]{acmart}

\usepackage{ulem}
\usepackage{multirow}
\usepackage{makecell}
\usepackage{tabularx}
\usepackage[T1]{fontenc}
\usepackage{array}
\usepackage{listings}
\usepackage{xspace} 
\usepackage{subfigure}
\usepackage{textcomp}
\usepackage{microtype} 
\usepackage{calc}
\usepackage{pifont}
\usepackage{tcolorbox}
\usepackage{graphicx}
\usepackage{subcaption}

\usepackage{adjustbox}
\usepackage{amsmath,amssymb,amsfonts}
\usepackage{algorithmic}
\usepackage{graphicx}
\usepackage{textcomp}
\usepackage{xcolor}
\usepackage{multirow}
\usepackage{mdframed}
\usepackage{balance}
\usepackage{tcolorbox}
\usepackage{makecell}
\usepackage{threeparttable}
\usepackage{colortbl}
\usepackage{subfigure}
\usepackage{hhline}
\usepackage{pifont}
\usepackage{listings}
\usepackage{enumitem}
\usepackage{booktabs}
\usepackage{ragged2e} 

\usepackage{booktabs} 
\usepackage{array} 
\usepackage{tabularx} 
\usepackage{threeparttable} 
\usepackage{multirow} 
\usepackage{amssymb}

\usepackage{tikz}
\usetikzlibrary{trees}

\usepackage[utf8]{inputenc}
\usepackage[T1]{fontenc}
\usepackage{lmodern}

\usepackage[edges]{forest}
\usepackage{xcolor}
\usepackage{tikz}
\usetikzlibrary{arrows.meta}

\definecolor{rootbg}{HTML}{F3E8FF}
\definecolor{l0bg}{HTML}{EAF2FF}
\definecolor{l1bg}{HTML}{EFFFF4}
\definecolor{l2bg}{HTML}{FFF8E6}
\definecolor{l3bg}{HTML}{FDEEEE}
\definecolor{leafbg}{HTML}{FFFFFB}

\newcommand{\Comment}[1]{}

\definecolor{ggray}{HTML}{eff0f0}
\definecolor{gggray}{HTML}{E8E8E8}
\definecolor{ggggray}{HTML}{BEBEBE}
\definecolor{myblue}{RGB}{255,255,255}
\definecolor{myyellow}{HTML}{FFF2CC}
\definecolor{myfinding}{HTML}{E7F1FA}
\definecolor{myanswer}{HTML}{FDDECE}

\usepackage[utf8]{inputenc}
\usepackage[english]{babel}

\newcounter{finding}
\newcommand{\finding}[1]{\refstepcounter{finding}
 	\vspace{1mm}
	\begin{mdframed}[linecolor=gray,roundcorner=12pt,backgroundcolor=gray!15,linewidth=3pt,innerleftmargin=2pt, leftmargin=0cm,rightmargin=0cm,topline=false,bottomline=false,rightline = false]
		\textbf{Finding \arabic{finding}:} #1
	\end{mdframed}
	\vspace{1mm}
}

\newcommand{\cmark}{\checkmark}
\newcommand{\xmark}{$\times$}
\newcommand{\nr}{\textsc{N/R}}
\newcolumntype{L}[1]{>{\raggedright\arraybackslash}p{#1}}
\newcolumntype{C}[1]{>{\centering\arraybackslash}p{#1}}

\title{Knowledge-Centric Automated Issue Resolution: A Systematic Survey and Taxonomy}
\author{Zhenxi Chen}
\email{chenzhx236@mail2.sysu.edu.cn}
\affiliation{%
  \institution{Sun Yat-sen University}
  \city{Zhuhai}
  \country{China}
}

\author{Mingwei Liu}
\authornote{Corresponding author.} 
\email{liumw26@mail.sysu.edu.cn}
\affiliation{%
  \institution{Sun Yat-sen University}
  \city{Zhuhai}
  \country{China}
}

\author{Zihao Wang}
\email{wangzh778@mail2.sysu.edu.cn}
\affiliation{%
  \institution{Sun Yat-sen University}
  \city{Zhuhai}
  \country{China}
}

\author{Zhanhui Ren}
\email{renzhh9@mail2.sysu.edu.cn}
\affiliation{%
  \institution{Sun Yat-sen University}
  \city{Zhuhai}
  \country{China}
}

\author{Leyan Hu}
\email{huly35@mail2.sysu.edu.cn}
\affiliation{%
  \institution{Sun Yat-sen University}
  \city{Zhuhai}
  \country{China}
}

\author{Xueying Du}
\email{xueyingdu21@m.fudan.edu.cn}
\affiliation{%
  \institution{Fudan University}
  \city{Shanghai}
  \country{China}
}

\author{Ying Wang}
\email{yingwang25@m.fudan.edu.cn}
\affiliation{%
  \institution{Fudan University}
  \city{Shanghai}
  \country{China}
}

\author{Chenxi Zhang}
\email{zhangchenxi@xidian.edu.cn}
\affiliation{%
  \institution{Xidian University}
  \city{Xi'an}
  \country{China}
}

\author{Chong Wang}
\email{chong.wang@ntu.edu.sg}
\affiliation{%
  \institution{Nanyang Technological University}
  \city{Singapore}
  \country{Singapore}
}

\author{Zhenchang Xing}
\email{zhenchang.xing@csiro.au}
\affiliation{%
  \institution{CSIRO Technology}
  \city{Canberra}
  \country{Australia}
}

\author{Haofen Wang}
\email{carter.whfcarter@gmail.com}
\affiliation{%
  \institution{Tongji University}
  \city{Shanghai}
  \country{China}
}

\author{Xin Peng}
\email{pengxin@fudan.edu.cn}
\affiliation{%
  \institution{Fudan University}
  \city{Shanghai}
  \country{China}
}

\author{Yanlin Wang}
\email{yanlin-wang@outlook.com}
\affiliation{%
  \institution{Sun Yat-sen University}
  \city{Zhuhai}
  \country{China}
}

\renewcommand{\shortauthors}{Z. Chen, M. Liu, Z. Wang, Z. Ren, L. Hu, X. Du, Y. Wang, C. Zhang, C. Wang, Z. Xing, H. Wang, X. Peng, and Y. Wang}

\begin{document}

\begin{abstract}
Automated issue resolution is a knowledge-intensive software maintenance task in which LLM-based agents must acquire, represent, use, and update knowledge from repositories, execution environments, historical artifacts, documents, tools, and prior repair processes. This paper presents a systematic survey of 133 studies on automated issue resolution from a knowledge-centric perspective. We introduce a three-layer taxonomy that organizes issue-resolution knowledge into the Background Knowledge Layer, the Repository Knowledge Layer, and the Procedural Knowledge Layer, and analyze the knowledge lifecycle along six dimensions: knowledge types, sources and extraction methods, representation formats, usage methods, workflow stages, and temporal evolution. Our analysis shows that current systems primarily rely on Existing Code Knowledge and Development Tool Knowledge, while Repository Evolution Knowledge and Experiential Knowledge are increasingly adopted, indicating a shift toward history-informed, feedback-aware, and experience-oriented issue resolution. In contrast, Design Architecture Knowledge and explicitly modeled Background Knowledge remain comparatively underexplored. These findings suggest that future systems should move beyond temporary context construction toward reusable, traceable, and evolvable software knowledge infrastructures that support more controllable, verifiable, and generalizable software engineering agents. The surveyed studies and related resources are maintained at \url{https://github.com/SYSUSELab/Awesome-Knowledge-for-AIR}.
\end{abstract}

\keywords{Issue Resolution, Knowledge, Systematic Literature Review}

\maketitle

\section{Introduction}

Issue resolution is a central activity in software maintenance, where developers must understand user-reported problems or feature requests, locate the relevant implementation context, modify one or more parts of an existing repository, and validate that the resulting patch resolves the issue without introducing regressions \cite{jiang2025agenticSurvey,li2026advancesFrontiers}. Unlike isolated code generation, issue resolution is grounded in real development artifacts: natural-language issue descriptions, evolving codebases, dependency configurations, tests, historical commits, and project-specific conventions. The introduction of SWE-bench made this task a concrete and reproducible research problem by requiring systems to resolve real GitHub issues using full repository snapshots and execution-based evaluation \cite{jimenez2023swe}. Since then, issue resolution has become a representative benchmark for autonomous software engineering, because it directly measures whether an AI system can perform the end-to-end reasoning and editing process needed in practical software maintenance \cite{yang2024swe,xia2025demystifying}.

The importance of automated issue resolution has grown alongside the rapid progress of large language models (LLMs) and software engineering agents. LLMs have moved beyond function-level code completion toward repository-level maintenance tasks, while agentic systems provide the planning, tool use, environment interaction, and feedback loops needed for long-horizon development work \cite{tao2024magis,zhang2024autocoderover,wang2024openhands,arora2024masai}. Recent surveys have shown that issue resolution now covers a broad ecosystem of benchmarks, training-free agent frameworks, training-based methods, empirical analyses, and practical tools \cite{jiang2025agenticSurvey,li2026advancesFrontiers}. This progress suggests a promising path toward improving software maintenance efficiency and quality. At the same time, repository-level issue resolution exposes a key challenge that is less visible in isolated code generation: agents must operate over large, evolving, and heterogeneous software contexts, including code structures, dependency configurations, execution environments, test feedback, historical changes, and project-specific conventions. Therefore, the effectiveness of automated issue resolution depends not only on the scaling of model capabilities, but also on how agents acquire, organize, represent, and apply task-relevant knowledge throughout the resolution process.

Despite these advances, issue resolution remains a formidable challenge that transcends self-contained code generation. An issue-resolution system must reason within large and unfamiliar codebases where the model's parametric knowledge is incomplete, stale, or mismatched with the target repository. A fundamental gap therefore exists between the implicit knowledge learned during pre-training and the explicit symbolic constraints embedded in software projects. Recent work on Code Digital Twin similarly emphasizes making tacit software knowledge, such as system functionalities, responsibility allocation, and design rationales, explicit and accessible to LLMs for software maintenance tasks \cite{peng2025code}. These constraints are distributed across code structures, dependency graphs, tests, documentation, issue discussions, pull requests, execution traces, and development histories. Without sufficient grounding in such heterogeneous knowledge, agents can misidentify the faulty location, overlook hidden architectural intent, violate interface contracts, or generate patches that pass local reasoning but fail repository-level validation \cite{reddy2025swerank,jiang2025cosil,yang2025kgcompass,wang2025aegis}.

Recent methods have therefore begun to enrich LLMs and agents with heterogeneous knowledge beyond model parameters. Some systems construct code graphs, dependency relations, or repository-aware knowledge graphs to support localization and repair \cite{ouyang2024repograph,yang2025kgcompass,chen2025prometheus}. Others use dynamic execution feedback, bug reproduction, test generation, and validation signals to constrain patch generation \cite{wang2025aegis,cheshkov2024exploring,liu2026dynamic}. Memory- and experience-driven agents further exploit historical trajectories, cross-task experience, or repository memories to guide future issue resolution \cite{chen2025swe,tang2025agent,wang2025improving}. Taken together, these developments show that automated issue resolution is increasingly shaped by how knowledge is extracted from heterogeneous sources, represented for models and tools, operationalized during agent interaction, and applied across localization, patch generation, and validation.

Nevertheless, the field still lacks a clear and systematic understanding of the knowledge strategies underlying automated issue resolution. Existing surveys organize this area primarily from the perspectives of benchmarks, workflow phases, agent architectures, training paradigms, and empirical findings \cite{jiang2025agenticSurvey,li2026advancesFrontiers}. These perspectives are valuable for mapping the overall development of the field, but they do not fully explain automated issue resolution from a knowledge-centric perspective: what types of knowledge current systems rely on, where such knowledge originates, how it is extracted from heterogeneous sources, how it is represented for LLMs, agents, and tools, how it is incorporated into resolution systems, and at which workflow stages it is applied. Moreover, as recent studies increasingly combine repository context, execution feedback, historical artifacts, tool interactions, and experiential trajectories, researchers still lack a unified taxonomy for comparing multi-source knowledge strategies and for identifying which knowledge types, sources, representations, usage methods, and application stages remain underexplored. This gap limits a deeper understanding of how knowledge supports issue comprehension, localization, patch generation, validation, and the broader evolution of automated software maintenance.

To address these gaps, this paper presents a systematic survey of 133 studies on automated issue resolution. We adopt a knowledge-centric lens to dissect, categorize, and analyze how knowledge supports coding agents throughout the issue resolution process. The main contributions of this work are as follows:

\begin{itemize}
\item We propose a knowledge-centric cognitive framework and a comprehensive three-layer taxonomy of knowledge used in automated issue resolution, classifying it into Background Knowledge, Repository Knowledge, and Procedural Knowledge. This framework clarifies how knowledge is grounded in software artifacts and development contexts, organized into functional knowledge systems, externalized into agent-accessible representations, operationalized through different usage methods, and applied across issue resolution stages.

\item We systematically analyze how knowledge is sourced, extracted, represented, operationalized, and applied across automated issue resolution systems. Specifically, we provide a coherent map of knowledge sources and extraction methods, representation formats, usage methods, and application stages, thereby revealing how heterogeneous software artifacts, execution feedback, repository histories, documents, expert guidance, and agent experiences are transformed into actionable support for LLM-based issue resolution.

\item We identify dominant knowledge usage patterns, recent trends, and emerging opportunities in automated issue resolution. Our analysis shows that current approaches have established a strong foundation around Existing Code Knowledge and Development Tool Knowledge, while Design Architecture Knowledge and L1 Background Knowledge, including Programming Language Knowledge, Technology Stack Knowledge, and Domain-specific Knowledge, are becoming important expansion directions for handling more complex, architecture-aware, multilingual, framework-dependent, and domain-specific issue-resolution tasks. We further analyze representative methods and high-performing SWE-bench Verified systems to discuss how recent work is moving from static code-centric context provision toward dynamic, feedback-aware, history-informed, and experience-oriented knowledge orchestration.
\end{itemize}

\section{Background}
\label{sec:background}
This section provides the necessary context for our survey. We first define automated issue resolution and introduce the major concepts that appear in current methods, including workflow stages, retrieval-augmented context construction, tool interaction, and agentic execution. We then discuss benchmarks for the issue-resolution task and clarify how our knowledge-centric survey differs from existing surveys in this area.

\subsection{Automated Issue Resolution}

\begin{figure}[htbp]
    \centering
    \includegraphics[width=0.9\linewidth]{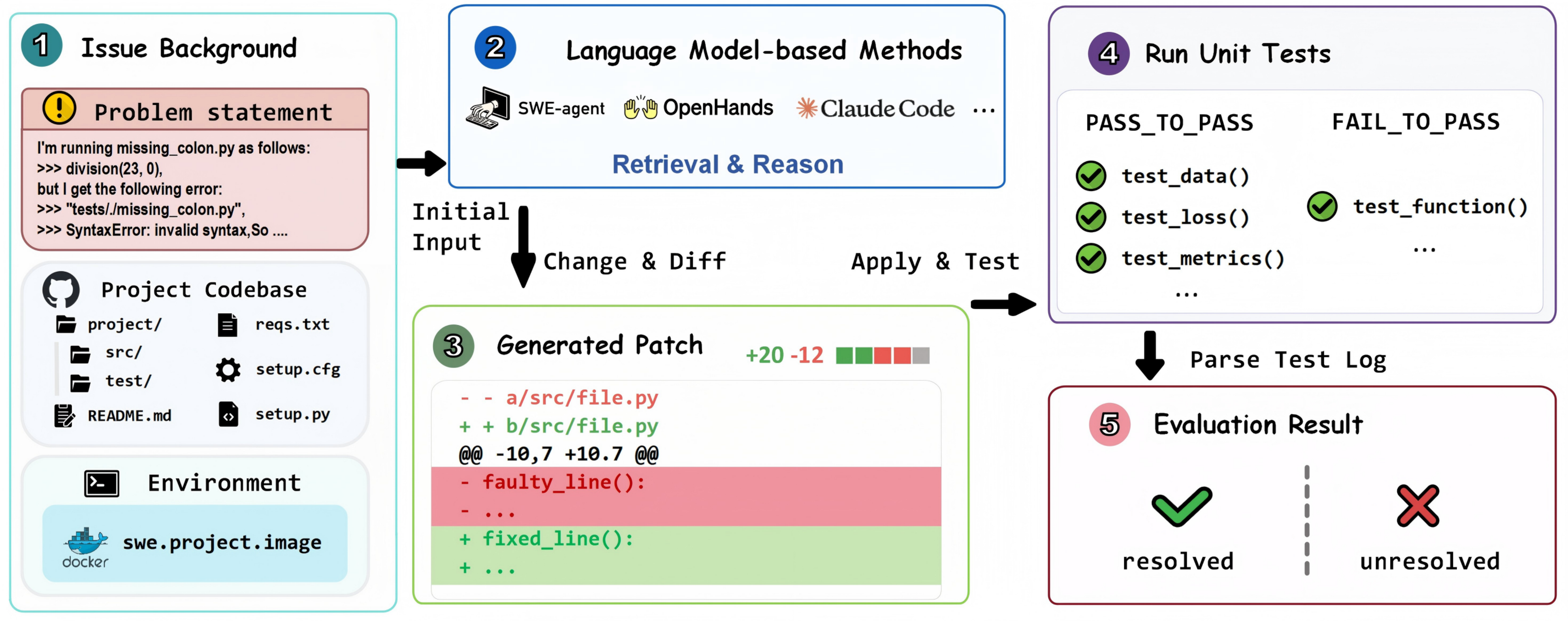}
    \caption{A typical workflow of automated issue resolution.}
    \label{fig:issue_resolution_overview}
\end{figure}

In modern software development, issue resolution refers to the process of addressing bug reports, feature requests, performance problems, or other maintenance tasks recorded in issue-tracking systems such as GitHub Issues. In the automated setting, the system is usually given an issue description and a repository snapshot, and it must produce a patch that can be applied to the repository and validated by tests or other checking mechanisms \cite{jimenez2023swe,jiang2025agenticSurvey}. This task is broader than conventional automated program repair (APR): it may involve not only bug fixing, but also feature implementation, refactoring, performance optimization, documentation-related changes, and multi-file repository evolution.

At a high level, Fig.~\ref{fig:issue_resolution_overview} illustrates the end-to-end workflow of automated issue resolution, from the issue description, project codebase, and execution environment, through retrieval and reasoning, to patch generation, unit-test execution, and final evaluation. The figure presents this process at a coarse-grained level; analytically, it can be further divided into several conceptual stages. First, issue comprehension interprets the natural-language report, which may be ambiguous, incomplete, or accompanied by logs and screenshots. Second, repository context construction collects task-relevant knowledge from the codebase, documents, dependency files, commit history, and issue discussions. Retrieval-augmented generation (RAG) is commonly used in this stage to retrieve files, functions, code snippets, summaries, or graph neighborhoods that can fit into the model context \cite{ouyang2024repograph,tawosi2025meta}. Third, localization identifies the files, classes, methods, or lines that are likely to require modification \cite{reddy2025swerank,jiang2025cosil}. Fourth, patch generation edits the localized code and produces one or more candidate patches. Finally, validation and selection use unit tests, linters, execution logs, reproduction cases, or LLM-based judging to filter incorrect patches and select the final submission \cite{wang2025aegis,cheshkov2024exploring}.

Current automated issue-resolution systems can be understood through two complementary technical paradigms. The first is a workflow-oriented paradigm, where the developer or framework designer explicitly decomposes the task into predefined modules such as localization, repair, and validation. Agentless-style pipelines and many RAG-based systems fall into this category: they reduce the model's search space through engineered prompts, retrieval rules, ranking procedures, and patch-selection heuristics \cite{xia2025demystifying,zhang2024autocoderover}. The advantage of this paradigm is controllability and lower exploratory cost, but its performance depends heavily on manually designed workflow knowledge and the quality of retrieved context.

The second is an agent-oriented paradigm, where an LLM-based agent interacts with a software environment more autonomously. Such agents can plan steps, invoke tools, inspect files, search code, edit repositories, run tests, read logs, and revise their strategy based on feedback \cite{yang2024swe,tao2024magis,wang2024openhands,arora2024masai}. In this paradigm, tools are not peripheral utilities but the interface through which the agent obtains external knowledge and verifies its actions. Typical tools include repository search, abstract syntax tree (AST) navigation, shell commands, version-control operations, test execution, build systems, static analyzers, and sometimes browser or multimodal interfaces. Compared with structured pipelines, pure agentic systems require less hand-specified step control but face greater challenges in context management, action selection, cost, and error recovery.

These paradigms are not mutually exclusive. Many state-of-the-art systems combine structured workflows with agentic tool use: they impose a high-level process while allowing the model to dynamically retrieve context, call tools, and refine patches through feedback. From the perspective of this survey, the key question behind both paradigms is not only which architecture is used, but what knowledge is made available to the model, how that knowledge is acquired, and how it is represented and applied during issue resolution.

\subsection{Benchmarks for the Issue-Resolution Task}

To facilitate rigorous and reproducible research on automated issue resolution, SWE-bench was introduced by Jimenez et al. \cite{jimenez2023swe}. Rather than evaluating isolated code snippets, SWE-bench formulates issue resolution as a repository-level task built from real GitHub issue--pull request pairs. Each task instance provides an issue description and the corresponding repository state before the fix, while the system must generate a patch that resolves the issue. A submission is considered successful when the patch can be applied and the associated tests pass, especially the fail-to-pass tests derived from the original developer fix.

The construction of SWE-bench is central to its realism. Its instances are mined from merged pull requests in popular open-source repositories, filtered to ensure that they are linked to real issues and include test changes, and then validated through execution to confirm that the tests fail before the patch and pass after the patch. This design makes SWE-bench substantially more demanding than earlier code-generation benchmarks: the input is large, the relevant context is often scattered across many files, the issue description may underspecify the root cause, and a correct patch must satisfy repository-level constraints rather than only local syntactic plausibility.

Following SWE-bench, the benchmark landscape has expanded rapidly. Later benchmarks refine data quality, broaden the task scope, or evaluate new dimensions of repository-level issue resolution. For example, SWE-bench Verified improves the reliability of evaluation instances, SWE-bench-java and multilingual variants extend evaluation beyond Python, SWE-bench Multimodal and CodeV introduce visual information for UI-related issues, and FEA-Bench focuses on repository-level feature implementation \cite{yang2024swe,zhang2024codev,li2025fea}. More recent benchmarks, such as SWE-bench Pro and SWE-Lancer, further emphasize more challenging and long-horizon issue-resolution instances, where agents are expected to handle more complex repository contexts, extended exploration processes, and realistic software-engineering constraints \cite{deng2025swe, miserendino2025swe}. These benchmarks collectively show that issue resolution is not a single narrow bug-fixing task, but a family of realistic software maintenance tasks with different languages, modalities, execution environments, and evaluation constraints.


\subsection{Comparison with Existing Surveys} 
The growth of benchmarks and methods has already motivated several related surveys. Existing issue-resolution surveys summarize the area from broad perspectives such as benchmark construction, agent frameworks, training-free and training-based methods, workflow phases, empirical analyses, and practical applications \cite{jiang2025agenticSurvey,li2026advancesFrontiers}. These surveys are valuable for mapping the overall field and explaining how agents, datasets, and training paradigms have evolved. However, their primary organizing dimensions are task resources and method architectures. 

Our survey is complementary to these works. Instead of asking only what benchmarks or agent frameworks exist, we focus on the knowledge mechanisms that enable automated issue resolution. Specifically, we study what kinds of knowledge systems use, where such knowledge is obtained and how it is extracted, how it is represented to LLMs, how it is utilized by LLMs during problem solving, and at which workflow stages it is applied. We also examine recent trends, representative methods, and high-performing SWE-bench Verified systems as descriptive lenses for understanding how knowledge is incorporated and organized in current approaches. This knowledge-centric perspective provides a systematic basis for comparing multi-source knowledge strategies and for identifying emerging knowledge sources, representation forms, usage mechanisms, and future research opportunities.


\section{Review Methodology}
\label{sec:methodology}
\begin{figure}
    \centering
    \includegraphics[width=0.9\linewidth]{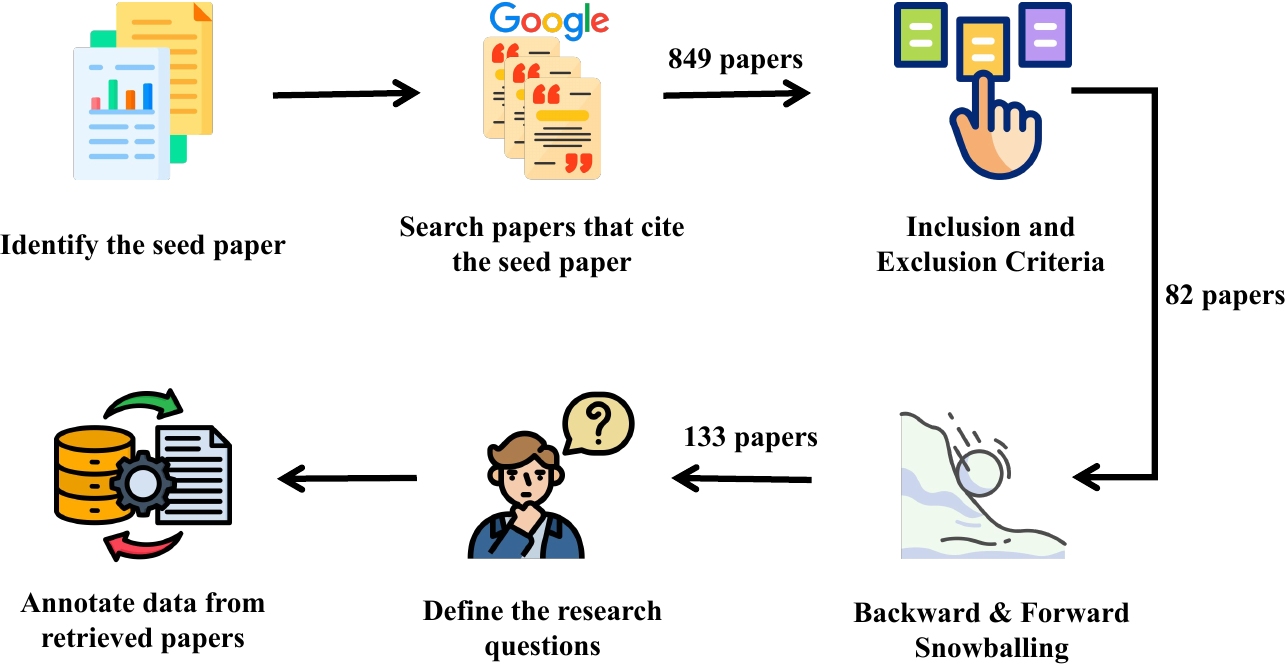}
    \caption{Overview of the process of paper collection and filtering.}
    \label{fig:method}
\end{figure}

To ensure a comprehensive and systematic review of automated issue resolution methods, we adopted a structured methodology illustrated in Fig.~\ref{fig:method}. Our process encompasses literature search, study selection, snowballing, research question definition, and knowledge coding.

\subsection{Literature Search and Selection Strategy}
We conducted a multi-stage literature search combining systematic citation querying with keyword verification. Initially, all papers citing the SWE-bench benchmark \cite{jimenez2023swe} were retrieved from Google Scholar, resulting in 849 studies. Applying the predefined inclusion and exclusion criteria (Section~\ref{sec:criteria}), 82 candidate papers were identified. To ensure comprehensive coverage, snowballing was applied to expand the candidate set to 133 studies (see Section~\ref{sec:snowballing}).

\subsection{Inclusion and Exclusion Criteria}
\label{sec:criteria}

\paragraph{\textbf{Inclusion Criteria}}
Papers were included if they:

\begin{itemize}
\item Address automated issue resolution at the repository or project level, including bug fixing, feature implementation, refactoring, performance optimization, or other maintenance-oriented issue tasks.
\item Propose methods, frameworks, or agents for automated issue resolution, with sufficient methodological details and empirical evaluation on relevant benchmarks.
\item Construct benchmarks or datasets for automated issue resolution, especially when such resources involve repository-level issues, executable evaluation, issue-fix pairs, or knowledge used for model training and performance evaluation.
\item Explicitly involve knowledge relevant to issue resolution, such as repository context, code structure, execution feedback, historical artifacts, documentation, tool interaction, expert guidance, or experiential trajectories.
\item Are full-length peer-reviewed articles or detailed technical preprints.
\end{itemize}

\paragraph{\textbf{Exclusion Criteria}}
Papers were excluded if they:

\begin{itemize}
\item Focus on unrelated tasks such as general code completion, code summarization, standalone test generation, or generic program analysis, unless these tasks are integrated into automated issue resolution methods or used to construct issue-resolution-related benchmarks or datasets.
\item Mention SWE-bench or automated issue resolution only superficially without proposing a relevant method, framework, agent, benchmark, or dataset.
\item Lack sufficient methodological, experimental, or dataset-construction detail for knowledge-oriented annotation.
\item Are duplicates, short abstracts, presentation-only materials, or non-technical reports.
\item Are inaccessible in full text or not written in English.
\end{itemize}

\subsection{Snowballing Procedure}
\label{sec:snowballing}
To ensure comprehensive coverage of relevant studies, we employed a systematic snowballing procedure, encompassing both backward and forward citation tracing:

\begin{enumerate}
    \item \textbf{Backward Snowballing:} For each of the 82 initially selected papers, reference lists were examined to identify prior studies relevant to automated issue resolution. Candidate references were added to the pool for further screening.
    
    \item \textbf{Forward Snowballing:} For each initially selected paper, papers that subsequently cited it were retrieved via Google Scholar. These citing papers were considered as potential candidates for inclusion.
    
    \item \textbf{Candidate Consolidation:} All papers identified through backward and forward snowballing were combined with the initial pool. Duplicate entries were removed to avoid redundancy.
    
    \item \textbf{Screening of Snowballed Papers:} Each snowballed paper underwent title and abstract screening, followed by full-text review if necessary. Inclusion and exclusion criteria were applied to determine relevance.
    
    \item \textbf{Expert Review and Verification:} A domain expert reviewed the snowballed set and suggested additional candidates, which were then screened and included according to the same criteria.
\end{enumerate}

Through this procedure, the candidate set expanded from 82 initially screened papers to 133 studies for subsequent research question coding and knowledge annotation.

\subsection{Collection Results and Statistics}
\label{sec:collection_stats}

\begin{figure}[htbp]
    \centering
    \begin{minipage}{0.55\linewidth}
        \centering
        \includegraphics[width=\linewidth]{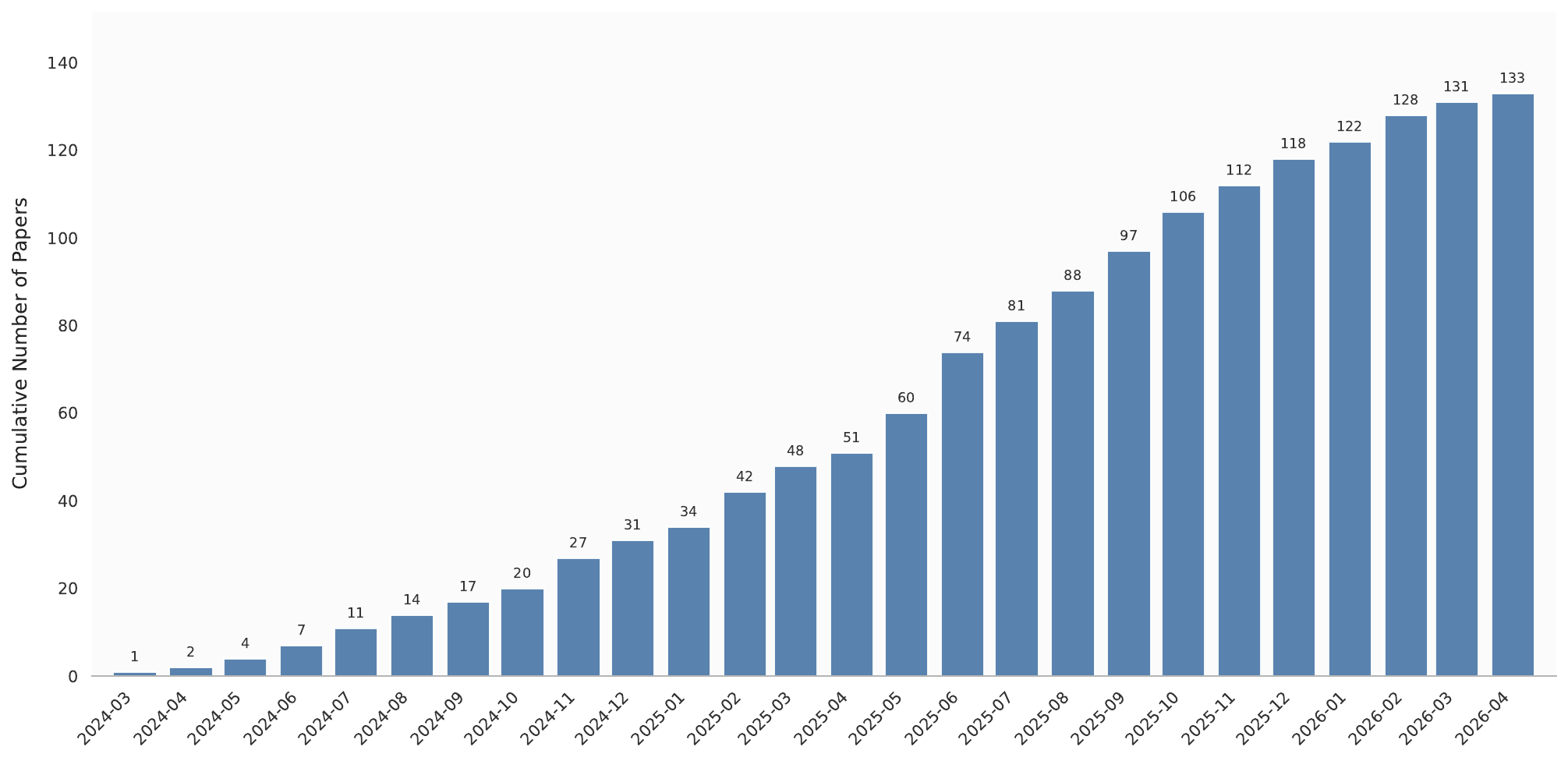}
        \caption{Monthly cumulative number of publications related to automated issue resolution.}
        \label{fig:monthly_publication_trend}
    \end{minipage}
    \hfill
    \begin{minipage}{0.40\linewidth}
        \centering
        \includegraphics[width=\linewidth]{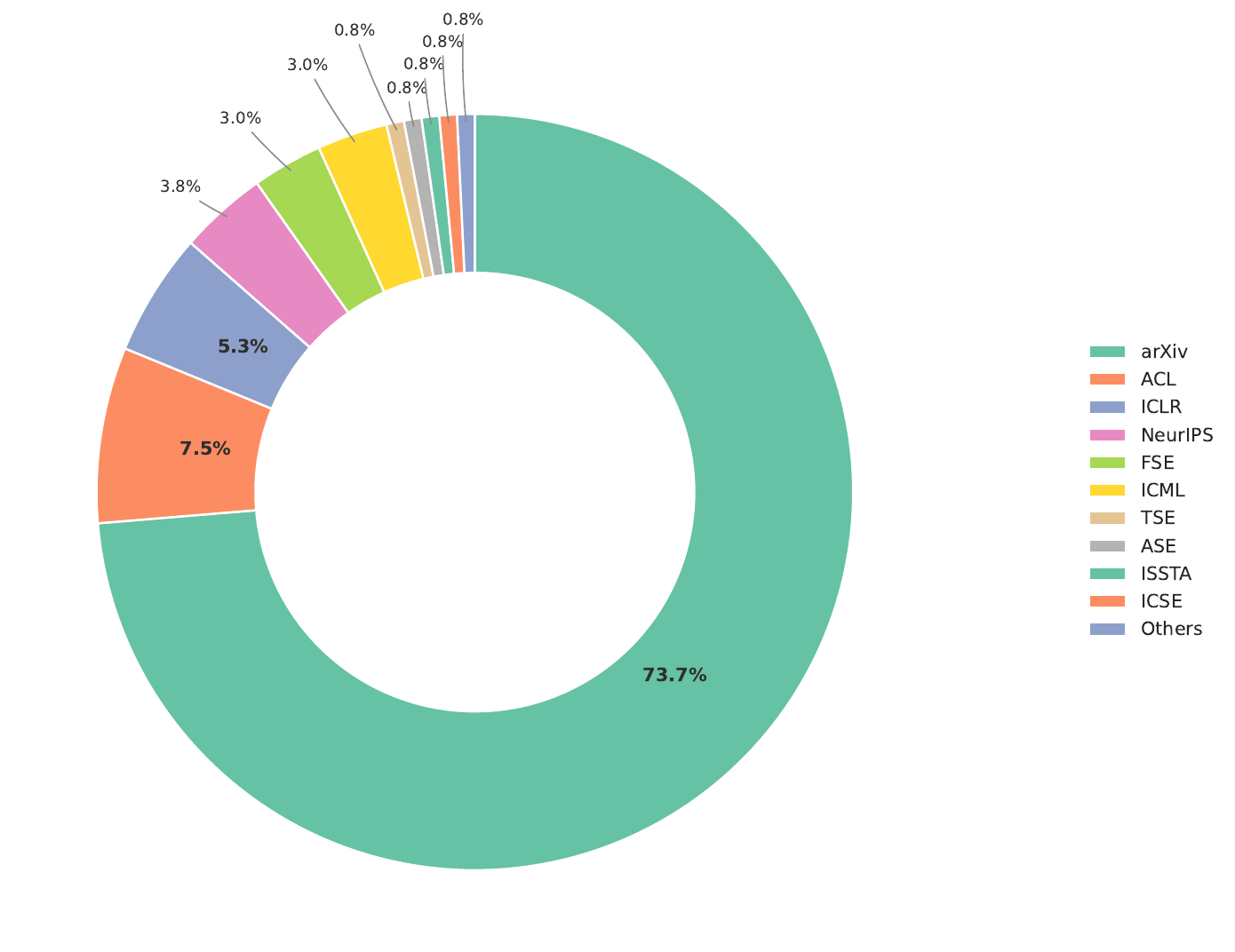}
        \caption{Distribution of publication venues for papers related to automated issue resolution.}
        \label{fig:venue_distribution}
    \end{minipage}
\end{figure}

Fig.~\ref{fig:monthly_publication_trend} illustrates the cumulative number of publications on automated issue resolution at a monthly granularity. The literature exhibits a progressive growth trajectory from March 2024 to April 2026, beginning with a limited number of publications per month and accelerating markedly after mid-2024. In particular, a pronounced surge is observed between May 2024 and October 2025, reflecting heightened research activity driven by the increasing adoption of Large Language Models and agent-oriented frameworks for repository-level issue resolution. This temporal pattern highlights the recency and rapid expansion of the field, underscoring the necessity of a systematic survey to synthesize and contextualize emerging contributions.

Fig.~\ref{fig:venue_distribution} presents the distribution of publication venues for the included studies. A substantial proportion of the works (73.7\%) were initially disseminated via arXiv, indicating that preprint publication remains a prevalent mode for rapid knowledge exchange in this nascent domain. The remaining studies are distributed across premier conferences and journals in artificial intelligence and software engineering, including ACL (7.5\%), ICLR (5.3\%), NeurIPS (3.8\%), ICML (3.0\%), FSE (3.0\%), TSE (0.8\%), ASE (0.8\%), ISSTA (0.8\%), and ICSE (0.8\%). This distribution evidences the interdisciplinary nature of automated issue resolution research, encompassing contributions from both AI/NLP and software engineering communities, and reflects a balance between rapid dissemination and formal peer-reviewed validation.

Collectively, these data provide a comprehensive overview of the temporal evolution and venue preferences within the field, establishing a contextual foundation for the subsequent knowledge-centric analysis.

\subsection{Data Coding and Taxonomy Construction}
\label{sec:data_coding}

To construct a knowledge-centric taxonomy grounded in the surveyed literature, we conducted an iterative qualitative coding process over the 133 selected studies. Rather than imposing a fully predefined taxonomy at the beginning, we used open coding to identify recurring knowledge-use phenomena in automated issue resolution, and then consolidated these observations through axial coding and selective coding. The resulting framework and coding schema are presented in the following subsections.

The coding process consisted of four stages.

\begin{enumerate}
\item \textbf{Open Coding.}
We first reviewed the selected studies to identify concrete knowledge-use phenomena as they appeared in the literature. At this stage, we did not force observations into a predefined category. Instead, we annotated recurring forms of knowledge used in issue resolution, such as retrieved code snippets, repository structures, test outputs, execution logs, historical commits, issue-fix pairs, pull requests, tool interactions, agent trajectories, reflective memories, architectural constraints, documentation, technology-stack information, and expert-designed workflows. This step ensured that the taxonomy was grounded in actual system designs and empirical practices rather than only in prior conceptual assumptions.

\item \textbf{Axial Coding.}
We then compared, merged, and organized the initial codes into higher-level conceptual categories. Codes related to language rules, domain semantics, and framework or platform constraints were grouped as background-oriented knowledge. Codes related to current repository artifacts, architectural constraints, and historical repository changes were grouped as repository-oriented knowledge. Codes related to tool use, execution feedback, repair trajectories, and reusable problem-solving experience were grouped as procedural knowledge. This stage reduced redundancy among open codes and produced the three-layer knowledge hierarchy used in this survey.

\item \textbf{Selective Coding.}
After the major knowledge categories became stable, we further aligned them with the analytical dimensions needed for a systematic review. For each knowledge instance, we examined not only what type of knowledge was used, but also where it came from, how it was extracted, how it was represented, how it was operationalized by LLM-based systems, and at which issue-resolution stage it was applied. This step led to the finalized coding schema covering six dimensions: knowledge types, knowledge sources and extraction methods, representation formats, usage methods, application stages, and temporal evolution.

\item \textbf{Full-Corpus Coding and Synthesis.}
Using the finalized schema, we coded all 133 studies in a consistent manner. Since a single study may use multiple knowledge types, sources, representations, usage methods, or application stages, we adopted multi-label coding where appropriate. The coded results were then synthesized through frequency analysis, cross-dimensional comparison, representative-work analysis, and temporal trend analysis. These results form the empirical basis for the RQ analyses reported in the remainder of the paper.
\end{enumerate}

This coding process makes the proposed taxonomy both inductively grounded and analytically consistent. Open coding allowed the categories to emerge from the literature, while axial and selective coding transformed scattered observations into a unified framework and reproducible coding schema. The following subsections present the resulting knowledge-centered cognitive framework, the operational definitions of the coding schema, and the research questions derived from this framework.

\subsection{A Knowledge-Centric Cognitive Framework and Coding Schema}
\label{sec:framework_schema}

\begin{figure}[htbp]
\centering
\includegraphics[width=0.9\linewidth]{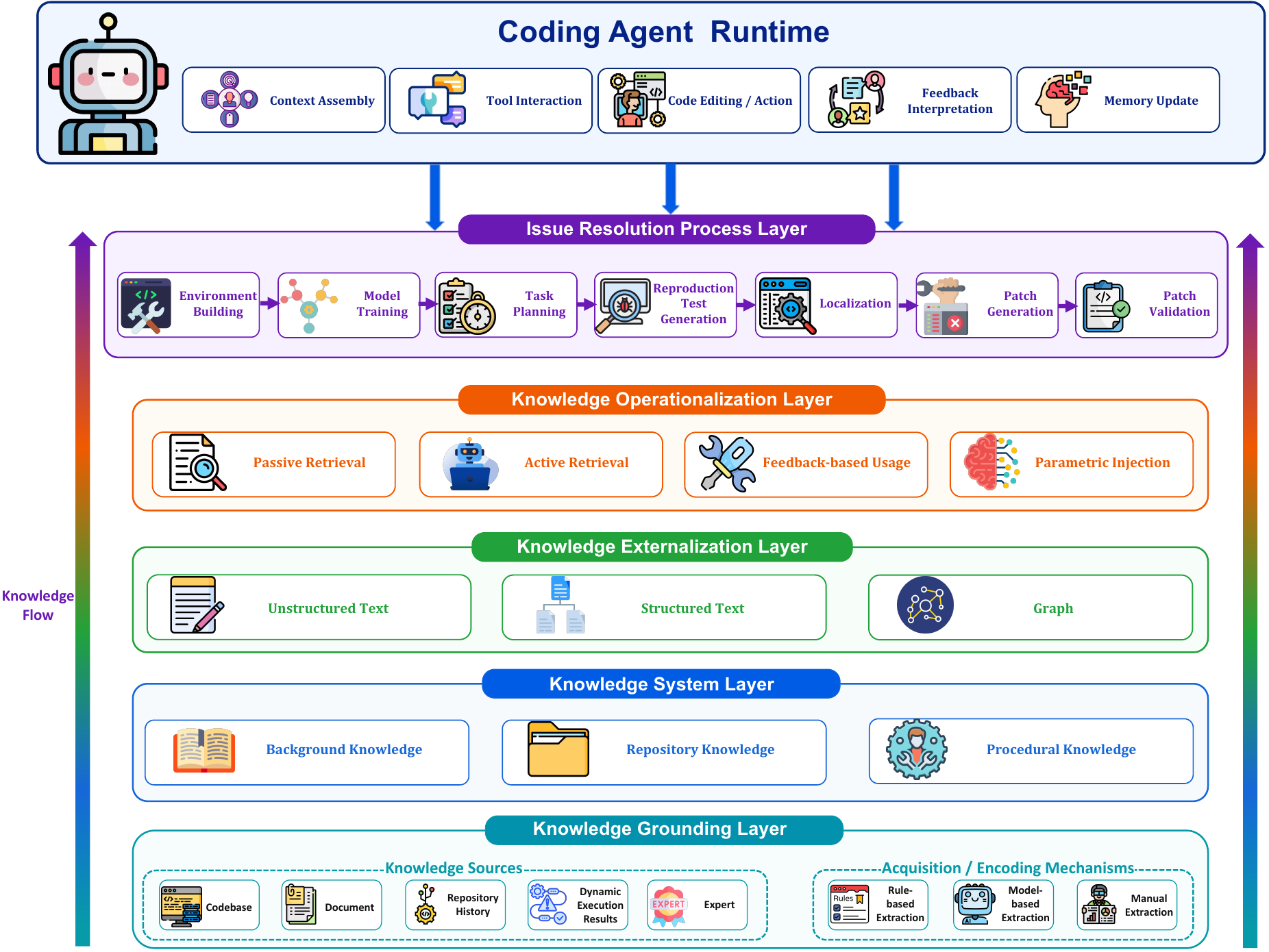}
\caption{A knowledge-centric cognitive framework and coding schema for coding agents in automated issue resolution.}
\label{fig:framework}
\end{figure}

The knowledge-centric cognitive framework in Fig.~\ref{fig:framework} is formed from the knowledge definitions, category distinctions, and coding dimensions adopted in this survey. Its layered organization follows how knowledge is identified, transformed, and used in automated issue resolution. Knowledge is first anchored in concrete sources, including software artifacts, execution environments, repository histories, documents, and expert guidance. It is then distinguished according to its functional relationship with the target task and repository, externalized into forms accessible to agents, operationalized through retrieval, feedback, or learning mechanisms, and finally applied to specific stages of the issue-resolution process. In this way, the framework reflects the knowledge concepts and analytical dimensions used in this survey, and provides the schema-level basis for subsequent literature coding and RQ analyses.

The layering of the framework is based on the functional role that knowledge plays in automated issue resolution. At the bottom, the Knowledge Grounding Layer explains where knowledge comes from and how raw artifacts are transformed into usable evidence, thereby capturing the acquisition basis of agent knowledge. The Knowledge System Layer then distinguishes knowledge according to its relationship with the target task and repository: Background Knowledge provides project-independent semantic and technical assumptions, Repository Knowledge grounds reasoning in the target codebase, architecture, and evolution, and Procedural Knowledge supports tool-mediated execution, feedback interpretation, and experience reuse. Above this layer, the Knowledge Externalization Layer describes how knowledge is represented in agent-accessible forms, the Knowledge Operationalization Layer explains how represented knowledge becomes actionable through retrieval, feedback, or learning mechanisms, and the Issue Resolution Process Layer specifies where knowledge is applied across the resolution workflow. This organization provides the rationale for the subsequent formal definitions of knowledge sources, knowledge types, representation formats, usage methods, and application stages.

\subsubsection{Knowledge Grounding Layer: Sources and Extraction Methods}
\label{sec:knowledge_grounding_schema}

The Knowledge Grounding Layer explains how knowledge is anchored in concrete software artifacts, execution environments, and development experience. In our coding schema, this layer is specified through two dimensions: knowledge sources and extraction methods. Knowledge sources describe where task-relevant knowledge originates, while extraction methods describe how raw materials are transformed into usable knowledge for automated issue resolution.

\noindent\textbf{Knowledge Sources.}
Knowledge sources refer to the origins from which automated issue-resolution systems acquire task-relevant information. In this survey, knowledge sources cover both project-internal artifacts and external or human-provided materials. They include the current codebase, documents, repository history, dynamic execution results, and expert input. Together, these sources form the grounding basis of issue-resolution knowledge: the codebase provides implementation-level evidence, documents provide semantic and explanatory context, repository history provides temporal development evidence, dynamic execution results provide behavior-grounded feedback, and expert input provides human-designed methodological guidance.

\begin{itemize}
    \item \textbf{Codebase.}
    Codebase refers to the source code, repository structure, dependency relations, configuration files, tests, and development interfaces embodied in a software project. It provides the most direct repository-grounded evidence for localization, patch generation, and consistency checking. It may also expose architectural signals, tool-facing environments, technology-stack clues, and project-specific development practices.

    \item \textbf{Document.}
    Document refers to textual or multimodal materials that describe software requirements, system behavior, usage rules, development conventions, or external environments. This category includes README files, API manuals, design specifications, tutorials, benchmark task descriptions, issue documents, screenshots with textual context, and human-written specifications. Documents often provide semantic explanations about what the system is expected to do and how developers should interact with it.

    \item \textbf{Repository History.}
    Repository History refers to historical development artifacts accumulated during software evolution. This category includes commits, pull requests, issue-fixing records, historical patches, test changes, version changes, issue-fix links, and other evolution traces. Compared with the codebase, which mainly reflects the current repository state, repository history provides temporal evidence about how the system has changed, why certain modifications were introduced, and how similar issues have been resolved before.

    \item \textbf{Dynamic Execution Results.}
    Dynamic Execution Results refer to runtime signals generated when an automated issue-resolution system interacts with an executable software environment. This category includes test outputs, runtime logs, error traces, command-line responses, tool observations, debugging information, and trajectory-level feedback. It provides behavior-grounded evidence about whether an intermediate action, generated test, or candidate patch is actually valid.

    \item \textbf{Expert.}
    Expert refers to human-designed or expert-like guidance injected into automated issue-resolution systems. This category includes curated workflows, handcrafted heuristics, expert feedback, process-level supervision, collaboration rules, evaluation criteria, and manually designed tool-use strategies. Unlike artifact-based sources, expert sources encode methodological knowledge about how software engineers diagnose issues, use tools, reason about constraints, and coordinate resolution actions.
\end{itemize}

\noindent\textbf{Extraction Methods.}
Extraction methods refer to the mechanisms through which raw materials from knowledge sources are transformed into usable knowledge. In automated issue resolution, extraction methods determine how heterogeneous artifacts are obtained, processed, organized, or inferred before they can be used by LLMs, agents, tools, or downstream reasoning components. We identify three extraction categories: rule-based extraction, model-based extraction, and manual extraction. These methods respectively emphasize deterministic processing, learned semantic inference, and human or expert judgment.

\begin{itemize}
    \item \textbf{Rule-based Extraction.}
    Rule-based Extraction refers to methods that acquire knowledge through predefined procedures, deterministic heuristics, static or dynamic analysis, parsers, tool-specific rules, or explicit processing pipelines. This method is especially suitable for knowledge embedded in artifacts with explicit structures, stable formats, or well-defined operational semantics, such as source files, call relations, execution logs, test outputs, and tool responses.

    \item \textbf{Model-based Extraction.}
    Model-based Extraction refers to methods that use learned models to infer, retrieve, summarize, rank, or transform knowledge from software artifacts. This category includes embedding models, LLMs, multimodal LLMs, reward models, and trained agent policies. It is often used when the target knowledge is semantically implicit, noisy, cross-modal, or difficult to formalize through handcrafted rules.

    \item \textbf{Manual Extraction.}
    Manual Extraction refers to the process of obtaining knowledge through human inspection, annotation, curation, expert-designed specifications, or qualitative judgment. This method is mainly used when the target knowledge requires expert interpretation, careful benchmark design, or manual validation, such as high-quality trajectory curation, environment construction, architectural constraint specification, or domain-specific annotation.
\end{itemize}

\subsubsection{Knowledge System Layer: Knowledge Type Taxonomy}
\label{sec:knowledge_type_taxonomy}

The Knowledge System Layer organizes issue-resolution knowledge into three categories used throughout this survey: Background Knowledge, Repository Knowledge, and Procedural Knowledge. This hierarchy is organized according to the functional relationship between knowledge and the issue-resolution process. Background Knowledge provides project-independent semantic and technical assumptions for understanding issues; Repository Knowledge grounds the agent in the target codebase, its architectural organization, and its historical evolution; and Procedural Knowledge supports tool-mediated execution, feedback interpretation, and experience reuse.

\noindent\textbf{L1: Background Knowledge.}
Background Knowledge refers to project-independent knowledge that provides the general semantic, technical, and language-level assumptions required to interpret issue descriptions and understand the broader problem context. It helps agents establish the semantic boundary of a task before detailed repository reasoning, connecting user-facing intent with programming semantics, domain rules, and technology-stack conventions. This layer includes Programming Language Knowledge, Domain-Specific Knowledge, and Technology Stack Knowledge.

\begin{itemize}
    \item \textbf{Programming Language Knowledge.}
    Programming Language Knowledge is knowledge of language syntax, type systems, standard semantics, runtime behavior, and language-specific idioms required to produce executable and idiomatic patches. It provides the basic formal rules under which generated code must operate. It is separated from repository knowledge because it constrains code changes across projects that use the same programming language.

    \item \textbf{Domain-Specific Knowledge.}
    Domain-Specific Knowledge refers to concepts, principles, rules, constraints, and correctness criteria that originate from a specialized application domain and hold independently of any particular software repository. It is required when issue resolution depends on understanding domain semantics that cannot be inferred solely from programming-language semantics, technology-stack conventions, or repository artifacts. Such knowledge may arise from specialized scientific or engineering domains, such as quantum physics and bioinformatics, where evaluating a patch requires domain-specific reasoning beyond general software knowledge.

    \item \textbf{Technology Stack Knowledge.}
    Technology Stack Knowledge is knowledge of external libraries, frameworks, platforms, APIs, dependency ecosystems, and runtime environments that shape how software systems are developed and maintained. It differs from Programming Language Knowledge because it focuses not on general language syntax or semantics, but on external technical constraints introduced by concrete software stacks, such as framework conventions, API changes, dependency versions, and platform-specific requirements.
\end{itemize}

\noindent\textbf{L2: Repository Knowledge.}
Repository Knowledge refers to project-specific knowledge that grounds the agent in the target codebase, its architectural organization, and its historical evolution. It provides the central bridge between general task understanding and concrete resolution actions. This layer connects the physical artifacts of a repository, such as files, functions, classes, tests, and configurations, with conceptual and temporal dimensions, including module responsibilities, design rationales, architectural constraints, historical decisions, and issue-fix trajectories. It includes Existing Code Knowledge, Design Architecture Knowledge, and Repository Evolution Knowledge.

\begin{itemize}
    \item \textbf{Existing Code Knowledge.}
    Existing Code Knowledge is knowledge of the concrete source code already present in the target repository, including files, functions, classes, methods, code snippets, interfaces, call sites, and structural dependencies among code entities. It provides the artifact-oriented backbone of issue resolution. Its role is not limited to exposing editable code fragments; it also serves as the physical substrate for repository-level reasoning, localization, patch generation, and consistency checking.

    \item \textbf{Design Architecture Knowledge.}
    Design Architecture Knowledge is high-level conceptual knowledge that captures the design intent, component responsibilities, architectural patterns, module-level organization, and design constraints of a software system. It explains not only how the system is structurally organized, but also why particular responsibilities, dependencies, and boundaries are arranged in a certain way. It helps agents preserve architectural assumptions and responsibility boundaries during issue resolution.

    \item \textbf{Repository Evolution Knowledge.}
    Repository Evolution Knowledge is temporal knowledge derived from the historical evolution of a software repository, including past issues, pull requests, commits, patches, version changes, issue-fix links, historical discussions, and development-process traces. It captures how the repository has changed over time, why certain changes were made, what constraints or conventions they introduced, and how previous modifications shaped the current repository state. It provides temporal traceability for repository-level reasoning.
\end{itemize}

\noindent\textbf{L3: Procedural Knowledge.}
Procedural Knowledge refers to process-level knowledge that enables agents to operationalize, validate, and refine knowledge during issue resolution. Unlike Background Knowledge, which provides project-independent assumptions, and Repository Knowledge, which grounds reasoning in the target repository, Procedural Knowledge focuses on how agents act upon the repository, interact with development environments, interpret feedback, and accumulate reusable problem-solving experience. This layer includes Development Tool Knowledge and Experiential Knowledge.

\begin{itemize}
    \item \textbf{Development Tool Knowledge.}
    Development Tool Knowledge is procedural knowledge about how to use software development tools, execution environments, and agent-specific tool interfaces during issue resolution. It does not primarily describe code entities, dependencies, or retrieved repository context themselves, but instead describes how agents operate on the repository through concrete tools, commands, feedback channels, and interaction protocols. It covers search tools, editors, shell commands, build systems, test runners, debuggers, static analyzers, validation tools, and tool-control strategies.

    \item \textbf{Experiential Knowledge.}
    Experiential Knowledge is reusable process-level knowledge distilled from prior or ongoing issue-resolution experiences, including successful trajectories, failed attempts, intermediate decisions, expert feedback, reasoning traces, tool-use patterns, validation outcomes, and repair strategies. It captures how previous problem-solving processes were carried out, why certain actions succeeded or failed, and what lessons can be reused to guide future localization, patch generation, tool use, validation, and self-correction.
\end{itemize}

To reduce category overlap during coding, we further clarify several boundaries. Programming Language Knowledge is separated from Technology Stack Knowledge because the former concerns general language-level rules, while the latter concerns concrete libraries, frameworks, platforms, APIs, dependency versions, and runtime ecosystems. Existing Code Knowledge is separated from Development Tool Knowledge because retrieved files, code snippets, call sites, and dependency relations describe what the repository contains, whereas tool knowledge describes how agents operate on the repository through tools and environments. Repository Evolution Knowledge is separated from Experiential Knowledge because historical issues, commits, pull requests, and patches are coded as repository-derived temporal evidence, while trajectories, memories, failed attempts, reflective summaries, and reusable repair strategies are coded as process-level experience. Design Architecture Knowledge is separated from Existing Code Knowledge because it captures design intent, responsibility allocation, architectural constraints, and conceptual organization rather than only concrete implementation artifacts.

\subsubsection{Knowledge Externalization Layer: Representation Formats}
\label{sec:knowledge_representation_schema}

The Knowledge Externalization Layer captures how organized knowledge is transformed into forms that can be accessed, retrieved, inspected, and reasoned over by agents. In current issue-resolution systems, externalized knowledge mainly appears in three dominant representation formats: Unstructured Text, Structured Text, and Graph. These formats differ in their degree of structural explicitness and support different forms of repository reasoning, context construction, and workflow control.

\begin{itemize}
    \item \textbf{Unstructured Text.}
    Unstructured Text refers to knowledge represented as natural-language descriptions, raw textual artifacts, or free-form records without explicit fields, schemas, or graph structures. This format preserves rich semantic context and is naturally compatible with LLM prompts. Examples include issue reports, documentation fragments, code comments, file contents, localized code snippets, terminal outputs, compilation errors, test logs, runtime exceptions, tool responses, candidate patches, and free-form reasoning trajectories.

    \item \textbf{Structured Text.}
    Structured Text refers to knowledge represented in textual forms with explicit fields, templates, lists, records, tables, checklists, rule templates, key-value summaries, JSON/YAML-like formats, or hierarchical organization. This format makes knowledge easier to parse, control, retrieve, compare, compose, and reuse. It preserves textual compatibility with LLMs while improving controllability and machine operability.

    \item \textbf{Graph.}
    Graph refers to knowledge represented through nodes and edges, where software entities, repository artifacts, issue elements, execution relations, dependencies, or reasoning states are explicitly connected. This format supports structural navigation, path-guided retrieval, multi-hop reasoning, dependency analysis, and architecture-aware planning. Examples include call graphs, dependency graphs, data-flow graphs, state-transition graphs, repository-aware knowledge graphs, and architecture-aware planning graphs.
\end{itemize}

These three formats are treated as the dominant representation categories observed in the current literature. Other emerging forms, such as DSL-like specifications, SDD-like structures, or skill-oriented procedural representations, are discussed later as potential extensions rather than as primary categories in the current coding schema.

\subsubsection{Knowledge Operationalization Layer: Usage Methods}
\label{sec:knowledge_usage_schema}

The Knowledge Operationalization Layer describes the mechanisms through which represented knowledge becomes actionable during issue resolution. In our coding schema, usage methods capture how knowledge moves from being stored or represented to being used for context construction, exploration, hypothesis revision, action selection, patch refinement, validation, and long-term improvement. We identify four usage methods: Passive Retrieval, Active Retrieval, Feedback-Based Usage, and Parametric Injection.

\begin{itemize}
    \item \textbf{Passive Retrieval.}
    Passive Retrieval refers to the usage method in which task-relevant knowledge is supplied to the LLM as external context, either by directly inserting selected information into the prompt or by retrieving relevant artifacts through predefined similarity-based, rule-based, or structured mechanisms before generation. The defining characteristic is that the LLM does not actively plan knowledge-seeking actions; instead, it consumes knowledge that has already been selected, retrieved, summarized, or organized by the system.

    \item \textbf{Active Retrieval.}
    Active Retrieval refers to the usage method in which the LLM or agent actively plans and triggers knowledge-seeking actions during issue resolution. Instead of only consuming pre-selected context, the agent may decide which files, code entities, historical artifacts, documents, memories, or external resources should be accessed according to its evolving understanding of the issue.

    \item \textbf{Feedback-Based Usage.}
    Feedback-Based Usage refers to the usage method in which the LLM obtains and applies knowledge by interacting with the development or execution environment. This method includes running code, generating tests, reading execution logs, observing runtime states, invoking non-search tools, and revising subsequent actions based on environment responses. It emphasizes dynamic feedback loops rather than retrieval of pre-existing artifacts.

    \item \textbf{Parametric Injection.}
    Parametric Injection refers to the usage method in which knowledge is internalized into model parameters, learned skills, agent policies, reward models, or long-term memory mechanisms. This method is typically realized through training, post-training, reinforcement learning, trajectory learning, or persistent memory construction. It allows knowledge to influence future issue resolution through learned model or agent capabilities rather than only through external context.
\end{itemize}

\subsubsection{Issue Resolution Process Layer: Application Stages}
\label{sec:issue_resolution_stage_schema}

The Issue Resolution Process Layer identifies where operationalized knowledge is applied in automated issue resolution. Following our stage definition, we code knowledge application across seven stages: Environment Building, Model Training, Task Planning, Reproduction Test Generation, Localization, Patch Generation, and Patch Validation. These stages provide a common analytical workflow rather than a strict execution order; not every system explicitly covers all stages, and some stages may be merged, skipped, or iterated in specific implementations.

\begin{itemize}
    \item \textbf{Environment Building.}
    Environment Building refers to constructing, configuring, or preparing executable repositories, dependencies, sandboxes, containers, build systems, test environments, and tool environments. This stage ensures that the issue-resolution system can interact with the target repository under an executable and tool-accessible setting.

    \item \textbf{Model Training.}
    Model Training refers to using issue-resolution data, historical fixes, trajectories, reward signals, execution feedback, or other training resources to improve model or agent capabilities. This stage includes pre-training, fine-tuning, reinforcement learning, reward modeling, trajectory learning, and other training processes that enhance issue-resolution performance.

    \item \textbf{Task Planning.}
    Task Planning refers to interpreting the issue, decomposing the task, selecting strategies, organizing workflows, assigning sub-tasks, or deciding how the agent should proceed before concrete repair actions. This stage provides the process-level structure for multi-step issue resolution and helps agents determine what information to inspect, which tools to use, and how to organize subsequent actions.

    \item \textbf{Reproduction Test Generation.}
    Reproduction Test Generation refers to generating, selecting, refining, or reusing tests to reproduce the reported issue. This stage provides executable evidence for understanding the bug, validating the existence of the reported behavior, and guiding subsequent patch generation.

    \item \textbf{Localization.}
    Localization refers to identifying issue-relevant files, classes, functions, methods, statements, code regions, dependency paths, or repository structures that may require inspection or modification. This stage narrows the search space from the full repository to candidate locations where the issue may originate or where the patch should be applied.

    \item \textbf{Patch Generation.}
    Patch Generation refers to editing repository artifacts and producing one or more candidate patches based on localized context, constraints, retrieved knowledge, and repair strategy. This stage transforms the agent's understanding of the issue and repository into concrete code modifications.

    \item \textbf{Patch Validation.}
    Patch Validation refers to evaluating candidate patches through tests, execution results, logs, static checks, verifier feedback, or other correctness signals. This stage judges whether the generated patch resolves the issue and whether it avoids introducing regressions or violating repository-level constraints.
\end{itemize}

\subsubsection{Coding Agent Runtime}
\label{sec:coding_agent_runtime}

The Coding Agent Runtime represents the situated execution context in which the above knowledge layers are finally used by an automated issue-resolution agent. After knowledge is grounded in software artifacts and development contexts, organized into knowledge systems, externalized into agent-accessible representations, operationalized through retrieval, feedback, or learning mechanisms, and associated with issue-resolution stages, it is ultimately consumed and acted upon during agent execution. In this runtime, knowledge is no longer treated as a static analytical category; rather, it becomes part of the agent's concrete problem-solving process.

Within this runtime, the agent assembles relevant knowledge into its working context, interacts with tools and environments, edits code or performs other repository-level actions, interprets feedback, and updates memory. Context assembly determines which knowledge enters the agent's limited working context. Tool interaction enables the agent to inspect repositories, query external information, execute commands, run tests, and gather dynamic evidence. Code editing or action execution transforms knowledge into concrete modifications or operational decisions. Feedback interpretation allows the agent to diagnose failures, validate intermediate results, and revise subsequent actions. Memory update further enables the agent to preserve useful experience that may support future issue-resolution tasks.

Therefore, the Coding Agent Runtime is not treated as an additional coding dimension in this survey. Instead, it describes the runtime setting in which the coded dimensions interact and become executable. The preceding layers define what knowledge is available, where it comes from, how it is represented, how it is operationalized, and where it is applied; the runtime layer explains how these dimensions are enacted together when an agent performs issue resolution. This completes the knowledge-centric framework and provides the basis for formulating the research questions in the next subsection.

\subsection{Research Questions}
\label{sec:rqs}

Based on the knowledge-centric cognitive framework and coding schema defined in Section~\ref{sec:framework_schema}, we formulate six research questions to systematically analyze knowledge utilization in automated issue resolution. Together, these questions operationalize the major analytical dimensions of the framework, covering knowledge types, sources and extraction methods, representation formats, usage methods, application stages, and temporal evolution. Rather than representing a strict sequential process, they provide complementary perspectives on how knowledge is grounded, organized, externalized, operationalized, applied, and evolved in automated issue-resolution systems. In this way, the research questions transform the proposed framework into an empirical review protocol.

\begin{itemize}
\item \textbf{RQ1:} What types of knowledge are used in issue resolution?
\item \textbf{RQ2:} What are the sources and extraction methods of knowledge?
\item \textbf{RQ3:} How is knowledge represented for automated issue resolution?
\item \textbf{RQ4:} How is knowledge used by LLMs in issue resolution systems?
\item \textbf{RQ5:} At which stages is knowledge applied in issue resolution?
\item \textbf{RQ6:} What recent trends and dominant patterns characterize knowledge usage in automated issue resolution?
\end{itemize}

\begin{figure}[htbp]
\centering
\includegraphics[width=0.9\linewidth]{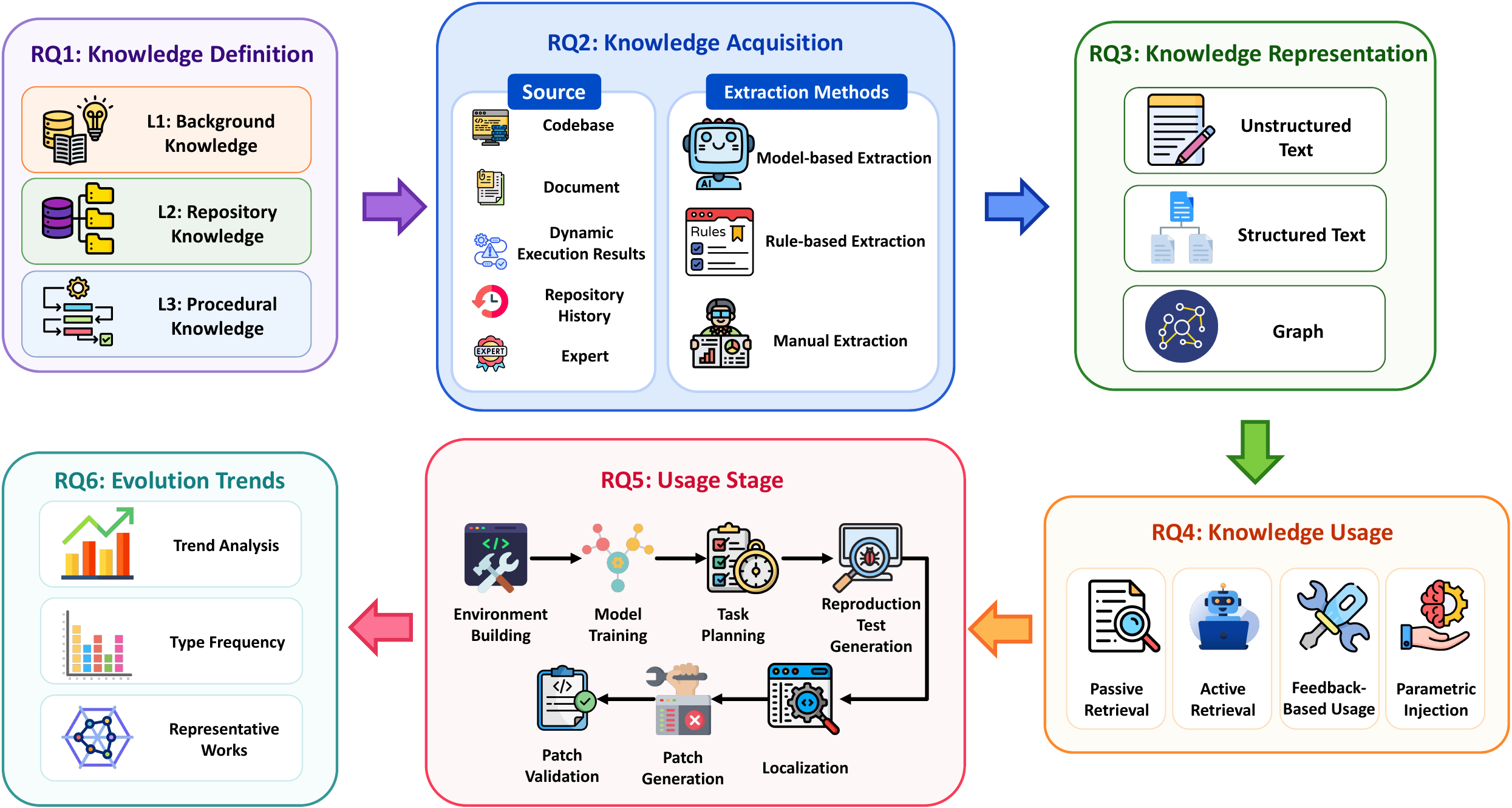}
\caption{Knowledge lifecycle overview for automated issue resolution.}
\label{fig:RQs}
\end{figure}

As illustrated in Fig.~\ref{fig:RQs}, the six research questions correspond to different analytical dimensions of the proposed framework. RQ1 focuses on the Knowledge System Layer and examines the distribution of knowledge types. RQ2 focuses on the Knowledge Grounding Layer and analyzes where knowledge originates and how it is extracted. RQ3 corresponds to the Knowledge Externalization Layer and studies how knowledge is represented for LLMs, agents, tools, and downstream reasoning components. RQ4 corresponds to the Knowledge Operationalization Layer and examines how knowledge is used by LLMs during issue resolution. RQ5 connects knowledge utilization with the Issue Resolution Process Layer by analyzing where knowledge is applied across the workflow. RQ6 further provides an evolutionary and synthetic perspective by examining temporal changes, representative systems, dominant patterns, and emerging directions across the preceding dimensions.

Together, these research questions enable a systematic analysis of knowledge use in automated issue resolution. Rather than treating knowledge as static context supplied to LLMs, they examine how knowledge is grounded, organized, externalized, used, applied, and evolved within coding-agent-based issue-resolution systems.







\section{RQ1: What types of knowledge are used in issue resolution?}
\label{sec:rq1}

\begin{figure}[htbp]
    \centering
    \includegraphics[width=0.5\linewidth]{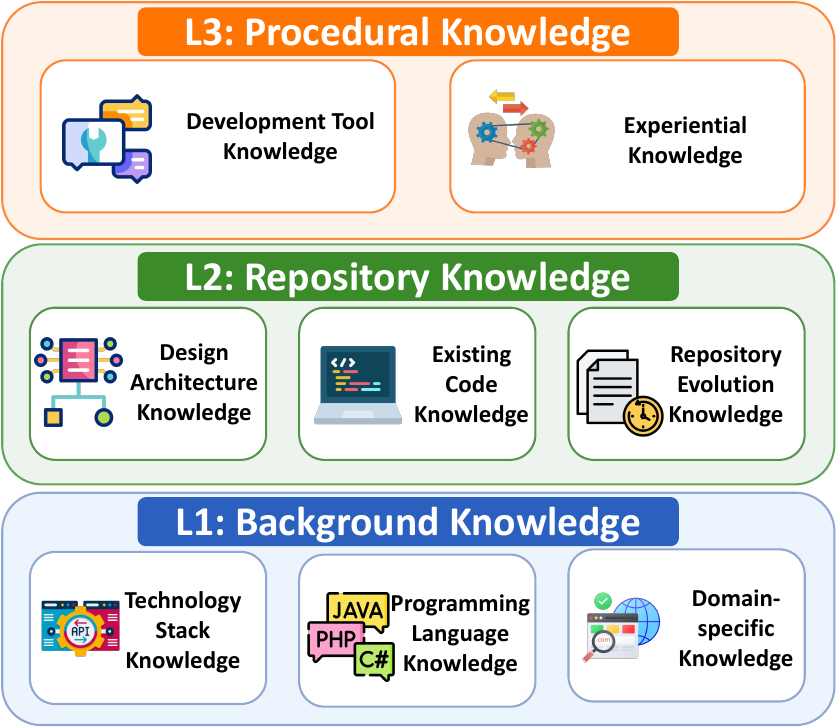}
    \caption{The three-layer hierarchy of knowledge in issue resolution.}
    \label{fig:knowledge_hierarchy}
\end{figure}

\begin{figure*}[htbp]
\centering
\scriptsize
\definecolor{leafbg}{HTML}{FFFFEC}

\newlength{\KGHierarchyWidth}
\setlength{\KGHierarchyWidth}{7.0cm}

\newcommand{\KGHierarchyBox}[2]{%
  \begin{varwidth}[t]{\KGHierarchyWidth}
    \normalfont\sffamily\tiny\raggedright
    \textbf{#1}\\[-1pt]
    #2%
  \end{varwidth}%
}


\newsavebox{\boxProgLang}
\sbox{\boxProgLang}{\KGHierarchyBox{Related Work}{\cite{reddy2025swerank+,xiang2026evaluating,xie2026arkeval}}}

\newsavebox{\boxDomainSpec}
\sbox{\boxDomainSpec}{\KGHierarchyBox{Related Work}{\cite{chen2026beyondswe}}}

\newsavebox{\boxTechStack}
\sbox{\boxTechStack}{\KGHierarchyBox{Related Work}{\cite{chen2026beyondswe,jain2025r2e,pan2024training,xie2026arkeval}}}

\newsavebox{\boxExistCode}
\sbox{\boxExistCode}{\KGHierarchyBox{Related Work}{\cite{aggarwal2025dars,ahmed2025otter,antoniades2024swe,arora2024masai,brown2024large,cao2025siadafix,chakraborty2025blaze,chang2025bridging,chen2024coder,chen2025locagent,chen2025old,chen2025prometheus,chen2025swe,chen2026beyondswe,cheshkov2024exploring,copet2025cwm,da2025agent,dai2025lita,ding2026swe,ehrlich2025codemonkeys,gandhi2025agents,gao2025trae,gautam2024supercoder2,golubev2025training,guo2025swe,gupta2025sacl,huang2025seeing,jain2025r2e,jiang2025cosil,jiang2025putting,kon2026swe,lei2024infant,lei2025infantagent,li2024codetree,li2025infcode,li2025patchpilot,li2025swe,lin2024llms,lin2025se,lindenbauer2025complexity,liu2024codexembed,liu2024codexgraph,liu2024marscode,liu2025context,liu2025graphlocator,liu2026architecture,liu2026dynamic,luo2026closing,ma2024lingma,ma2024repository,ma2025alibaba,ma2025sorft,ma2025thinking,ma2025tool,mu2025experepair,nashid2025issue2test,ouyang2024repograph,ouyang2025reasoningbank,pabba2025semagent,pan2024training,phan2024hyperagent,pimpale2025forecasting,raghavendra2026agentic,rastogi2025devstral,reddy2025swerank,reddy2025swerank+,rombaut2024watson,ruan2025specrover,shi2025aime,shum2025swe,sohrabizadeh2025nemotron,song2026swe,soni2025coding,soni2026swe,sonwane2025bugpilot,sun2025scaling,suresh2024cornstack,tang2025boosting,tang2025co,tang2025synfix,tao2024magis,tao2025code,tao2026swe,tawosi2025meta,vaghasiya2025corethink,vinh2025repeton,wang2024openhands,wang2025aegis,wang2025extracting,wang2025improving,wang2025mcts,wang2025practitioner,wang2025seamlessflow,wang2025swe,wei2025swe,wei2025toward,wong2025confucius,wu2025git,xia2025demystifying,xia2025live,xiang2026empowering,xiang2026evaluating,xiao2025improving,xie2025swe,xie2026arkeval,xiong2025think,yang2024swe,yang2025kgcompass,yang2025kimi,yang2025lingxi,yang2025swe,yu2025building,yu2025orcaloca,yu2025utboost,yu2026does,yuan2026swe,zainullina2025guided,zeng2025satori,zeng2025skywork,zhang2024autocoderover,zhang2024codev,zhang2024diversity,zhang2025cast,zhang2025darwin,zhang2025one,zhou2025tom,zhu2025training,dong2025infcode}}}

\newsavebox{\boxRepoEvol}
\sbox{\boxRepoEvol}{\KGHierarchyBox{Related Work}{\cite{ahmed2025otter,cao2025siadafix,chakraborty2025blaze,chang2025bridging,chen2024coder,copet2025cwm,guo2025swe,jain2025r2e,li2025patchpilot,lindenbauer2025complexity,liu2024codexembed,luo2026closing,ma2024lingma,ma2024repository,ma2025sorft,ma2025thinking,ma2025tool,reddy2025swerank,reddy2025swerank+,sohrabizadeh2025nemotron,soni2026swe,tang2025co,tao2024magis,tao2025code,tao2026swe,wang2025improving,wang2025mcts,wang2025seamlessflow,wang2025swe,wei2025swe,wei2025toward,xie2025swe,xie2026arkeval,yang2025kgcompass,yang2025kimi,yang2025lingxi,yang2025swe,yu2025utboost,yu2026does,zainullina2025guided,zeng2025satori,zeng2025skywork}}}

\newsavebox{\boxDesignArch}
\sbox{\boxDesignArch}{\KGHierarchyBox{Related Work}{\cite{cao2025siadafix,chen2026beyondswe,huang2025seeing,li2025swe,liu2026architecture,luo2026closing,nashid2025issue2test,raghavendra2026agentic,wu2025git,yang2025lingxi,yu2026does,zhou2025tom,zhu2025training}}}

\newsavebox{\boxDevTool}
\sbox{\boxDevTool}{\KGHierarchyBox{Related Work}{\cite{aggarwal2025dars,ahmed2025otter,antoniades2024swe,arora2024masai,brown2024large,cao2025siadafix,chen2024coder,chen2025old,chen2025prometheus,chen2025swe,chen2026beyondswe,cheshkov2024exploring,copet2025cwm,da2025agent,dai2025lita,ding2026swe,ehrlich2025codemonkeys,gandhi2025agents,gao2025trae,gautam2024supercoder2,golubev2025training,guo2025swe,jain2025r2e,jiang2025putting,kon2026swe,lei2024infant,lei2025infantagent,li2025infcode,li2025patchpilot,li2025swe,lin2024llms,lin2025se,lindenbauer2025complexity,liu2024marscode,liu2025context,liu2025graphlocator,liu2026architecture,liu2026dynamic,luo2026closing,ma2024lingma,ma2025alibaba,ma2025thinking,ma2025tool,mu2025experepair,nashid2025issue2test,ouyang2025reasoningbank,pabba2025semagent,pan2024training,pan2025codecor,phan2024hyperagent,pimpale2025forecasting,raghavendra2026agentic,rastogi2025devstral,ruan2025specrover,shi2025aime,shum2025swe,sohrabizadeh2025nemotron,song2026swe,soni2025coding,soni2026swe,sonwane2025bugpilot,sun2025scaling,tang2025boosting,tang2025co,tang2025synfix,tao2024magis,tao2026swe,vaghasiya2025corethink,vinh2025repeton,wang2024openhands,wang2025aegis,wang2025practitioner,wang2025seamlessflow,wang2025swe,wei2025toward,wong2025confucius,wu2025git,xia2025demystifying,xia2025live,xiang2026empowering,xiang2026evaluating,xiao2025improving,xie2025swe,xie2026arkeval,yang2024swe,yang2025kgcompass,yang2025kimi,yang2025lingxi,yang2025swe,yu2025building,yu2025orcaloca,yu2025utboost,yuan2026swe,zainullina2025guided,zeng2025satori,zeng2025skywork,zhang2024autocoderover,zhang2025darwin,zhang2025one,zhang2025sealign,zhou2025tom,zhu2025training,dong2025infcode}}}

\newsavebox{\boxExperiential}
\sbox{\boxExperiential}{\KGHierarchyBox{Related Work}{\cite{aggarwal2025dars,antoniades2024swe,cao2025siadafix,chen2024coder,chen2025locagent,chen2025swe,chen2026beyondswe,copet2025cwm,da2025agent,ding2026swe,gandhi2025agents,gao2025trae,golubev2025training,jain2025r2e,kon2026swe,lei2024infant,lei2025infantagent,li2025patchpilot,lin2024llms,lin2025se,lindenbauer2025complexity,liu2025context,ma2024lingma,ma2025alibaba,ma2025thinking,ma2025tool,mu2025experepair,ouyang2025reasoningbank,pan2024training,phan2024hyperagent,pimpale2025forecasting,rastogi2025devstral,reddy2025swerank+,rombaut2024watson,shum2025swe,song2026swe,sonwane2025bugpilot,su2025learn,tang2025agent,tang2025boosting,tang2025co,tao2026swe,wang2024openhands,wang2025aegis,wang2025practitioner,wang2025swe,wei2025toward,wong2025confucius,wu2025git,xia2025live,xiao2025improving,xie2025swe,yang2024swe,yang2025kimi,yang2025lingxi,yang2025swe,yu2025building,yu2025orcaloca,yuan2026swe,zainullina2025guided,zeng2025satori,zeng2025skywork,zhang2024diversity,zhang2025darwin,zhang2025one,zhang2025sealign,zhou2025tom,zhu2025training}}}

\begin{forest}
for tree={
    grow'=0,
    anchor=west,
    child anchor=west,
    parent anchor=east,
    font=\sffamily\scriptsize,
    inner sep=2.2pt,  
    s sep=2.8pt,      
    l sep=2.8mm,      
    edge={draw=black!80},
    align=left,
},
where level=0{
    draw=purple!50,
    thick,
    rounded corners=4pt,
    fill=purple!6,
    font=\sffamily\bfseries\scriptsize,
    align=center,
    inner xsep=4pt,
    inner ysep=2pt,
    fit=band
}{},
where level=1{
    draw=blue!40,
    thick,
    rounded corners=4pt,
    fill=blue!6,
    font=\sffamily\bfseries\scriptsize,
    align=center,
    inner xsep=3pt,
    inner ysep=1.5pt,
    fit=band
}{},
where level=2{
    draw=orange!60,         
    thick,                  
    rounded corners=4pt,    
    fill=orange!10,         
    font=\sffamily\bfseries\scriptsize, 
    align=center,           
    inner xsep=4pt,         
    inner ysep=2pt,         
    tier=knowledgename      
}{},
where level=3{
    draw=orange!25,
    rounded corners=3pt,
    fill=leafbg,
    font=\sffamily\tiny,
    align=left,
    inner xsep=3pt,         
    tier=relatedwork        
}{}
[Knowledge Hierarchy\\for Issue Resolution
    [{L1: Background\\Knowledge}
        [{Programming Language\\Knowledge}
            [{\usebox{\boxProgLang}}]
        ]
        [{Domain-specific\\Knowledge}
            [{\usebox{\boxDomainSpec}}]
        ]
        [{Technology Stack\\Knowledge}
            [{\usebox{\boxTechStack}}]
        ]
    ]
    [{L2: Repository\\Knowledge}
        [{Existing Code\\Knowledge}
            [{\usebox{\boxExistCode}}]
        ]
        [{Repository Evolution\\Knowledge}
            [{\usebox{\boxRepoEvol}}]
        ]
        [{Design Architecture\\Knowledge}
            [{\usebox{\boxDesignArch}}]
        ]
    ]
    [{L3: Procedural\\Knowledge}
        [{Development Tool\\Knowledge}
            [{\usebox{\boxDevTool}}]
        ]
        [{Experiential\\Knowledge}
            [{\usebox{\boxExperiential}}]
        ]
    ]
]
\end{forest}

\caption{
Knowledge hierarchy for automated issue resolution.
}
\label{fig:knowledge_hierarchy_tree}
\end{figure*}
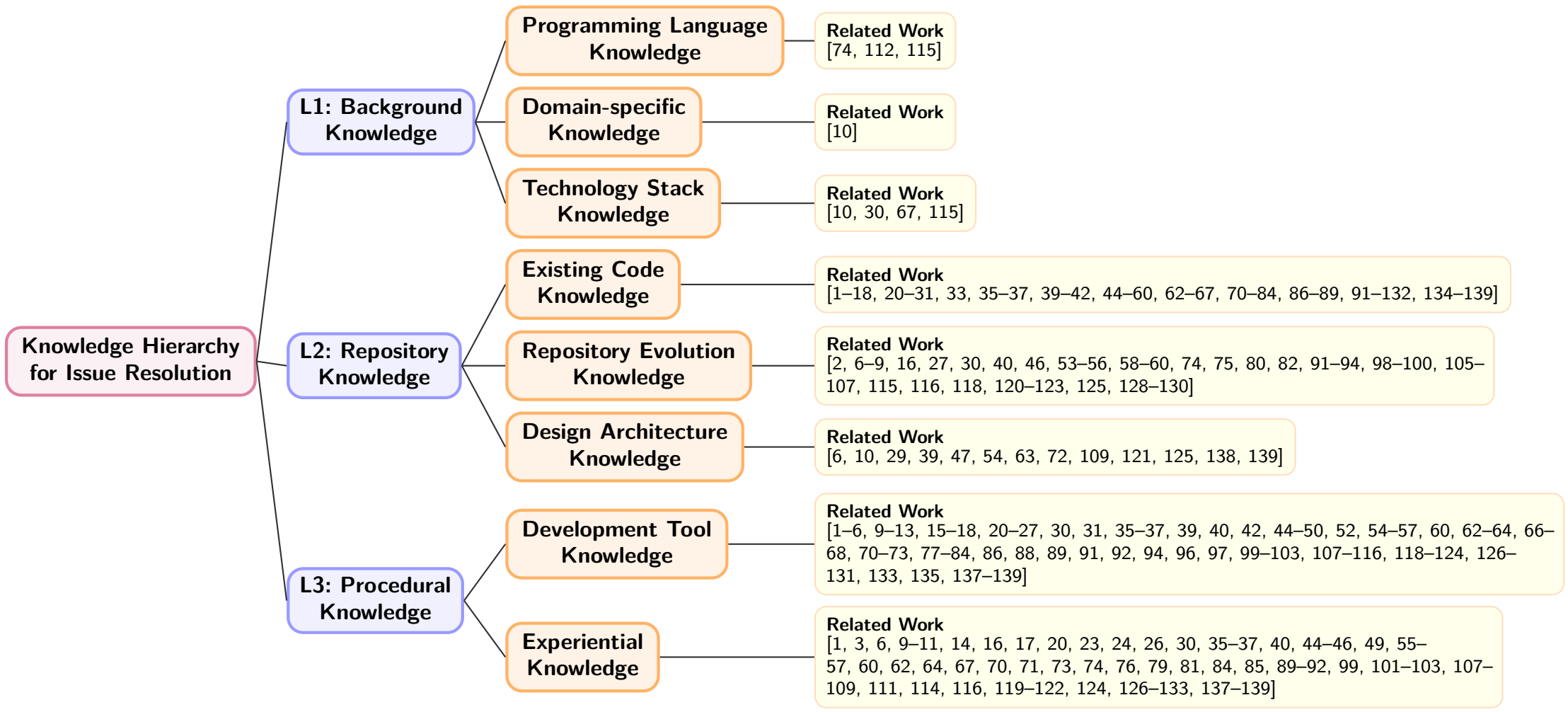

Following the knowledge-type taxonomy defined in Section~\ref{sec:knowledge_type_taxonomy}, this section analyzes how different types of knowledge are instantiated and distributed in existing automated issue-resolution studies. Automated issue resolution is a knowledge-intensive task in which agents must understand issue descriptions, locate relevant repository context, respect project-specific constraints, interact with development environments, and revise candidate patches based on feedback. Existing studies often describe these elements in fragmented terms, such as retrieved code snippets, execution feedback, historical patches, tool observations, or agent trajectories. 
The three-layer knowledge hierarchy is illustrated in Fig.~\ref{fig:knowledge_hierarchy}, while Fig.~\ref{fig:knowledge_hierarchy_tree} further maps each knowledge type to the corresponding studies.
Accordingly, we organize the following analysis around these three knowledge layers and examine how the knowledge types within each layer are reflected in representative studies. This qualitative analysis further provides context for the distribution results reported in Section~\ref{sec:knowledge_type_distribution}.



\subsection{\textbf{L1: Background Knowledge Layer}}

Among the three knowledge layers, L1 is currently adopted more selectively, but it becomes important when issue resolution cannot be completed through repository navigation, code inspection, or tool feedback alone. The studies coded under this layer show that agents sometimes need external semantic and technical assumptions to determine what the issue means and what kinds of patches are valid. This is especially visible in tasks involving specialized application domains, multilingual or low-resource programming ecosystems, dependency migration, and platform-specific development environments.

These studies reveal a common challenge of \textbf{semantic boundary construction}. Issue reports often describe symptoms, requirements, or expected behaviors in natural language, while valid patches must satisfy programming rules, domain assumptions, and technology-stack constraints. Before agents inspect repository-specific evidence, they may need to clarify the semantic boundary of the task: what the reported behavior refers to, which external constraints are relevant, and what kinds of solutions should be considered acceptable. The following analysis discusses how this challenge appears in studies involving Domain-Specific Knowledge, Programming Language Knowledge, and Technology Stack Knowledge.

\subsubsection{Domain-Specific Knowledge.}

Domain-Specific Knowledge appears in issue-resolution settings where the correctness of a patch depends on specialized application semantics rather than only on repository context or general programming competence. This knowledge is observed in only a small portion of the surveyed studies, but it becomes crucial when agents must understand domain-dependent requirements and judge semantic correctness beyond syntactic or test-level validity. For instance, BeyondSWE \cite{chen2026beyondswe} explicitly evaluates this capability through its DomainFix task, which challenges agents to integrate code manipulation with genuine scientific reasoning by applying expert knowledge from specialized fields such as quantum physics and bioinformatics. This setting shows that some issue-resolution tasks cannot be solved solely through repository navigation or general programming ability; instead, agents must understand the domain semantics behind the expected behavior. The limited adoption of this category indicates that most current methods still prioritize general-purpose repository repair, leaving domain-aware issue resolution as an underexplored but important direction.

\subsubsection{Programming Language Knowledge.}

Programming Language Knowledge is mainly reflected in studies that examine multilingual, low-resource, or strongly constrained programming ecosystems. These studies show that language-level constraints can directly affect whether a generated patch is executable, idiomatic, and semantically valid. For instance, SWERANK+ \cite{reddy2025swerank+} highlights the need for issue localization methods to generalize across multiple programming languages, since real-world repositories often contain heterogeneous language environments and language-specific code semantics. Rust-SWE-bench \cite{xiang2026evaluating} further shows that repository-level issue resolution in Rust is constrained by the language's strict type system, trait semantics, ownership, borrowing, and lifetime rules, making many failures arise not only from missing repository context but also from insufficient understanding of Rust-specific programming semantics. Similarly, ArkEval \cite{xie2026arkeval} demonstrates the difficulty of automated repair for ArkTS, a low-resource and statically constrained extension of TypeScript, where general-purpose models often transfer permissive TypeScript idioms that violate ArkTS's ahead-of-time compilation requirements and declarative UI rules. These studies indicate that programming-language-aware issue resolution is necessary when patch correctness depends on language-specific constraints that cannot be fully inferred from the local repository context alone.

\subsubsection{Technology Stack Knowledge.}

Technology Stack Knowledge is mainly reflected in tasks where issue resolution depends on external libraries, frameworks, platforms, APIs, dependency ecosystems, and runtime environments. In these settings, agents must reason about external technical ecosystems that shape how a repository should be modified and maintained. BeyondSWE \cite{chen2026beyondswe} provides a representative example through its dependency-driven migration setting, where agents must understand upstream library ecosystems and version-specific API changes, such as migrations from NumPy 1.x to 2.0 or Pydantic v1 to v2. ArkEval \cite{xie2026arkeval} further involves technology stack knowledge from the HarmonyOS ecosystem, including ArkTS development conventions, platform-specific compilation requirements, declarative UI frameworks, and mobile application runtime constraints. These studies indicate that technology-stack knowledge becomes necessary when issue resolution is constrained by external technical ecosystems rather than only by the target repository or the programming language itself.

\subsection{\textbf{L2: Repository Knowledge Layer}}

L2 is the most central and consistently adopted layer in current automated issue-resolution studies. In contrast to L1, which provides external semantic and technical assumptions, Repository Knowledge grounds agents in the target software project itself. The surveyed studies show that repository-level issue resolution rarely depends on the nearest suspicious code fragment alone. A correct patch usually requires agents to understand how implementation artifacts are organized, how code entities are used across the repository, what architectural assumptions should be preserved, and how previous changes have shaped the current repository state.

Across existing studies, this layer is mainly instantiated through three forms of repository grounding. Existing Code Knowledge provides direct access to the current implementation and supports localization, patch generation, and consistency checking. Design Architecture Knowledge introduces higher-level constraints about module responsibilities, design intent, and architectural boundaries. Repository Evolution Knowledge provides temporal evidence from historical issues, commits, pull requests, patches, and development traces. Together, these knowledge types allow agents to reason beyond isolated code snippets and connect issue descriptions with implementation details, architectural conventions, and historical development patterns.

\subsubsection{Existing Code Knowledge.}

Existing Code Knowledge is the most widely observed knowledge type in the surveyed literature, reflecting the fact that automated issue resolution must first identify what parts of the repository are relevant to the reported issue. Existing studies do not use code merely as raw textual context; instead, they access and organize code at different granularities and functional roles. Based on the observed practices, Existing Code Knowledge is mainly instantiated through three complementary forms: local implementation context, interface and usage context, and structural and dependency context. These forms respectively support locating concrete edit candidates, maintaining compatibility with existing usages, and reasoning over cross-file or cross-module relations.

\textbf{Local implementation context.}
Many systems first use existing code to obtain suspicious files, classes, functions, methods, and surrounding code regions that are directly relevant to the reported issue. This form provides the most immediate repository evidence for localization and patch generation, because it identifies the concrete implementation fragments that the agent needs to inspect or edit. For example, Agentless \cite{xia2025demystifying} follows a hierarchical localization process from files to classes, functions, and fine-grained edit locations, while AutoCodeRover \cite{zhang2024autocoderover} retrieves class signatures and method implementation details through AST-based code search APIs. These studies show that local implementation context is usually the first layer of repository grounding, narrowing a large codebase into issue-relevant edit candidates.

\textbf{Interface and usage context.}
Another recurring use of existing code is to inspect how code entities are exposed, invoked, and reused within the repository, including function signatures, class interfaces, parameters, return values, references, call sites, and project-specific usage conventions. Compared with local implementation context, this form focuses less on the internal body of a suspicious code fragment and more on whether a candidate modification remains consistent with how the code is used elsewhere. Agentic Rubrics \cite{raghavendra2026agentic} generates repository-grounded verification criteria by inspecting relevant code paths, interfaces, and project conventions, while Devstral \cite{rastogi2025devstral} trains coding-agent models on trajectories in which agents inspect and edit project-specific code contexts through file editing and execution tools. Such knowledge helps agents avoid locally plausible but globally inconsistent patches, especially when changes may affect callers, downstream components, or implicit interface contracts.

\textbf{Structural and dependency context.}
A third form of existing-code use focuses on relationships among code entities, such as class--method structures, function-call relations, reference links, import dependencies, inheritance relations, and cross-file dependencies. This form becomes important when the relevant implementation logic is scattered across multiple files or modules, making issue resolution dependent on repository-level navigation rather than isolated code inspection. LingmaAgent \cite{ma2025alibaba} constructs repository knowledge graphs that summarize hierarchical repository structures and reference relationships across functions and classes; CoSIL \cite{jiang2025cosil} dynamically constructs module and function call graphs during issue localization; and DARS \cite{aggarwal2025dars} incorporates repository search, localization, and edit-location information into its branching and patch-generation process. RepoLens \cite{wang2025extracting} further abstracts fine-grained functionalities from existing code into issue-relevant concerns, helping agents locate scattered implementation logic in large repositories. InfCode-C++ combines AST-based structural search with semantic code-intent retrieval to localize issue-relevant code regions in large C++ repositories \cite{dong2025infcode}.

Together, these studies show that Existing Code Knowledge is not limited to raw code snippets. Instead, it supports a layered understanding of the current repository implementation: local implementation context identifies concrete edit candidates, interface and usage context constrains how changes should remain compatible with existing uses, and structural and dependency context supports cross-file reasoning over repository-wide relations. This explains why Existing Code Knowledge forms the core repository-grounded basis for localization, patch generation, and consistency checking in automated issue resolution.

\subsubsection{Design Architecture Knowledge.}

Design Architecture Knowledge is less frequently adopted than Existing Code Knowledge, but it becomes important when issue resolution must preserve design intent, component responsibilities, architectural patterns, module-level organization, or implicit system constraints. The surveyed studies show that architecture-aware knowledge helps agents reason about how implementation elements should cooperate and what assumptions should be maintained during repair. This is especially relevant for repository-level issues where a locally plausible edit may still violate cross-module responsibilities, latent design constraints, or the intended decomposition of the system.

Several recent studies illustrate how architectural knowledge can strengthen repository-level reasoning. SWE-Shield \cite{yu2026does} constructs a benchmark based on implicit project-specific design constraints mined from historical pull requests to assess whether agent-generated patches truly comply with architectural conventions rather than merely passing functional tests. From a human-agent collaboration perspective, ToM-SWE \cite{zhou2025tom} maintains a persistent mental model of developer intents and coding preferences to constrain the agent's behavior, ensuring the code is shaped to align with the user's implicit architectural expectations. Furthermore, RPG-Encoder \cite{luo2026closing} explicitly reconstructs the latent functional architecture of a repository into a unified planning graph, bidirectionally linking high-level semantic design intents with low-level structural dependencies to guide precise and architecturally-aware navigation.

These studies suggest that Design Architecture Knowledge provides a conceptual complement to code-level repository grounding. While Existing Code Knowledge helps agents identify what code exists and where modifications may be applied, architecture-aware knowledge helps agents reason about why certain components should be modified, how responsibilities are distributed, and whether a candidate patch remains consistent with broader system organization. Its relatively limited adoption also indicates that architectural intent remains difficult to extract, represent, and validate automatically in current issue-resolution systems.

\subsubsection{Repository Evolution Knowledge.}

Repository Evolution Knowledge is increasingly used to provide temporal grounding for automated issue resolution. In the surveyed studies, this knowledge is mainly instantiated through three forms: historical issue-fix data, structured repository-history representations, and development-process traces. These forms respectively support learning from previous repairs, connecting historical artifacts with current code entities, and reusing historical development activities as training or reasoning signals. Consistent with the taxonomy boundary in Section~\ref{sec:knowledge_type_taxonomy}, the analysis here treats historical issues, commits, pull requests, and patches as repository-derived temporal evidence, while runtime trajectories, reflective memories, and reusable repair strategies are discussed under Experiential Knowledge.

\textbf{Historical issue-fix data.}
The first form constructs training or retrieval data from historical issue-fix pairs, pull requests, commits, and code modifications. This form uses past repository evolution to learn the mapping between natural-language issues and changed code locations, thereby supporting issue localization, patch generation, and model training. BLAZE builds cross-project and cross-language bug localization data from historical bugs to improve bug localization across projects and languages \cite{chakraborty2025blaze}. BugCerberus learns hierarchical localization patterns from bug reports and bug-related contextual information at file, function, and statement levels \cite{chang2025bridging}. ReSAT constructs localization and code-edit training data from issues and corresponding pull requests to improve repository-level issue resolution \cite{ma2024repository}. SoRFT decomposes issue resolution into localization and editing subtasks using ground-truth historical fixes as supervision and reward signals \cite{ma2025sorft}. SWE-Rank curates issue-code pairs from public GitHub repositories to train efficient retrievers and rerankers for issue localization \cite{reddy2025swerank}, and SWE-Rank+ extends this historical issue-code training paradigm to multilingual and multi-turn issue localization settings \cite{reddy2025swerank+}. These studies show that historical issue-fix data provides static, reusable supervision for learning where and how repositories have been modified in response to previous issues.

\textbf{Structured repository-history representations.}
The second form organizes repository history into structured or graph-based representations that connect historical artifacts with current code entities. Compared with using historical issue-fix data as isolated training examples, this form emphasizes the relational structure among issues, pull requests, commits, files, classes, functions, and code changes. KGCompass links issues, pull requests, files, classes, and functions in a repository-aware knowledge graph to support path-guided localization and repair \cite{yang2025kgcompass}. RPG-Encoder incrementally updates repository representations from commit diffs so that semantic and structural knowledge remains synchronized with code evolution \cite{luo2026closing}. These studies indicate that structured repository-history representations can transform scattered historical artifacts into navigable temporal knowledge, helping agents reason over how past changes relate to the current repository state.

\textbf{Development-process traces.}
The third form uses development-process traces and historical software-engineering activities as training or reasoning signals. This form goes beyond static historical artifacts by capturing how developers or agents interact with repositories during real or simulated development processes, including repository understanding, tool use, localization, editing, and validation. Lingma SWE-GPT learns from real-world code submission activities and development processes, including repository understanding, fault localization, and patch generation \cite{ma2024lingma}. CodeXEmbed incorporates diverse code retrieval tasks across repositories and programming languages to improve retrieval over evolving and heterogeneous codebases \cite{liu2024codexembed}. CWM is trained on Python execution traces and agentic interactions with repository environments, capturing dynamic software-development trajectories beyond static code \cite{copet2025cwm}. ToolTrain improves repo deep search by training models to use repository retrieval tools for multi-hop navigation during issue localization \cite{ma2025tool}. Nemotron-CORTEXA improves localization and solution diversity by optimizing how agents retrieve and use relevant repository context in real-world software engineering tasks \cite{sohrabizadeh2025nemotron}. SIADAFIX also reflects this category when it uses issue-description optimization and expert-inspired repair workflows to guide adaptive repair based on accumulated repair patterns rather than isolated code context alone \cite{cao2025siadafix}. These studies show that development-process traces provide procedural and temporal evidence about how issue resolution is actually carried out, making repository evolution useful not only as historical data, but also as behavioral guidance for future agents.

Together, these studies show that Repository Evolution Knowledge provides temporal and historical grounding for automated issue resolution. Historical issue-fix data enables systems to learn from previous issue--patch mappings, structured repository-history representations connect past artifacts with current code entities, and development-process traces capture how repair activities unfold over time. This knowledge therefore complements Existing Code Knowledge: while Existing Code Knowledge describes the current repository state, Repository Evolution Knowledge explains how that state was reached and how similar problems have been resolved before, thereby supporting more historically informed localization, patch generation, training, and reasoning.

\subsection{\textbf{L3: Procedural Knowledge Layer}}

L3 is widely observed in agent-based issue-resolution systems because automated issue resolution is increasingly treated as an iterative process of exploration, action, feedback, and adaptation. The surveyed studies show that agents must not only understand issue descriptions and repository artifacts, but also operate on the repository through tools, interpret intermediate observations, revise decisions, and reuse prior solving experience. In this sense, the Procedural Knowledge Layer captures how issue-resolution knowledge becomes executable during agent-environment interaction.

Across existing studies, this layer is mainly instantiated through two forms of process-level support. Development Tool Knowledge is reflected in how agents use search utilities, editors, shell commands, build systems, test runners, debuggers, static analyzers, validation tools, and other repository-facing interfaces to inspect, modify, execute, and validate software projects. Experiential Knowledge is reflected in how systems preserve or internalize successful trajectories, failed attempts, reasoning traces, tool-use patterns, feedback signals, and repair strategies. Together, these two knowledge types enable automated issue resolution to move beyond one-shot patch generation toward tool-supported, feedback-driven, and experience-guided problem solving.

\subsubsection{Development Tool Knowledge}

Development Tool Knowledge is most visible in studies where agents rely on development tools and execution environments to conduct issue resolution as an interactive process. In these systems, tools are not merely auxiliary interfaces for retrieving information; they provide the operational channel through which agents search repositories, inspect files, edit code, run commands, execute tests, observe failures, and validate candidate patches. Existing studies mainly use this knowledge in two forms: testing and validation tool knowledge, and environment-interaction and tool-control knowledge. The former supports issue reproduction and patch verification through executable feedback, while the latter supports repository exploration, editing, planning, action scheduling, and tool-mediated interaction.

\textbf{Testing and validation tool knowledge.}
The first form concerns how agents use testing and validation tools to reproduce issues, expose failures, and evaluate candidate patches. This form is important because it allows issue-resolution systems to obtain executable feedback rather than relying only on static code inspection or model-generated reasoning. Otter generates tests from issue descriptions to validate SWE patches and uses rule-based analysis to check and repair generated tests \cite{ahmed2025otter}. TestPrune prunes and reuses regression tests to guide issue reproduction and patch validation \cite{chen2025old}. Conversational test-suite-based repair uses failing test information as an interactive signal in patch generation dialogs \cite{cheshkov2024exploring}. UTBoost augments SWE-Bench evaluation by automatically generating additional unit tests to expose incorrect patches that pass insufficient original tests \cite{yu2025utboost}. AutoCodeRover uses spectrum-based fault localization and available test suites to sharpen the repair context and validate generated patches \cite{zhang2024autocoderover}. Large Language Monkeys shows that repeated sampling becomes more effective in coding tasks when generated candidates can be filtered by automatic verifiers such as unit tests \cite{brown2024large}. Agent-RLVR uses unit-test rewards and environment feedback to train software engineering agents in verifiable environments \cite{da2025agent}. CWM further treats Python execution traces and agentic Docker-environment interactions as tool-mediated observations for learning code execution and debugging behavior \cite{copet2025cwm}. These studies show that testing and validation tools provide a crucial feedback channel for grounding patch generation in executable evidence.

\textbf{Environment-interaction and tool-control knowledge.}
The second form concerns how agents decide which tools to use, when to use them, and how to coordinate tool actions during long-horizon issue resolution. Compared with testing and validation tool knowledge, this form focuses more on repository exploration, action scheduling, editing, planning, navigation, search control, and workflow management. DARS enhances coding agents by branching trajectories at key decision points based on previous execution feedback and alternative tool actions \cite{aggarwal2025dars}. SWE-Search integrates Monte Carlo Tree Search with iterative refinement, value estimation, and agent feedback to guide exploration in software engineering tasks \cite{antoniades2024swe}. MASAI decomposes issue resolution into specialized sub-agents with distinct objectives and tool-use strategies \cite{arora2024masai}. SIADAFIX designs adaptive repair workflows and fast-thinking components that select suitable repair modes and guide bug-fix agents \cite{cao2025siadafix}. CodeR uses a multi-agent framework with predefined task graphs to organize tool-mediated repository exploration and issue resolution \cite{chen2024coder}. Lita studies a lightweight agent scaffold that retains only essential autonomous tools to evaluate coding capabilities with minimal hand-crafted workflow design \cite{dai2025lita}. SWE-Replay recycles prior agent trajectories and branches from critical intermediate steps to reduce the cost of test-time scaling \cite{ding2026swe}. OrcaLoca integrates priority-based scheduling, action decomposition, and context pruning to control LLM-guided localization actions \cite{yu2025orcaloca}. SWE-MiniSandbox replaces heavyweight per-task containers with lightweight isolated workspaces for scalable reinforcement learning of SWE agents \cite{yuan2026swe}. Guided search strategies use learned action-value estimators to guide exploration in non-serializable environments such as Docker-based SWE tasks \cite{zainullina2025guided}. Darwin G\"odel Machine allows agents to modify their own coding-agent tools and workflows, such as code editing tools and long-context management mechanisms, and validates improvements through benchmarks \cite{zhang2025darwin}. RepoNavigator simplifies repository-level navigation through a single execution-aware ``jump-to-definition'' tool trained with reinforcement learning \cite{zhang2025one}. These studies indicate that tool-control knowledge is essential for transforming issue resolution from one-shot generation into an interactive process of searching, editing, executing, observing, and revising.

Together, these studies show that Development Tool Knowledge has become a core form of procedural support in automated issue resolution. Testing and validation tool knowledge provides executable feedback for reproducing issues and checking candidate patches, while environment-interaction and tool-control knowledge enables agents to operate effectively within complex repository environments. This knowledge therefore supports issue resolution as an interactive and feedback-driven process, where agents must not only reason about code, but also decide how to use tools, manage actions, interpret observations, and adapt their strategies throughout the resolution workflow.

\subsubsection{Experiential Knowledge}

Experiential Knowledge is increasingly used in studies that attempt to reuse lessons from prior or ongoing issue-resolution processes. Instead of treating each issue as an isolated trial, these systems preserve, retrieve, summarize, or internalize previous solving attempts, including successful trajectories, failed attempts, intermediate decisions, expert feedback, reasoning traces, tool-use patterns, validation outcomes, and repair strategies. Consistent with the taxonomy boundary in Section~\ref{sec:knowledge_type_taxonomy}, this analysis focuses on process-level experience produced or reused during issue-resolution execution, while repository-derived historical artifacts such as issues, commits, pull requests, and patches are discussed under Repository Evolution Knowledge.

Existing studies primarily leverage this knowledge in two forms: experience memories and knowledge banks, and training trajectories and post-training data. The former stores and retrieves reusable problem-solving lessons during future issue resolution, while the latter distills agent experiences into model capabilities through training, fine-tuning, or reinforcement learning. These two forms show how process-level experience can support future localization, patch generation, tool use, validation, and self-correction.

\textbf{Experience memories and knowledge banks.}
The first form explicitly constructs experience memories or knowledge banks to guide future issue resolution. This type of knowledge preserves prior successful and failed experiences as reusable demonstrations, reflective insights, reasoning patterns, or workflow-level guidance. EXPEREPAIR organizes historical repair trajectories into episodic memory for concrete demonstrations and semantic memory for abstract reflective insights \cite{mu2025experepair}. SWE-Exp distills successful and failed agent trajectories into a multi-faceted experience bank covering problem comprehension and code modification strategies \cite{chen2025swe}. ReasoningBank extracts generalizable reasoning strategies from self-judged successful and failed experiences and retrieves them to support future agent interactions \cite{ouyang2025reasoningbank}. AGENT KB builds a shared cross-domain experience base that provides workflow-level guidance and execution-level refinement patterns \cite{tang2025agent}. Watson further recovers and inspects agents' implicit reasoning traces, making past decision processes observable and reusable for diagnosis and correction \cite{rombaut2024watson}. These studies show that experience memories and knowledge banks allow agents to reuse prior problem-solving lessons without relying solely on the model's parametric knowledge or the current repository context.

\textbf{Training trajectories and post-training data.}
The second form distills experience into training trajectories, post-training data, or specialized agent capabilities. Compared with explicit memory-based reuse, this form incorporates experiential knowledge into the model or agent policy itself, enabling systems to internalize patterns of planning, localization, editing, validation, and self-correction. SWE-Gym trains SWE agents and verifiers from sampled agent-environment trajectories \cite{pan2024training}. Devstral fine-tunes models on coding-agent trajectories for multi-step issue resolution tasks \cite{rastogi2025devstral}. SWE-Master combines teacher-trajectory synthesis, long-horizon supervised fine-tuning, reinforcement learning, and execution feedback to improve agentic issue resolution \cite{song2026swe}. SWE-Lego, BugPilot, Learn-by-interact, Repairity, and Co-PatcheR similarly leverage validated trajectories or reasoning traces to improve issue resolution capabilities \cite{tao2026swe, sonwane2025bugpilot, su2025learn, tang2025boosting, tang2025co}. These studies indicate that experiential knowledge can be converted from external traces into more persistent agent capabilities, allowing models to better reproduce effective resolution behaviors across tasks.

Together, these studies demonstrate that Experiential Knowledge captures accumulated lessons about how agents solve, fail, refine, and improve across issue-resolution tasks. Experience memories and knowledge banks support explicit retrieval and reuse of prior problem-solving lessons, while training trajectories and post-training data transform past experiences into internalized model or agent capabilities. This knowledge therefore enables automated issue resolution to move beyond isolated trial-and-error toward memory-guided, feedback-aware, and experience-driven problem solving.

\subsection{Distribution of Knowledge Types across Layers}
\label{sec:knowledge_type_distribution}
\begin{figure}[htbp]
    \centering
    \includegraphics[width=0.9\linewidth]{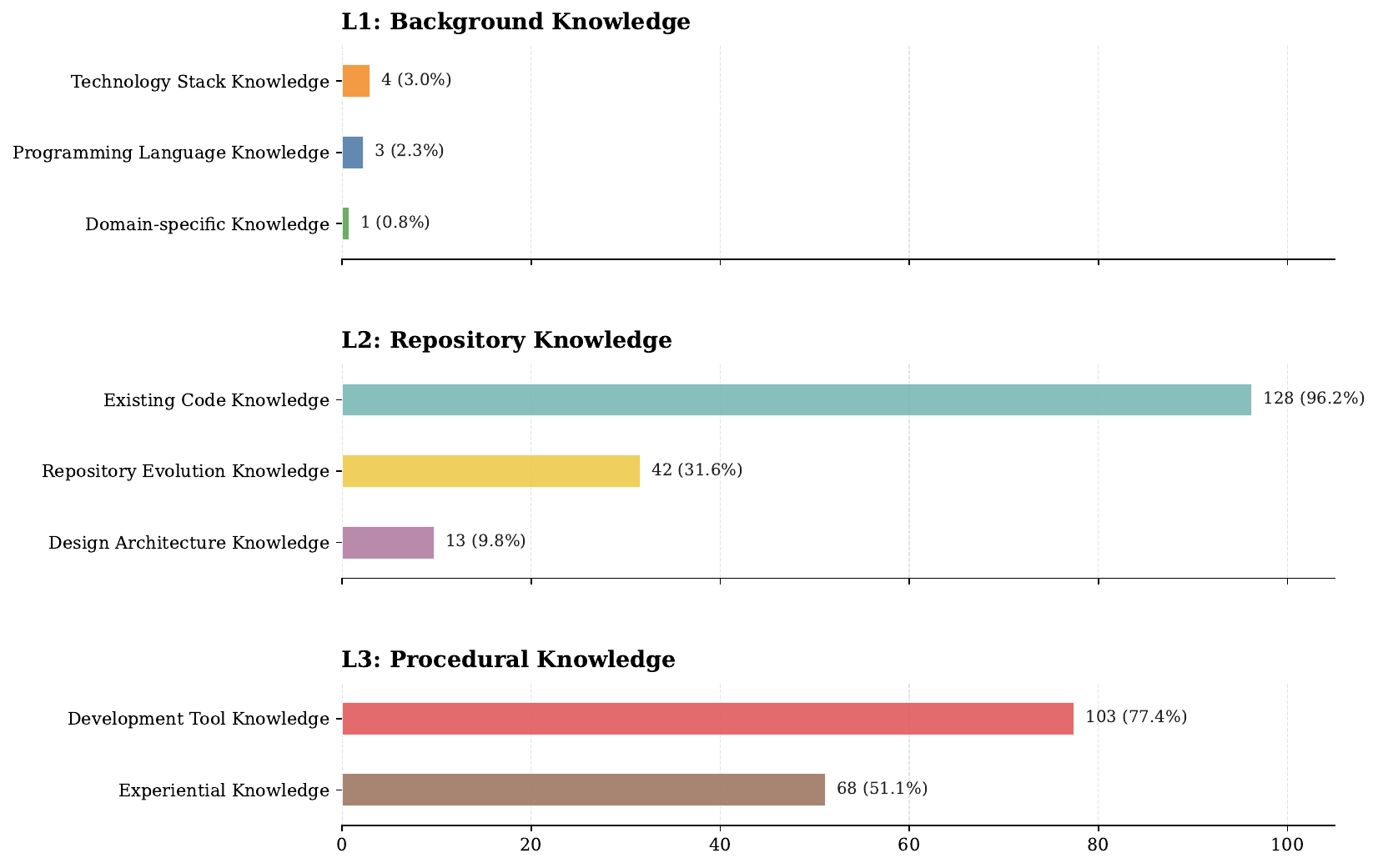}
    \caption{Percentage distribution of knowledge types across knowledge layers.}
    \label{fig:level_knowledge_percentages}
\end{figure}

Fig.~\ref{fig:level_knowledge_percentages} quantifies the distribution of knowledge types across the three-layer hierarchy and shows how current automated issue resolution systems have developed a strong foundation around repository-grounded and execution-oriented knowledge. Existing Code Knowledge is the most widely used knowledge type, appearing in 96.2\% of the analyzed studies. This high proportion is expected because issue resolution fundamentally requires agents to understand the current implementation before they can locate relevant components, generate patches, and check whether the modification is consistent with the surrounding code. By providing concrete information about files, functions, classes, interfaces, call sites, and code dependencies, Existing Code Knowledge directly supports the core reasoning process of identifying what should be inspected and modified. Development Tool Knowledge is also highly adopted, appearing in 77.4\% of the studies. This indicates that automated issue resolution has increasingly moved from static code reading toward interactive repository operation, where agents use search tools, editors, build systems, test runners, execution environments, and feedback channels to explore, modify, and validate software projects. Together, these two knowledge types form the current practical basis of automated issue resolution: Existing Code Knowledge grounds the agent in the repository state, while Development Tool Knowledge enables the agent to act on that repository and obtain feedback.

The distribution also shows that process-aware and history-aware knowledge is becoming an important direction for extending issue resolution capabilities. Experiential Knowledge appears in 51.1\% of the studies, suggesting that many recent methods have begun to reuse prior trajectories, failed attempts, repair strategies, reasoning traces, and feedback signals to improve future problem solving. This type of knowledge is valuable because issue resolution is often a long-horizon task: agents need not only code context, but also guidance on how to plan, explore, revise, and recover from incorrect intermediate decisions. Repository Evolution Knowledge appears in 31.6\% of the studies, showing that historical issues, commits, pull requests, issue-fix links, and prior patches are increasingly used to provide temporal evidence for localization, repair, retrieval, and training. These two knowledge types expand issue resolution from reasoning over the current repository snapshot to learning from previous development and repair processes. Their growing adoption suggests a promising shift toward agents that can benefit from accumulated repository history and reusable problem-solving experience.

Design Architecture Knowledge, which appears in 9.8\% of the studies, represents another valuable expansion direction. Although its current adoption is more selective, it can provide high-level guidance about module responsibilities, component boundaries, design intent, and architectural constraints. Such knowledge is especially helpful for repository-level issues where the correct patch must respect not only the local implementation, but also the broader organization of the software system. For example, when an issue involves cross-module behavior or implicit design conventions, architecture knowledge can help agents understand why a component should be modified in one location rather than another, and how a local change may affect the overall system structure. Its relatively lower frequency reflects the fact that architectural knowledge is often implicit, distributed, and harder to extract automatically than code snippets or test feedback. Therefore, methods that can infer, represent, and use architectural intent from code, document, history, and developer preferences have strong potential to improve the reliability of future issue resolution systems.

L1 Background Knowledge, including Technology Stack Knowledge, Programming Language Knowledge, and Domain-Specific Knowledge, currently appears in a smaller portion of the studies, with 3.0\%, 2.3\%, and 0.8\%, respectively. Rather than indicating limited usefulness, this suggests that background-aware issue resolution is still an emerging research space. These knowledge types are important for extending automated issue resolution beyond general repository navigation. Programming Language Knowledge helps agents follow language-specific syntax, type systems, runtime behavior, and idioms; Technology Stack Knowledge supports reasoning about frameworks, libraries, APIs, dependency versions, and platform constraints; and Domain-Specific Knowledge helps agents understand requirements governed by specialized application semantics. These forms of knowledge are less frequently used partly because they are more difficult to collect, standardize, and evaluate: they often come from external documents, domain rules, library ecosystems, expert assumptions, or project-specific usage conventions. However, as issue resolution benchmarks expand to multilingual, framework-dependent, multimodal, and domain-specific settings, these background knowledge types are likely to become increasingly important.

Overall, the distribution suggests that knowledge use in automated issue resolution is developing from a code-centric and tool-mediated foundation toward a more comprehensive software knowledge structure. Current systems have already established a strong physical and operational grounding through Existing Code Knowledge and Development Tool Knowledge: the former anchors agents in concrete repository artifacts, while the latter enables agents to act on the repository through search, editing, execution, testing, and feedback. However, complex issue resolution requires more than inspecting the current code snapshot and operating development tools. Experiential Knowledge and Repository Evolution Knowledge extend agents with process memory and temporal traceability, allowing them to reuse prior trajectories, understand how the repository has changed, and reason from previous issues, commits, pull requests, and repair attempts. Meanwhile, Design Architecture Knowledge and L1 Background Knowledge point to the still underdeveloped conceptual and semantic layers of automated issue resolution, including architectural rationale, responsibility allocation, programming-language constraints, technology-stack conventions, and domain semantics. Therefore, future progress may depend not only on retrieving more code or invoking more tools, but also on constructing persistent, traceable, and evolvable knowledge structures that connect implementation artifacts, historical decisions, architectural intent, external constraints, and process-level experience. Such expansion can make automated issue resolution agents more context-aware, generalizable, and capable of handling complex real-world maintenance tasks.

\finding{
The knowledge-type distribution shows that current automated issue resolution is built on a strong repository-operational foundation. Existing Code Knowledge and Development Tool Knowledge constitute the dominant basis of current systems, indicating that agents primarily rely on concrete repository artifacts and tool-mediated interactions to locate, modify, execute, and validate code. At the same time, the increasing use of Repository Evolution Knowledge and Experiential Knowledge suggests that issue-resolution systems are beginning to move beyond static repository snapshots toward temporal traceability and reusable process experience. However, Design Architecture Knowledge and L1 Background Knowledge remain comparatively underexplored, showing that current systems still have limited support for architectural intent, programming-language constraints, technology-stack conventions, and domain semantics.

This finding suggests that the next step for automated issue-resolution agents is not simply to retrieve more code or invoke more tools, but to construct richer software knowledge infrastructures. Future systems need to connect implementation artifacts with historical evolution, architectural rationale, external semantic constraints, and accumulated repair experience, so that agents can reason over both the current repository state and the broader knowledge context in which software changes are made. Such a knowledge-centric foundation may help software agents become more context-aware, controllable, and generalizable when handling complex, long-lived, and evolving software projects.
}



\section{RQ2: What are the sources and extraction methods of knowledge?}
\label{sec:rq2}

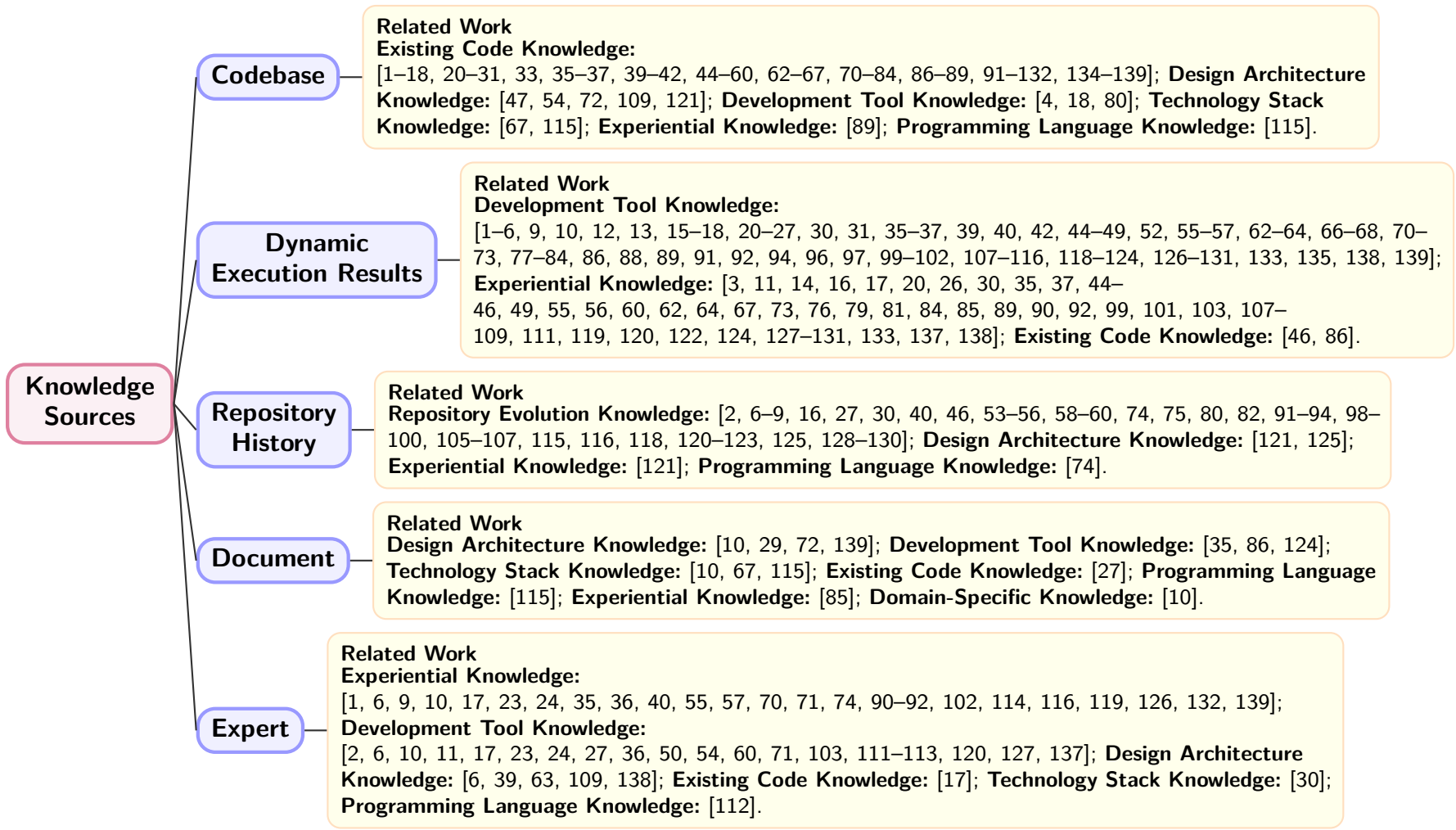
\begin{figure*}[htbp]
\centering
\scriptsize
\definecolor{leafbg}{HTML}{FFFFEC}

\newlength{\KGSourceWidth}
\setlength{\KGSourceWidth}{8.0cm}

\newcommand{\KGSourceBox}[2]{%
  \begin{varwidth}[t]{\KGSourceWidth}
    \normalfont\sffamily\tiny\raggedright
    \textbf{#1}\\[-1pt]
    #2%
  \end{varwidth}%
}


\newsavebox{\boxSourceCodebase}
\sbox{\boxSourceCodebase}{%
\KGSourceBox{Related Work}{%
\textbf{Existing Code Knowledge:} \cite{aggarwal2025dars, ahmed2025otter, antoniades2024swe, arora2024masai, brown2024large, cao2025siadafix, chakraborty2025blaze, chang2025bridging, chen2024coder, chen2025locagent, chen2025old, chen2025prometheus, chen2025swe, chen2026beyondswe, cheshkov2024exploring, copet2025cwm, da2025agent, dai2025lita, ding2026swe, ehrlich2025codemonkeys, gandhi2025agents, gao2025trae, gautam2024supercoder2, golubev2025training, guo2025swe, gupta2025sacl, huang2025seeing, jain2025r2e, jiang2025cosil, jiang2025putting, kon2026swe, lei2024infant, lei2025infantagent, li2024codetree, li2025infcode, li2025patchpilot, li2025swe, lin2024llms, lin2025se, lindenbauer2025complexity, liu2024codexembed, liu2024codexgraph, liu2024marscode, liu2025context, liu2025graphlocator, liu2026architecture, liu2026dynamic, luo2026closing, ma2024lingma, ma2024repository, ma2025alibaba, ma2025sorft, ma2025thinking, ma2025tool, mu2025experepair, nashid2025issue2test, ouyang2024repograph, ouyang2025reasoningbank, pabba2025semagent, pan2024training, phan2024hyperagent, pimpale2025forecasting, raghavendra2026agentic, rastogi2025devstral, reddy2025swerank, reddy2025swerank+, rombaut2024watson, ruan2025specrover, shi2025aime, shum2025swe, sohrabizadeh2025nemotron, song2026swe, soni2025coding, soni2026swe, sonwane2025bugpilot, sun2025scaling, suresh2024cornstack, tang2025boosting, tang2025co, tang2025synfix, tao2024magis, tao2025code, tao2026swe, tawosi2025meta, vaghasiya2025corethink, vinh2025repeton, wang2024openhands, wang2025aegis, wang2025extracting, wang2025improving, wang2025mcts, wang2025practitioner, wang2025seamlessflow, wang2025swe, wei2025swe, wei2025toward, wong2025confucius, wu2025git, xia2025demystifying, xia2025live, xiang2026empowering, xiang2026evaluating, xiao2025improving, xie2025swe, xie2026arkeval, xiong2025think, yang2024swe,  yang2025kgcompass, yang2025kimi, yang2025lingxi, yang2025swe, yu2025building, yu2025orcaloca, yu2025utboost, yu2026does, yuan2026swe, zainullina2025guided, zeng2025satori, zeng2025skywork, zhang2024autocoderover, zhang2024codev, zhang2024diversity, zhang2025cast, zhang2025darwin, zhang2025one, zhou2025tom, zhu2025training,dong2025infcode};
\textbf{Design Architecture Knowledge:} \cite{liu2026architecture, luo2026closing, raghavendra2026agentic, wu2025git, yang2025lingxi};
\textbf{Development Tool Knowledge:} \cite{arora2024masai, dai2025lita, sohrabizadeh2025nemotron};
\textbf{Technology Stack Knowledge:} \cite{pan2024training, xie2026arkeval};
\textbf{Experiential Knowledge:} \cite{tang2025boosting};
\textbf{Programming Language Knowledge:} \cite{xie2026arkeval}.%
}}

\newsavebox{\boxSourceDynamicExecution}
\sbox{\boxSourceDynamicExecution}{%
\KGSourceBox{Related Work}{%
\textbf{Development Tool Knowledge:} \cite{aggarwal2025dars, ahmed2025otter, antoniades2024swe, arora2024masai, brown2024large, cao2025siadafix, chen2024coder, chen2025old, chen2025prometheus, chen2026beyondswe, cheshkov2024exploring, copet2025cwm, da2025agent, dai2025lita, ding2026swe, ehrlich2025codemonkeys, gandhi2025agents, gao2025trae, gautam2024supercoder2, golubev2025training, guo2025swe, jain2025r2e, jiang2025putting, kon2026swe, lei2024infant, lei2025infantagent, li2025infcode, li2025patchpilot, li2025swe, lin2024llms, lin2025se, lindenbauer2025complexity, liu2024marscode, liu2025context, liu2026architecture, liu2026dynamic, ma2024lingma, ma2025alibaba, ma2025thinking, mu2025experepair, nashid2025issue2test, ouyang2025reasoningbank, pabba2025semagent, pan2024training, pan2025codecor, phan2024hyperagent, pimpale2025forecasting, raghavendra2026agentic, rastogi2025devstral, ruan2025specrover, shi2025aime, shum2025swe, sohrabizadeh2025nemotron, song2026swe, soni2025coding, soni2026swe, sonwane2025bugpilot, sun2025scaling, tang2025boosting, tang2025co, tang2025synfix, tao2024magis, tao2026swe, vaghasiya2025corethink, vinh2025repeton, wang2025aegis, wang2025practitioner, wang2025seamlessflow, wang2025swe, wei2025toward, wong2025confucius, wu2025git, xia2025demystifying, xia2025live, xiang2026empowering, xiang2026evaluating, xiao2025improving, xie2025swe, xie2026arkeval, yang2024swe,  yang2025kgcompass, yang2025kimi, yang2025lingxi, yang2025swe, yu2025building, yu2025orcaloca, yu2025utboost, yuan2026swe, zainullina2025guided, zeng2025satori, zeng2025skywork, zhang2024autocoderover, zhang2025darwin, zhang2025sealign, zhou2025tom, zhu2025training,dong2025infcode};
\textbf{Experiential Knowledge:} \cite{antoniades2024swe, chen2025locagent, chen2025swe, copet2025cwm, da2025agent, ding2026swe, golubev2025training, jain2025r2e, kon2026swe, lei2024infant, lin2024llms, lin2025se, lindenbauer2025complexity, liu2025context, ma2024lingma, ma2025thinking, ma2025tool, mu2025experepair, ouyang2025reasoningbank, pan2024training, rastogi2025devstral, rombaut2024watson, shum2025swe, song2026swe, sonwane2025bugpilot, su2025learn, tang2025agent, tang2025boosting, tao2026swe, wang2024openhands, wang2025practitioner, wang2025swe, wei2025toward, wong2025confucius, wu2025git, xia2025live, yang2024swe, yang2025kimi, yang2025swe, yu2025building, yuan2026swe, zainullina2025guided, zeng2025satori, zeng2025skywork, zhang2025darwin, zhang2025one, zhang2025sealign, zhou2025tom};
\textbf{Existing Code Knowledge:} \cite{lindenbauer2025complexity, sun2025scaling}.%
}}

\newsavebox{\boxSourceRepositoryHistory}
\sbox{\boxSourceRepositoryHistory}{%
\KGSourceBox{Related Work}{%
\textbf{Repository Evolution Knowledge:} \cite{ahmed2025otter, cao2025siadafix, chakraborty2025blaze, chang2025bridging, chen2024coder, copet2025cwm, guo2025swe, jain2025r2e, li2025patchpilot, lindenbauer2025complexity, liu2024codexembed, luo2026closing, ma2024lingma, ma2024repository, ma2025sorft, ma2025thinking, ma2025tool, reddy2025swerank, reddy2025swerank+, sohrabizadeh2025nemotron, soni2026swe, tang2025co, tao2024magis, tao2025code, tao2026swe, wang2025improving, wang2025mcts, wang2025seamlessflow, wang2025swe, wei2025swe, wei2025toward, xie2025swe, xie2026arkeval,  yang2025kgcompass, yang2025kimi, yang2025lingxi, yang2025swe, yu2025utboost, yu2026does, zainullina2025guided, zeng2025satori, zeng2025skywork};
\textbf{Design Architecture Knowledge:} \cite{yang2025lingxi, yu2026does};
\textbf{Experiential Knowledge:} \cite{yang2025lingxi};
\textbf{Programming Language Knowledge:} \cite{reddy2025swerank+}.%
}}

\newsavebox{\boxSourceDocumentation}
\sbox{\boxSourceDocumentation}{%
\KGSourceBox{Related Work}{%
\textbf{Design Architecture Knowledge:} \cite{chen2026beyondswe, huang2025seeing, raghavendra2026agentic, zhu2025training};
\textbf{Development Tool Knowledge:} \cite{kon2026swe, sun2025scaling, yu2025building};
\textbf{Technology Stack Knowledge:} \cite{chen2026beyondswe, pan2024training, xie2026arkeval};
\textbf{Existing Code Knowledge:} \cite{guo2025swe};
\textbf{Programming Language Knowledge:} \cite{xie2026arkeval};
\textbf{Experiential Knowledge:} \cite{su2025learn};
\textbf{Domain-Specific Knowledge:} \cite{chen2026beyondswe}.%
}}

\newsavebox{\boxSourceExpert}
\sbox{\boxSourceExpert}{%
\KGSourceBox{Related Work}{%
\textbf{Experiential Knowledge:} \cite{aggarwal2025dars, cao2025siadafix, chen2024coder, chen2026beyondswe, da2025agent, gandhi2025agents, gao2025trae, kon2026swe, lei2025infantagent, li2025patchpilot, ma2024lingma, ma2025alibaba, phan2024hyperagent, pimpale2025forecasting, reddy2025swerank+, tang2025agent, tang2025co, tao2026swe, wang2025aegis, xiao2025improving, xie2025swe, yang2024swe, yu2025orcaloca, zhang2024diversity, zhu2025training};
\textbf{Development Tool Knowledge:} \cite{ahmed2025otter, cao2025siadafix, chen2025swe, chen2026beyondswe, da2025agent, gandhi2025agents, gao2025trae, guo2025swe, lei2025infantagent, liu2025graphlocator, luo2026closing, ma2025tool, pimpale2025forecasting, wang2024openhands, xia2025live, xiang2026empowering, xiang2026evaluating, yang2025swe, yuan2026swe, zhang2025one};
\textbf{Design Architecture Knowledge:} \cite{cao2025siadafix, li2025swe, nashid2025issue2test, wu2025git, zhou2025tom};
\textbf{Existing Code Knowledge:} \cite{da2025agent};
\textbf{Technology Stack Knowledge:} \cite{jain2025r2e};
\textbf{Programming Language Knowledge:} \cite{xiang2026evaluating}.%
}}

\begin{forest}
for tree={
    grow'=0,
    anchor=west,
    child anchor=west,
    parent anchor=east,
    font=\sffamily\scriptsize,
    inner sep=2.2pt,
    s sep=2.8pt,
    l sep=1.8mm,
    edge={draw=black!80},
    align=left,
},
where level=0{
    draw=purple!50,
    thick,
    rounded corners=4pt,
    fill=purple!6,
    font=\sffamily\bfseries\scriptsize,
    align=center,
    inner xsep=4pt,
    inner ysep=2pt,
    fit=band
}{},
where level=1{
    draw=blue!40,
    thick,
    rounded corners=4pt,
    fill=blue!6,
    font=\sffamily\bfseries\scriptsize,
    align=center,
    inner xsep=3pt,
    inner ysep=1.5pt,
    fit=band
}{},
where level=2{
    draw=orange!25,
    rounded corners=3pt,
    fill=leafbg,
    font=\sffamily\tiny,
    align=left,
    inner xsep=3pt, 
}{},
[Knowledge\\Sources
    [{Codebase}
        [{\usebox{\boxSourceCodebase}}]
    ]
    [{Dynamic\\Execution Results}
        [{\usebox{\boxSourceDynamicExecution}}]
    ]
    [{Repository\\History}
        [{\usebox{\boxSourceRepositoryHistory}}]
    ]
    [{Document}
        [{\usebox{\boxSourceDocumentation}}]
    ]
    [{Expert}
        [{\usebox{\boxSourceExpert}}]
    ]
]
\end{forest}

\caption{
Knowledge sources in automated issue resolution.
Each knowledge-source box lists the corresponding knowledge types and associated studies.
}
\label{fig:knowledge_source_tree}
\end{figure*}

\begin{figure*}[htbp]
\centering
\scriptsize
\definecolor{leafbg}{HTML}{FFFFEC}

\newlength{\KGExtractMethodWidth}
\setlength{\KGExtractMethodWidth}{8.0cm} 

\newcommand{\KGExtractMethodBox}[2]{%
  \begin{varwidth}[t]{\KGExtractMethodWidth}
    \normalfont\sffamily\tiny\raggedright
    \textbf{#1}\\[-1pt]
    #2%
  \end{varwidth}%
}


\newsavebox{\boxExtractManual}
\sbox{\boxExtractManual}{%
\KGExtractMethodBox{Related Work}{%
\textbf{Experiential Knowledge:} \cite{aggarwal2025dars,cao2025siadafix,chen2024coder,chen2026beyondswe,gandhi2025agents,gao2025trae,lei2025infantagent,li2025patchpilot,ma2025alibaba,phan2024hyperagent,pimpale2025forecasting,reddy2025swerank+,tang2025agent,tao2026swe,wang2024openhands,wang2025aegis,xiao2025improving,yang2024swe,yu2025orcaloca,zhang2024diversity,zhu2025training};
\textbf{Development Tool Knowledge:} \cite{ahmed2025otter,cao2025siadafix,chen2025swe,chen2026beyondswe,gandhi2025agents,gao2025trae,kon2026swe,lei2025infantagent,liu2025graphlocator,luo2026closing,ma2025tool,pimpale2025forecasting,sun2025scaling,wang2024openhands,xia2025live,xiang2026empowering,xiang2026evaluating,yang2025swe,yu2025building,yuan2026swe,zhang2025one};
\textbf{Design Architecture Knowledge:} \cite{cao2025siadafix,wu2025git};
\textbf{Technology Stack Knowledge:} \cite{jain2025r2e,pan2024training};
\textbf{Repository Evolution Knowledge:} \cite{xie2026arkeval};
\textbf{Programming Language Knowledge:} \cite{xiang2026evaluating}.%
}}

\newsavebox{\boxExtractModel}
\sbox{\boxExtractModel}{%
\KGExtractMethodBox{Related Work}{%
\textbf{Existing Code Knowledge:} \cite{ahmed2025otter,brown2024large,chen2025old,chen2026beyondswe,da2025agent,ehrlich2025codemonkeys,gao2025trae,gautam2024supercoder2,guo2025swe,gupta2025sacl,huang2025seeing,jiang2025cosil,jiang2025putting,li2024codetree,li2025swe,lindenbauer2025complexity,liu2024codexembed,liu2024marscode,liu2026architecture,luo2026closing,ma2025alibaba,mu2025experepair,nashid2025issue2test,ouyang2024repograph,pabba2025semagent,raghavendra2026agentic,reddy2025swerank,rombaut2024watson,ruan2025specrover,sohrabizadeh2025nemotron,soni2026swe,sonwane2025bugpilot,tang2025synfix,tao2024magis,tao2025code,tao2026swe,tawosi2025meta,wang2025aegis,wang2025extracting,wei2025swe,wei2025toward,xia2025demystifying,xiang2026evaluating,xiao2025improving,xie2026arkeval,xiong2025think,yang2025swe,zhang2024diversity,dong2025infcode};
\textbf{Experiential Knowledge:} \cite{chen2025locagent,chen2025swe,copet2025cwm,da2025agent,golubev2025training,jain2025r2e,kon2026swe,lin2024llms,lin2025se,lindenbauer2025complexity,liu2025context,ma2024lingma,ma2025thinking,mu2025experepair,ouyang2025reasoningbank,pan2024training,rombaut2024watson,song2026swe,su2025learn,tang2025agent,tang2025boosting,tang2025co,wang2025practitioner,wang2025swe,wong2025confucius,wu2025git,xia2025live,xie2025swe,yang2025kimi,yang2025lingxi,yang2025swe,yuan2026swe,zeng2025satori,zeng2025skywork,zhang2025darwin,zhang2025one,zhou2025tom};
\textbf{Development Tool Knowledge:} \cite{arora2024masai,chen2025prometheus,chen2026beyondswe,da2025agent,gao2025trae,guo2025swe,lin2024llms,lindenbauer2025complexity,liu2026architecture,liu2026dynamic,ma2025alibaba,nashid2025issue2test,pabba2025semagent,raghavendra2026agentic,ruan2025specrover,sohrabizadeh2025nemotron,sonwane2025bugpilot,tao2026swe,wei2025toward,wong2025confucius,xiang2026evaluating,xiao2025improving,yang2025swe,yu2025orcaloca,zhang2025sealign,dong2025infcode};
\textbf{Design Architecture Knowledge:} \cite{chen2026beyondswe,huang2025seeing,li2025swe,liu2026architecture,luo2026closing,nashid2025issue2test,raghavendra2026agentic,wu2025git,yang2025lingxi,yu2026does,zhou2025tom,zhu2025training};
\textbf{Repository Evolution Knowledge:} \cite{chang2025bridging,chen2024coder,lindenbauer2025complexity,luo2026closing,wang2025improving,wei2025toward,xie2026arkeval,yang2025lingxi,yu2026does};
\textbf{Technology Stack Knowledge:} \cite{chen2026beyondswe,xie2026arkeval};
\textbf{Programming Language Knowledge:} \cite{xie2026arkeval};
\textbf{Domain-Specific Knowledge:} \cite{chen2026beyondswe}.%
}}

\newsavebox{\boxExtractRule}
\sbox{\boxExtractRule}{%
\KGExtractMethodBox{Related Work}{%
\textbf{Existing Code Knowledge:} \cite{aggarwal2025dars,ahmed2025otter,antoniades2024swe,arora2024masai,brown2024large,cao2025siadafix,chakraborty2025blaze,chang2025bridging,chen2024coder,chen2025locagent,chen2025old,chen2025prometheus,chen2025swe,chen2026beyondswe,cheshkov2024exploring,copet2025cwm,da2025agent,dai2025lita,ding2026swe,gandhi2025agents,gao2025trae,gautam2024supercoder2,golubev2025training,gupta2025sacl,huang2025seeing,jain2025r2e,jiang2025cosil,jiang2025putting,kon2026swe,lei2024infant,lei2025infantagent,li2025infcode,li2025patchpilot,li2025swe,lin2024llms,lin2025se,lindenbauer2025complexity,liu2024codexgraph,liu2024marscode,liu2025context,liu2025graphlocator,liu2026architecture,liu2026dynamic,luo2026closing,ma2024lingma,ma2024repository,ma2025alibaba,ma2025sorft,ma2025thinking,ma2025tool,nashid2025issue2test,ouyang2024repograph,ouyang2025reasoningbank,pan2024training,phan2024hyperagent,pimpale2025forecasting,rastogi2025devstral,reddy2025swerank,reddy2025swerank+,ruan2025specrover,shi2025aime,shum2025swe,sohrabizadeh2025nemotron,song2026swe,soni2025coding,soni2026swe,sonwane2025bugpilot,sun2025scaling,suresh2024cornstack,tang2025boosting,tang2025co,tang2025synfix,tao2024magis,tao2025code,tao2026swe,tawosi2025meta,vaghasiya2025corethink,vinh2025repeton,wang2024openhands,wang2025aegis,wang2025extracting,wang2025improving,wang2025mcts,wang2025practitioner,wang2025seamlessflow,wang2025swe,wei2025swe,wong2025confucius,wu2025git,xia2025demystifying,xia2025live,xiang2026empowering,xiang2026evaluating,xie2025swe,xie2026arkeval,xiong2025think,yang2024swe,yang2025kgcompass,yang2025kimi,yang2025lingxi,yang2025swe,yu2025building,yu2025orcaloca,yu2025utboost,yu2026does,yuan2026swe,zainullina2025guided,zeng2025satori,zeng2025skywork,zhang2024autocoderover,zhang2024codev,zhang2024diversity,zhang2025cast,zhang2025darwin,zhang2025one,zhou2025tom,zhu2025training,dong2025infcode};
\textbf{Development Tool Knowledge:} \cite{aggarwal2025dars,ahmed2025otter,antoniades2024swe,arora2024masai,brown2024large,cao2025siadafix,chen2024coder,chen2025old,cheshkov2024exploring,copet2025cwm,da2025agent,dai2025lita,ding2026swe,ehrlich2025codemonkeys,gandhi2025agents,gautam2024supercoder2,golubev2025training,guo2025swe,jain2025r2e,jiang2025putting,kon2026swe,lei2024infant,lei2025infantagent,li2025infcode,li2025patchpilot,li2025swe,lin2025se,lindenbauer2025complexity,liu2024marscode,liu2025context,liu2026architecture,liu2026dynamic,ma2024lingma,ma2025thinking,mu2025experepair,nashid2025issue2test,ouyang2025reasoningbank,pabba2025semagent,pan2024training,pan2025codecor,phan2024hyperagent,pimpale2025forecasting,rastogi2025devstral,ruan2025specrover,shi2025aime,shum2025swe,sohrabizadeh2025nemotron,song2026swe,soni2025coding,soni2026swe,sonwane2025bugpilot,sun2025scaling,tang2025boosting,tang2025co,tang2025synfix,tao2024magis,tao2026swe,vaghasiya2025corethink,vinh2025repeton,wang2025aegis,wang2025practitioner,wang2025seamlessflow,wang2025swe,wu2025git,xia2025demystifying,xia2025live,xiang2026empowering,xiang2026evaluating,xie2025swe,xie2026arkeval,yang2024swe,yang2025kgcompass,yang2025kimi,yang2025lingxi,yang2025swe,yu2025building,yu2025orcaloca,yu2025utboost,yuan2026swe,zainullina2025guided,zeng2025satori,zeng2025skywork,zhang2024autocoderover,zhang2025darwin,zhu2025training};
\textbf{Repository Evolution Knowledge:} \cite{ahmed2025otter,cao2025siadafix,chakraborty2025blaze,chang2025bridging,copet2025cwm,guo2025swe,jain2025r2e,li2025patchpilot,liu2024codexembed,luo2026closing,ma2024lingma,ma2024repository,ma2025sorft,ma2025thinking,ma2025tool,reddy2025swerank,reddy2025swerank+,sohrabizadeh2025nemotron,soni2026swe,tang2025co,tao2024magis,tao2025code,tao2026swe,wang2025improving,wang2025mcts,wang2025seamlessflow,wang2025swe,wei2025swe,xie2025swe,xie2026arkeval,yang2025kgcompass,yang2025kimi,yang2025swe,yu2025utboost,yu2026does,zainullina2025guided,zeng2025satori,zeng2025skywork};
\textbf{Experiential Knowledge:} \cite{antoniades2024swe,ding2026swe,lei2024infant,liu2025context,ma2025tool,rastogi2025devstral,shum2025swe,sonwane2025bugpilot,tao2026swe,wei2025toward,yu2025building,zainullina2025guided,zhang2025darwin,zhang2025sealign,zhou2025tom};
\textbf{Programming Language Knowledge:} \cite{reddy2025swerank+,xie2026arkeval};
\textbf{Technology Stack Knowledge:} \cite{xie2026arkeval};
\textbf{Design Architecture Knowledge:} \cite{liu2026architecture}.%
}}

\begin{forest}
for tree={
    grow'=0,
    anchor=west,
    child anchor=west,
    parent anchor=east,
    font=\sffamily\scriptsize,
    inner sep=2.2pt,
    s sep=2.8pt,
    l sep=1.8mm,
    edge={draw=black!80},
    align=left,
},
where level=0{
    draw=purple!50,
    thick,
    rounded corners=4pt,
    fill=purple!6,
    font=\sffamily\bfseries\scriptsize,
    align=center,
    inner xsep=4pt,
    inner ysep=2pt,
    fit=band
}{},
where level=1{
    draw=blue!40,
    thick,
    rounded corners=4pt,
    fill=blue!6,
    font=\sffamily\bfseries\scriptsize,
    align=center,
    inner xsep=3pt,
    inner ysep=1.5pt,
    fit=band
}{},
where level=2{
    draw=orange!25,
    rounded corners=3pt,
    fill=leafbg,
    font=\sffamily\tiny,
    align=left,
    inner xsep=3pt, 
}{},
[Knowledge\\Extraction\\Methods
    [{Rule-based\\Extraction}
        [{\usebox{\boxExtractRule}}]
    ]
    [{Model-based\\Extraction}
        [{\usebox{\boxExtractModel}}]
    ]
    [{Manual\\Extraction}
        [{\usebox{\boxExtractManual}}]
    ]
]
\end{forest}

\caption{
Knowledge extraction methods in automated issue resolution.
Each extraction-method box lists the corresponding knowledge types and associated studies.
}
\label{fig:knowledge_extraction_method_tree}
\end{figure*}
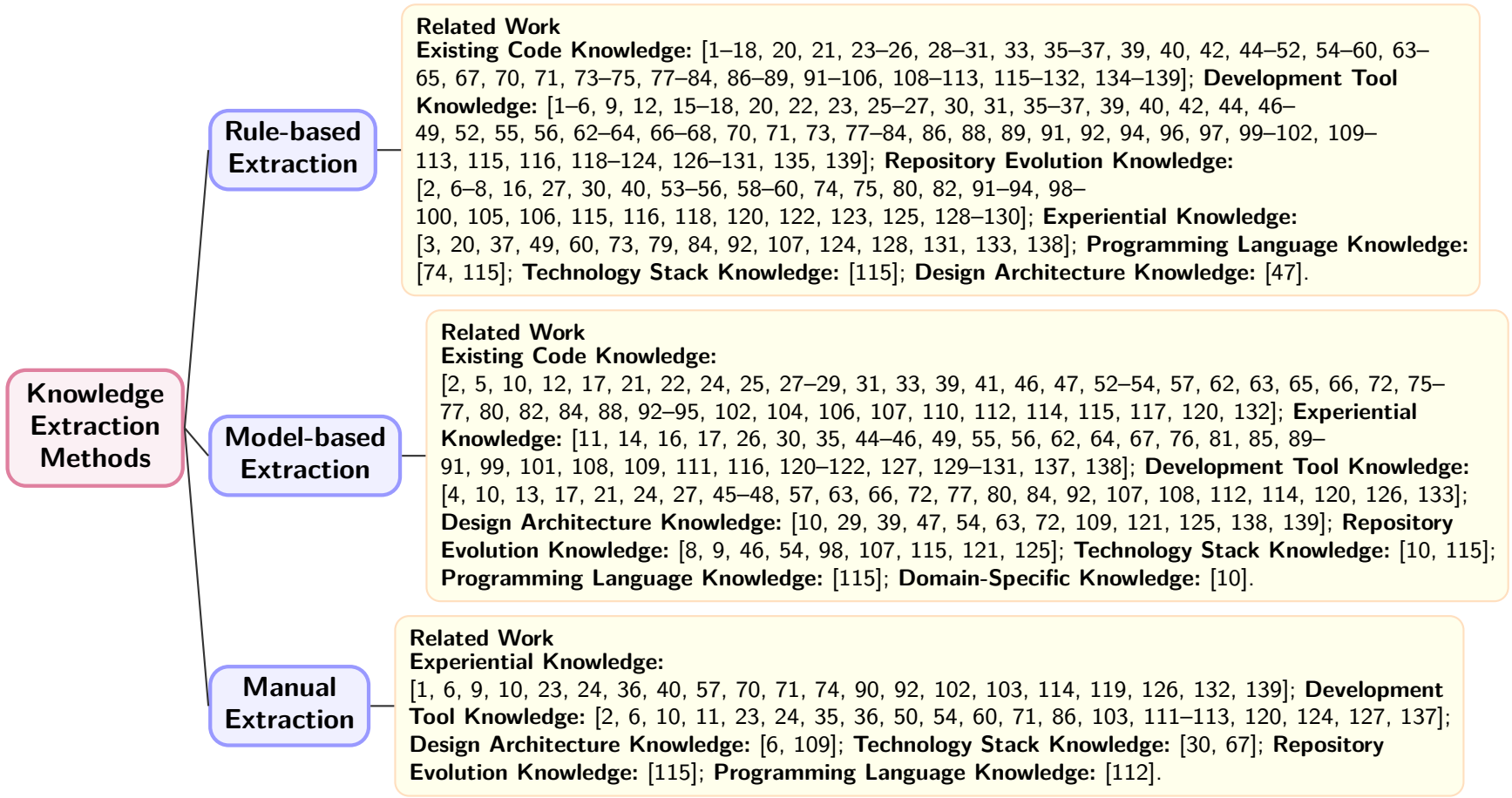

Following the knowledge-grounding schema defined in Section~\ref{sec:knowledge_grounding_schema}, this section analyzes how automated issue-resolution systems acquire knowledge from heterogeneous sources and transform it into usable forms. The central challenge is \textbf{knowledge fragmentation}: task-relevant evidence is distributed across source code, runtime feedback, repository history, documents, and expert guidance, while different sources require different extraction mechanisms. As shown in Fig.~\ref{fig:knowledge_source_tree} and Fig.~\ref{fig:knowledge_extraction_method_tree}, we therefore examine both the sources from which knowledge is obtained and the methods through which raw artifacts are converted into actionable knowledge. The following analysis focuses on representative studies and acquisition patterns, rather than redefining the categories introduced in the methodology.

\subsection{Knowledge Sources}

Based on the knowledge-source schema defined in Section~\ref{sec:knowledge_grounding_schema}, this subsection analyzes how existing automated issue-resolution systems acquire task-relevant knowledge from different sources. As summarized in Fig.~\ref{fig:knowledge_source_tree}, the surveyed studies draw knowledge from five major sources: Codebase, Dynamic Execution Results, Repository History, Document, and Expert. These sources form a heterogeneous acquisition space that combines static repository artifacts, runtime feedback, historical development records, documentary materials, and human-designed guidance.

Rather than functioning as isolated inputs, these sources often complement each other during issue resolution. Codebase provides direct repository-grounded evidence for localization and modification. Dynamic Execution Results provide behavioral feedback for validation, tool adaptation, and experience accumulation. Repository History supplies temporal evidence about how the project has evolved and how similar issues were previously resolved. Documents provide explicit semantic explanations, project conventions, ecosystem knowledge, and task specifications. Expert sources introduce methodological guidance, human-inspired workflows, and constraint-aware reasoning strategies. Together, these sources show that automated issue resolution is moving from single-source code retrieval toward multi-source knowledge grounding.

\subsubsection{Codebase}

Among all knowledge sources, Codebase is the most consistently used source in current automated issue-resolution studies. This reflects a practical requirement of repository-level repair: before an agent can generate a patch, it must identify relevant implementation artifacts, inspect how code entities are organized and reused, and understand the repository context in which the issue occurs. The surveyed studies show that codebase-derived knowledge is not limited to raw source files. Instead, it supports multiple forms of repository grounding, including implementation-level evidence, structural relations, architectural signals, tool-facing environments, and project-specific development conventions.

\textbf{Repository Content and Structural Context.}
The most common use of the codebase is to retrieve, summarize, or reorganize relevant files, functions, classes, code snippets, and their relations. This use directly supports issue localization and patch generation by helping agents identify which implementation fragments are related to the reported issue and where potential modifications should be made. For example, CodexGraph \cite{liu2024codexgraph} connects LLM agents with graph database interfaces extracted from repositories, enabling structured access to code entities and their relations. MarsCode Agent \cite{liu2024marscode} combines code analysis with agentic bug-fixing workflows to support repository-level issue resolution. Other methods, including Context Retrieval \cite{liu2025context}, SACL \cite{gupta2025sacl}, CoSIL \cite{jiang2025cosil}, GraphLocator \cite{liu2025graphlocator}, Agentless \cite{xia2025demystifying}, and long-context training approaches \cite{golubev2025training}, further exploit code snippets, semantic retrieval, call relations, repository graphs, or interaction trajectories to improve localization and patch generation. These studies show that codebase-derived implementation evidence remains the primary entry point for understanding and modifying a target repository.

\textbf{Repository Organization and Architectural Signals.}
Beyond direct code retrieval, several studies use the codebase to infer higher-level repository organization and architectural signals. This direction focuses on module responsibilities, cross-file dependencies, component boundaries, architectural awareness, and repository-specific design conventions. Architecture-aware methods \cite{liu2026architecture}, RPG-Encoder \cite{luo2026closing}, Agentic Rubrics \cite{raghavendra2026agentic}, and Git Context Controller \cite{wu2025git} extract or infer such constraints from code structures, dependency relations, versioned development context, or codebase-specific verification criteria. These studies indicate that the codebase can support not only local modification, but also system-level consistency checking, helping agents judge whether a candidate patch aligns with the broader organization of the repository.

\textbf{Tool-Oriented Repository Interaction.}
The codebase also serves as the operational environment on which development tools act. In many agent-based systems, agents do not simply read repository files; they interact with the repository through search, editing, execution, debugging, and validation tools. Representative systems such as MASAI \cite{arora2024masai}, Lita \cite{dai2025lita}, and Nemotron-CORTEXA \cite{sohrabizadeh2025nemotron} show how tool-mediated repository interaction supports information gathering, action planning, and iterative problem solving. This line of work suggests that the codebase functions as both a knowledge source and an action space, where agents acquire information by operating on repository artifacts rather than only retrieving them.

\textbf{Project Ecosystem and Development Practices.}
Some studies further use the codebase to infer broader project-level knowledge, including dependencies, programming conventions, platform constraints, and reusable interaction patterns. SWE-Gym \cite{pan2024training} constructs executable environments with real codebases, dependencies, and tests, making the repository a source for learning project-specific development settings and agent-environment interaction patterns. ArkEval \cite{xie2026arkeval} highlights the need to understand ArkTS and HarmonyOS-specific development conventions, showing that codebases may embody language-level and platform-specific constraints that are not captured by generic code retrieval alone. Experience-oriented methods such as Repairity \cite{tang2025boosting} reuse prior codebase interaction patterns to guide future issue resolution. These studies indicate that codebase-derived knowledge can extend from implementation details to technology-stack awareness, language-specific constraints, and reusable problem-solving experience.

Overall, the codebase functions as a multi-layered knowledge substrate for automated issue resolution. It provides implementation evidence for localization and patch generation, structural and architectural signals for repository-wide reasoning, tool-facing environments for interactive problem solving, and project-specific cues for understanding development constraints. This explains why codebase-derived knowledge remains the dominant foundation of current issue-resolution systems.

\subsubsection{Dynamic Execution Results}

Dynamic Execution Results are mainly used in studies that treat issue resolution as an interactive and behavior-grounded process. Instead of relying only on static repository inspection, these systems obtain feedback from tests, logs, error traces, command-line responses, verifier outputs, tool observations, and environment interactions. Such runtime signals help agents judge whether an intermediate action, generated test, or candidate patch is actually valid. The surveyed studies show that this source supports three recurring forms of runtime grounding: adapting tool actions, accumulating reusable experience, and managing long-horizon context.

\textbf{Runtime Feedback for Tool Adaptation.}
Runtime feedback is commonly transformed into actionable signals for revising intermediate decisions during issue resolution. Test outputs, error messages, verifier results, command-line responses, and tool observations help agents decide whether to continue, revise, branch, or discard a trajectory. For example, DARS \cite{aggarwal2025dars} branches and re-samples agent actions according to execution feedback from previous attempts, while Otter \cite{ahmed2025otter} uses generated test execution and rule-based repair signals to validate whether fail-to-pass tests correctly reproduce issues. SWE-Search \cite{antoniades2024swe} incorporates iterative feedback into Monte Carlo Tree Search to guide trajectory refinement, and MASAI \cite{arora2024masai} relies on tool-mediated observations collected by specialized sub-agents during localization, patching, and testing. Similarly, inference-scaling and adaptive-repair methods such as Large Language Monkeys \cite{brown2024large}, SIADAFIX \cite{cao2025siadafix}, and guided search strategies \cite{zainullina2025guided} use execution or verifier feedback to select, revise, or rank candidate trajectories. These studies show that dynamic execution results provide direct behavioral evidence for tool adaptation, enabling agents to resolve issues through repeated acting, observing, and adjusting.

\textbf{Trajectory Accumulation and Experience Distillation.}
Beyond immediate tool adaptation, dynamic execution results are also accumulated and distilled into reusable experience. Instead of treating execution feedback as a one-time validation signal, these methods preserve successful and failed attempts, reward signals, environment-interaction traces, and reasoning trajectories to train agents, construct memories, build reward models, or guide future reasoning. EXPEREPAIR \cite{mu2025experepair}, ReasoningBank \cite{ouyang2025reasoningbank}, Devstral \cite{rastogi2025devstral}, SWE-RM \cite{shum2025swe}, SWE-Master \cite{song2026swe}, SWE-Exp \cite{chen2025swe}, CWM \cite{copet2025cwm}, and Agent-RLVR \cite{da2025agent} accumulate execution trajectories, reward signals, successful and failed attempts, or environment-interaction traces for agent training, experience-bank construction, reward modeling, and future reasoning guidance. These studies indicate that runtime feedback can be transformed from immediate behavioral evidence into longer-term problem-solving experience, helping agents learn how to explore repositories, recover from errors, refine patches, and improve long-horizon decision making.

\textbf{Execution-Guided Context Management.}
In long-horizon settings, dynamic execution results further support context management. Runtime observations do not merely indicate whether a patch passes or fails; they also help agents decide which observations, code fragments, tool outputs, or intermediate results should remain available for later reasoning. For instance, \cite{lindenbauer2025complexity} shows that masking old observations can manage long agent contexts efficiently, while \cite{sun2025scaling} folds completed sub-trajectories into concise summaries so that agents can preserve relevant code-related outcomes without carrying full execution histories. These studies indicate that execution feedback can support context retention and compression under limited context windows, especially when issue resolution requires multi-step exploration across files, commands, tests, and intermediate observations.

Overall, Dynamic Execution Results provide the behavioral grounding needed for adaptive issue resolution. They help agents revise tool actions, validate candidate patches, accumulate reusable experience, and manage long interaction histories. This source is therefore especially important for agentic systems that rely on iterative exploration and feedback interpretation rather than one-shot patch generation.

\subsubsection{Repository History}

Repository History is used when systems need temporal evidence about how a project has evolved and how similar problems were previously addressed. Compared with the current codebase, repository history provides evidence from commits, pull requests, issue-fixing records, historical patches, test changes, and other evolution traces. The surveyed studies use this source not only to mine past changes, but also to construct training data, infer design constraints, and reuse prior repair experience.

\textbf{Historical Change Mining.}
Historical change mining uses past software changes and issue-fixing records to capture how repositories evolve over time. This direction helps agents understand which code locations were previously modified, what kinds of issues occurred, and how similar problems were resolved. Satori-SWE \cite{zeng2025satori} formulates software engineering improvement as an evolutionary test-time scaling process, while Skywork-SWE \cite{zeng2025skywork} scales training data from real-world repository tasks and validated trajectories. Other studies, such as KGCompass \cite{yang2025kgcompass}, Kimi-Dev \cite{yang2025kimi}, Lingxi \cite{yang2025lingxi}, UTBoost \cite{yu2025utboost}, MCTS-Refined CoT \cite{wang2025mcts}, SeamlessFlow \cite{wang2025seamlessflow}, SWE-Dev \cite{wang2025swe}, SWE-RL \cite{wei2025swe}, and Self-play SWE-RL \cite{wei2025toward}, further exploit historical issues, pull requests, patches, tests, trajectories, or software evolution data to support localization, repair, training, evaluation, and reinforcement learning. These studies show that historical change mining provides direct temporal evidence for connecting current issues with prior repository modifications.

\textbf{Evolution-Aware Training and Evaluation.}
Repository history is also widely used to construct training data, reward signals, benchmark instances, or evaluation resources. Compared with directly mining historical changes for localization or repair, this direction focuses more on scaling supervision and evaluation from past development activities. For example, Skywork-SWE \cite{zeng2025skywork} scales training data from real-world repository tasks and validated trajectories, while SWE-RL \cite{wei2025swe} and Self-play SWE-RL \cite{wei2025toward} use historical software engineering tasks and feedback signals to support reinforcement learning for issue-resolution agents. UTBoost \cite{yu2025utboost} also reflects this direction by using test-related historical information to strengthen evaluation beyond the original benchmark tests. These studies indicate that repository history can serve as a reusable resource for building realistic training and evaluation settings, because it records not only final patches but also the development context in which those patches were produced and validated.

\textbf{History-Derived Design Constraints.}
Another important role of repository history is to expose design and architecture constraints from past development decisions. Pull requests, code reviews, issue discussions, and accepted patches often encode project-specific architectural conventions, coding styles, component responsibilities, and maintainability requirements. Lingxi \cite{yang2025lingxi} abstracts design patterns and coding practices from historical issue-fixing pairs, showing that prior repairs can provide guidance about how a repository tends to structure acceptable solutions. Similarly, \cite{yu2026does} shows that design constraints mined from real pull requests can reveal quality requirements that are not captured by pass-rate-based evaluation alone. These studies suggest that repository history can make implicit design expectations more observable, helping agents judge whether a generated patch is consistent with the project's long-term architectural and maintainability practices.

\textbf{Historical Repair Experience.}
More recent studies further transform prior issue-fixing processes into reusable repair guidance. Rather than using history only as raw data about code changes, this direction extracts procedural knowledge about how previous issues were analyzed, localized, planned, fixed, and validated. Lingxi \cite{yang2025lingxi} is representative in this direction, as it extracts procedural knowledge from previous fixes and uses similar historical experiences to guide future issue resolution. This direction is especially useful for long-horizon issue resolution because it allows agents to reuse prior development strategies, repair plans, and decision patterns when facing similar repository contexts or issue types. In addition, historical repair data can also support language-aware reasoning; for example, SWE-Rank+ \cite{reddy2025swerank+} leverages historical repair data to improve understanding of language-specific repair patterns and solution ranking.

Overall, Repository History complements static repository artifacts by adding temporal traceability. It helps agents reason from accumulated development evidence, accepted repair practices, historical software evolution, and prior issue-resolution processes. This source is therefore important for moving agents beyond the current code snapshot toward historically informed localization, patch generation, training, and evaluation.

\subsubsection{Document}

Documents are mainly used when systems need explicit semantic explanations that cannot be reliably inferred from source code or execution feedback alone. In the surveyed studies, documentary materials support requirement interpretation, project guidance, tool workflows, ecosystem awareness, and domain-specific reasoning. This source is especially useful when issue resolution depends on understanding what the software is expected to do, how contributors should interact with the project, which external APIs or platforms constrain the solution, or what task-specific assumptions should guide the final patch.

\textbf{Specification and Requirement Interpretation.}
Documents often help agents interpret software requirements, expected behavior, and task-level specifications. This direction is important when the issue cannot be solved only by inspecting source code, because the agent must understand what the software is intended to achieve before deciding how the repository should be modified. BeyondSWE \cite{chen2026beyondswe} extends software engineering tasks beyond single-repository bug fixing and emphasizes scenarios such as dependency migration and document-to-repository generation, where agents must understand specifications, official documentation, or human-written requirements. Seeing is Fixing \cite{huang2025seeing} retrieves relevant project documents to interpret visual issue scenarios and connect GUI symptoms with executable code contexts. SWE-Factory \cite{guo2025swe} also reflects this category by processing issue descriptions, environment setup requirements, grading information, and validation signals to construct issue-resolution benchmark instances. These studies show that documents help agents connect natural-language requirements, issue descriptions, and expected system behavior with concrete repository changes.

\textbf{Project Guidance and Verification Rules.}
Documentary materials are also used to extract project-level guidance and verification rules. This direction focuses on how documents describe repository conventions, contribution rules, project-specific expectations, and criteria for judging whether a patch is acceptable. Agentic Rubrics \cite{raghavendra2026agentic} uses repository documents such as README files, contributing guidelines, and pull request descriptions to construct context-grounded verification rubrics. Similarly, \cite{zhu2025training} leverages synthetic project descriptions, task specifications, and test definitions to generate training environments for versatile coding agents. These studies indicate that documents can provide explicit project guidance that helps agents evaluate candidate solutions beyond local code correctness, especially when patch quality depends on project conventions, task definitions, or repository-specific acceptance criteria.

\textbf{Tool Workflow Instructions.}
In agentic settings, documents can further guide tool interaction, environment operation, and long-horizon workflows. Here, documents do not merely describe what the software should do; they also provide instructions or demonstrations about how agents should interact with tools, environments, or multi-step development processes. SWE-Protégé \cite{kon2026swe} incorporates expert guidance into software engineering trajectories, showing how documented or human-provided guidance can shape agent behavior during issue resolution. Context-Folding \cite{sun2025scaling} manages long-horizon reasoning by folding tool-intensive sub-trajectories into concise summaries, while \cite{yu2025building} optimizes multi-turn tool-assisted coding agents through trajectory-level preference learning. These studies suggest that documents can serve as workflow-level guidance for organizing tool use, summarizing intermediate processes, and improving agent-environment coordination.

\textbf{Ecosystem- and Language-Specific Documents.}
Documents are also used to obtain knowledge about external technical ecosystems and programming-language constraints. This role becomes important when issue resolution depends on frameworks, libraries, APIs, dependency versions, runtime settings, or language-specific development rules rather than only on the target repository. BeyondSWE \cite{chen2026beyondswe} includes tasks that require external knowledge about domains, dependencies, and official documentation. SWE-Gym \cite{pan2024training} constructs training environments with task descriptions, dependencies, runtime settings, and tests, making documents a source for understanding executable project settings. ArkEval \cite{xie2026arkeval} highlights the importance of ArkTS and HarmonyOS-specific documents for evaluating automated repair in a low-resource programming ecosystem. These studies show that documents can provide the external technical context needed for agents to handle framework-dependent, platform-specific, or language-constrained issue-resolution tasks.

\textbf{Instructional and Domain Resources.}
Instructional documents and domain-specific resources can be transformed into reusable knowledge for agent adaptation and specialized reasoning. Learn-by-interact \cite{su2025learn} synthesizes agent-environment interaction data from document and tutorials, turning static instructional resources into reusable demonstrations for future agent adaptation. BeyondSWE \cite{chen2026beyondswe} also explicitly includes domain-specific issue-resolution settings that require agents to consult and integrate specialized external information. This direction shows that documents can provide more than project-level explanations: they can encode domain rules, expert assumptions, tutorials, and task-specific procedures that help agents reason about requirements beyond general programming and repository navigation.

Overall, Document complements code-centric and history-centric sources by providing explicit semantic and normative context. It helps agents understand expected behavior, project conventions, workflow instructions, external technical constraints, and domain-specific assumptions. This source is therefore important for issue-resolution tasks where correctness depends not only on what the repository contains, but also on what the software is expected to do and which external rules should constrain the final patch.

\subsubsection{Expert}

Expert sources are used when automated issue-resolution systems require high-level methodological guidance rather than only artifact-based evidence. In the surveyed studies, expert guidance is incorporated through curated workflows, handcrafted heuristics, expert feedback, process-level supervision, collaboration rules, evaluation criteria, and manually designed tool-use strategies. Unlike sources that mainly provide artifacts to be mined, expert sources shape how agents diagnose issues, coordinate tools, reason about constraints, and structure long-horizon resolution actions.

\textbf{Expert-Guided Resolution Strategies.}
Expert guidance is often transformed into reusable resolution strategies, process supervision, and feedback mechanisms. This direction focuses on how human-like guidance can help agents plan issue resolution, avoid ineffective behaviors, learn from failed attempts, and follow more disciplined task-solving procedures. Agent-RLVR \cite{da2025agent} introduces agent guidance inspired by human pedagogy, using strategic plans and feedback on failed attempts to steer agents toward successful trajectories. SWE-PRM \cite{gandhi2025agents} designs a taxonomy-guided process reward model to detect inefficient agent behaviors such as looping, redundant exploration, and premature termination. SWE-Protégé \cite{kon2026swe} frames software maintenance as an expert--protégé collaboration process, where a small model learns when to request help from a stronger expert and how to follow through on expert advice. Other systems such as Co-PatcheR \cite{tang2025co}, SWE-Lego \cite{tao2026swe}, and AEGIS \cite{wang2025aegis} also embody expert experience by decomposing issue resolution into specialized stages, designing validated training recipes, or imposing rule-based reproduction procedures. These studies show that expert-guided strategies can convert human issue-resolution experience into reusable procedural knowledge for improving agent planning, correction, and long-horizon problem solving.

\textbf{Expert-Designed Tool Coordination.}
Expert sources also influence how agents use, coordinate, or learn from tools. This direction focuses on tool-use principles designed by human experts, including how to organize tool-mediated trajectories, how to construct agent scaffolds, how to select among candidate solutions, and how to support structure-aware repository navigation. Agent-RLVR \cite{da2025agent} and SWE-PRM \cite{gandhi2025agents} use expert-style feedback to guide tool-mediated trajectories. Trae Agent \cite{gao2025trae} designs modular agents for generation, pruning, and selection in large ensemble spaces. RPG-Encoder \cite{luo2026closing} introduces a unified repository representation as an expert-designed interface for structure-aware navigation. ToolTrain \cite{ma2025tool} trains models to use repository retrieval tools for deep issue localization, and \cite{pimpale2025forecasting} analyzes agent capability elicitation through optimized scaffolds, tools, and prompts. These studies indicate that expert-designed tool coordination helps agents move beyond arbitrary tool invocation toward more purposeful, structured, and efficient interaction with repository environments.

\textbf{Human-Inspired Constraint Reasoning.}
Expert-like reasoning patterns further help impose design, architecture, testing, workflow, and human-intent constraints on issue resolution. This direction reflects how experienced developers evaluate whether a code change is appropriate not only by checking local correctness, but also by considering maintainability, project conventions, user intent, and broader system organization. SIADAFIX \cite{cao2025siadafix} translates the fast-and-slow thinking paradigm of human software experts into adaptive workflow rules for issue resolution. SWE-Debate \cite{li2025swe} organizes specialized agents with different reasoning perspectives to consolidate localization and solution plans. Issue2Test \cite{nashid2025issue2test} uses project-specific guidelines and iterative refinement principles to generate reproducing tests. Git Context Controller \cite{wu2025git} borrows the concepts of commit, branch, merge, and context from version control to structure long-horizon agent memory. ToM-SWE \cite{zhou2025tom} models user goals, preferences, and constraints to better align issue-resolution behavior with human intent. These studies show that expert sources can provide constraint-aware reasoning mechanisms that help agents produce code changes aligned with project design, testing requirements, and human expectations.

\textbf{Expert-Informed Context and Environment Understanding.}
Expert guidance is also used to improve how agents understand relevant code contexts, technology stacks, programming-language constraints, and execution environments. In this setting, expert sources help agents identify what information should be inspected and which external assumptions should constrain the resolution process. Agent-RLVR \cite{da2025agent} uses expert guidance to direct agents toward relevant code contexts. R2E-Gym \cite{jain2025r2e} incorporates procedural environment construction and hybrid verification strategies that reflect expert understanding of software execution environments. Furthermore, \cite{xiang2026evaluating} identifies Rust-specific challenges such as type and trait semantics, motivating expert-informed strategies for repository-level Rust issue resolution. These studies suggest that expert sources can complement artifact-based sources by clarifying how agents should interpret repository context, execution settings, technology-stack requirements, and language-specific constraints during issue resolution.

Overall, Expert sources complement artifact-based sources by providing methodological and constraint-aware guidance. They help agents plan issue resolution, coordinate tools, avoid inefficient behaviors, reason about project and human constraints, and focus on relevant code and environment information. This source therefore supports a more disciplined, controllable, and human-aligned form of automated issue resolution.

\subsection{Knowledge Extraction Methods}

Based on the knowledge-extraction schema defined in Section~\ref{sec:knowledge_grounding_schema}, this subsection analyzes how existing automated issue-resolution systems transform raw software artifacts, runtime signals, historical records, documents, and expert inputs into usable knowledge. As summarized in Fig.~\ref{fig:knowledge_extraction_method_tree}, the surveyed studies mainly rely on three extraction methods: Rule-based Extraction, Model-based Extraction, and Manual Extraction. These methods differ not only in their technical mechanisms, but also in the kinds of knowledge they are most suitable for acquiring.

Across the surveyed literature, rule-based extraction is commonly used when the target knowledge is embedded in explicit software structures, stable formats, tool outputs, or well-defined operational rules. Model-based extraction is increasingly adopted when the knowledge is semantically implicit, noisy, cross-modal, or difficult to capture through handcrafted procedures. Manual extraction remains important when knowledge requires expert judgment, qualitative interpretation, careful benchmark design, or human-curated process experience. Together, these extraction methods show that automated issue resolution depends on a hybrid knowledge-acquisition process, where deterministic analysis, learned semantic inference, and expert curation complement one another.

\subsubsection{Rule-based Extraction}

Rule-based Extraction is widely used in studies that need controllable and reproducible ways to acquire knowledge from structured or semi-structured software artifacts. It is especially suitable for code structures, tool outputs, execution traces, historical links, and constraint patterns that can be captured through predefined procedures, parsers, static or dynamic analysis, or tool-specific rules. The surveyed studies show that rule-based methods mainly support four extraction patterns: structure-based repository extraction, rule-based runtime feedback extraction, linking-based historical extraction, and constraint-based knowledge extraction.

\textbf{Structure-based Repository Extraction.}
Structure-based repository extraction is commonly used to obtain code-related knowledge from files, modules, classes, functions, code spans, call graphs, dependency relations, and other explicit repository structures. By following structural signals, systems can narrow the search space and identify issue-relevant code locations before patch generation. For example, CoSIL searches module-level and function-level call graphs to narrow down issue-relevant code locations \cite{jiang2025cosil}. Other systems organize repository contexts through predefined file-, function-, or context-selection rules \cite{jiang2025putting,kon2026swe,lei2024infant}. These methods show that structure-based extraction provides a controllable way to transform large and complex codebases into structured contexts that agents can inspect, retrieve, and modify.

\textbf{Rule-based Runtime Feedback Extraction.}
Runtime feedback extraction uses predefined rules to process tool outputs and execution observations. In this setting, raw signals from testing, debugging, execution, search, editing, and validation tools are converted into usable feedback for subsequent localization, patch refinement, or validation. Agent frameworks often rely on explicit tool interfaces, command protocols, debugging rules, execution traces, and context-management operations to determine how feedback should be interpreted and incorporated into the next step. For example, MarsCode Agent integrates tool-supported procedures for bug reproduction, localization, patch generation, and validation \cite{liu2024marscode}. CAT maintains structured context through explicit context-management operations \cite{liu2025context}. Dynamic-analysis-based methods expose runtime states through predefined tracing or debugging interfaces \cite{xiang2026empowering,xiang2026evaluating}. Similar rule-guided retrieval and filtering mechanisms are also adopted in efficient issue-resolution workflows such as SWE-Fixer \cite{xie2025swe}. These studies indicate that rule-based runtime feedback extraction helps agents turn low-level execution and tool signals into actionable evidence for iterative issue resolution.

\textbf{Linking-based Historical Extraction.}
Linking-based historical extraction connects historical artifacts, software evolution records, and prior agent trajectories with current issue-resolution tasks. Repository-aware approaches extract historical knowledge from issues, pull requests, commits, patches, and code changes by aligning software evolution records with code entities \cite{yang2025kgcompass,chakraborty2025blaze,chang2025bridging}. In this way, historical artifacts become usable temporal evidence for understanding how similar issues were previously resolved or which code locations were frequently involved in related changes. Agent-oriented systems further reuse prior trajectories, archived trials, or interaction histories as experience-oriented guidance, enabling agents to refine search strategies, resume from intermediate states, or avoid repeated ineffective actions \cite{antoniades2024swe,ding2026swe,lei2024infant,liu2025context,ma2025tool}. These studies show that linking-based extraction makes repository history and agent trajectories operationally useful by connecting past development or interaction records to the current issue-resolution process.

\textbf{Constraint-based Knowledge Extraction.}
Constraint-based extraction applies predefined rules or domain-specific analysis to obtain knowledge that guides issue resolution beyond local code context. This pattern is useful for extracting programming-language rules, technology-stack conventions, architectural constraints, and other background assumptions when they can be identified through stable syntactic, semantic, or structural patterns. For example, multilingual and language-aware localization methods use rule-guided processing to handle language-specific solution patterns or ranking signals \cite{reddy2025swerank+}. ArkEval highlights the importance of extracting ArkTS- and HarmonyOS-specific constraints in a low-resource programming ecosystem \cite{xie2026arkeval}. Architecture-aware methods also use structural rules to identify design constraints and repository-level architectural signals \cite{liu2026architecture}. These studies suggest that rule-based extraction can support broader constraint awareness when the relevant knowledge is encoded in explicit patterns, predefined analysis procedures, or domain-specific rules.

Overall, rule-based extraction provides a controllable and interpretable foundation for knowledge acquisition in automated issue resolution. It is particularly effective when knowledge is grounded in explicit structures, stable formats, or well-defined operational rules. By parsing repository structures, processing runtime feedback, linking historical artifacts, and extracting explicit constraints, rule-based methods help agents obtain reliable knowledge before or during interactive issue resolution.

\subsubsection{Model-based Extraction}

Model-based Extraction is increasingly adopted in studies where the target knowledge is difficult to obtain through fixed rules alone. Learned models, including embedding models, LLMs, multimodal LLMs, reward models, and trained agent policies, are used to infer, retrieve, summarize, rank, or transform knowledge from heterogeneous software artifacts. This method is especially useful when knowledge is semantically implicit, noisy, cross-file, cross-modal, or weakly structured. The surveyed studies mainly use model-based extraction for semantic repository understanding, trajectory summarization, and heterogeneous knowledge inference.

\textbf{Representation-based Semantic Extraction.}
Representation-based semantic extraction uses learned representations to obtain semantically relevant knowledge from software artifacts. Instead of relying only on exact lexical matching, handcrafted patterns, or explicit structural rules, this mechanism encodes code, repository structures, dependencies, and historical artifacts into semantic spaces for retrieval, matching, ranking, and repository understanding. This is especially useful when the relationship between an issue description and the relevant code context is indirect or semantically distant. For example, CodeXEmbed learns code embeddings for multilingual and multi-task retrieval \cite{liu2024codexembed}. RepoGraph and architecture-aware methods construct model-assisted repository representations to capture code entities, dependencies, and architectural contexts \cite{ouyang2024repograph,liu2026architecture}. Other studies further extract conceptual or repository-evolution knowledge to bridge the semantic gap between issue descriptions and relevant code locations \cite{wang2025extracting,chen2024coder,luo2026closing,wang2025improving}. These studies show that representation-based extraction helps agents obtain issue-relevant repository knowledge even when relevant artifacts are scattered, indirectly expressed, or difficult to locate through surface-level matching.

\textbf{Summarization-based Trajectory Extraction.}
Model-based extraction is also used to summarize, filter, and distill agent trajectories, prior attempts, tool interactions, and historical resolution processes. Raw trajectories are often long, noisy, and difficult to reuse directly, so models are used to convert them into reasoning memories, demonstrations, reflective insights, training trajectories, or feedback signals. Several studies distill raw interaction histories into experience-oriented knowledge, such as reasoning memories, resolution demonstrations, reflective insights, or training trajectories \cite{ma2025thinking,mu2025experepair,ouyang2025reasoningbank,wang2025practitioner,wang2025swe,wong2025confucius}. Other methods use learned models to identify useful actions, reduce noisy trajectories, or transform tool-use processes into actionable feedback for future agent behavior \cite{xiao2025improving,yu2025orcaloca,zhang2025sealign}. These studies indicate that summarization-based extraction can convert long-horizon interaction processes into compact, reusable, and generalizable knowledge for planning, tool use, error recovery, and future issue resolution.

\textbf{Inference-based Heterogeneous Extraction.}
Inference-based heterogeneous extraction uses LLMs or multimodal models to acquire knowledge from implicit, weakly structured, or cross-modal sources. This pattern appears when the target knowledge cannot be directly obtained through deterministic rules, such as domain-specific requirements, architectural intent, language-specific constraints, technology-stack conventions, or visual issue information. In these cases, models help interpret heterogeneous inputs, infer latent constraints, and connect high-level task semantics with repository-level actions. BeyondSWE emphasizes cross-repository, domain, document, and specification knowledge \cite{chen2026beyondswe}. GUIRepair uses multimodal LLMs to interpret visual issue information and connect visual symptoms with executable code contexts \cite{huang2025seeing}. RPG-Encoder and architecture-aware methods extract higher-level repository intent and architectural constraints from repository structures and development contexts \cite{luo2026closing,liu2026architecture}. ArkEval focuses on ArkTS-specific language and technology-stack knowledge in a low-resource programming ecosystem \cite{xie2026arkeval}. These studies show that inference-based extraction extends automated issue resolution beyond explicit artifact processing by enabling agents to acquire implicit design intent, external technical knowledge, specialized domain assumptions, programming-language constraints, and multimodal issue evidence.

Overall, model-based extraction expands the knowledge boundary of automated issue resolution. It allows agents to acquire semantically relevant repository context, distill reusable experience from long interaction histories, and infer implicit or heterogeneous knowledge that cannot be easily captured by handcrafted rules. This makes model-based extraction especially important for moving issue-resolution systems from explicit artifact processing toward deeper semantic understanding, experience reuse, and cross-modal reasoning.

\subsubsection{Manual Extraction}

Manual Extraction appears in studies where automatic rules or learned models alone are insufficient for obtaining reliable knowledge. Human effort is used to inspect, annotate, curate, or specify knowledge that requires expert judgment, qualitative interpretation, or careful task and environment design. Rather than serving as a large-scale extraction mechanism, manual extraction often provides high-quality knowledge for experience construction, tool and environment design, benchmark creation, and high-level constraint specification.

\textbf{Curation-based Experience Extraction.}
Curation-based experience extraction manually organizes agent behaviors, failure cases, successful trajectories, and task-solving patterns into reusable process knowledge. This mechanism is useful when the value of a trajectory depends not only on its final outcome, but also on the quality of intermediate decisions, exploration strategies, error recovery, and process-level reasoning. For example, AGENT KB leverages manually crafted examples and cross-domain experiences to support agentic problem solving \cite{tang2025agent}, while SWE-Lego and related training-oriented studies curate high-quality trajectories or task instances to improve software engineering agents \cite{tao2026swe,zhu2025training}. Other works manually inspect or organize agent behaviors to identify inefficiencies, diverse expertise, or recurring failure patterns, thereby informing trajectory reduction, agent collaboration, or issue reproduction strategies \cite{wang2025aegis,xiao2025improving,zhang2024diversity}. These studies show that curation-based extraction can transform raw agent activities into reusable process knowledge that supports planning, collaboration, and long-horizon issue resolution.

\textbf{Design-based Tool and Environment Extraction.}
Manual extraction is also reflected in the design of agent-accessible interfaces, executable environments, sandboxes, navigation tools, and workflow settings. In this setting, human effort encodes expert understanding of how agents should interact with development tools, debugging commands, execution traces, and repository environments. For instance, ADI introduces an agent-centric debugging interface with high-level navigation commands and execution traces \cite{xiang2026empowering}, while RustForger incorporates Rust-specific dynamic tracing and environment setup based on observed limitations of existing agents \cite{xiang2026evaluating}. Other studies construct executable training environments, lightweight sandboxes, or simplified navigation tools to provide agents with curated tool-interaction feedback \cite{yu2025building,yuan2026swe,zhang2025one}. These studies indicate that design-based extraction helps convert expert understanding of software environments into structured tool-use settings that agents can learn from and operate within.

\textbf{Expert-judgment-based Constraint Extraction.}
Expert judgment is further used to identify or specify high-level constraints that are difficult to infer reliably from raw artifacts alone. This includes architectural assumptions, workflow constraints, language-specific requirements, technology-stack settings, benchmark construction criteria, and other expert-defined conditions that shape the issue-resolution process. SIADAFIX uses expert-designed complexity assessment and workflow orchestration to guide adaptive resolution modes \cite{cao2025siadafix}, while Git Context Controller introduces a version-control-inspired memory structure to manage long-horizon architectural contexts \cite{wu2025git}. Similarly, R2E-Gym and SWE-Gym involve careful construction of executable environments and verifiers that reflect technology-stack requirements \cite{jain2025r2e,pan2024training}, and ArkEval manually curates and standardizes ArkTS issue-resolution tasks to capture language- and ecosystem-specific knowledge \cite{xie2026arkeval}. These studies show that expert-judgment-based extraction is valuable for knowledge that requires qualitative interpretation, domain familiarity, or careful task and evaluation design.

Overall, manual extraction provides human-curated knowledge where fully automatic extraction may be insufficient. It turns agent behaviors and trajectories into reusable experience, embeds expert understanding into tool interfaces and executable settings, and captures architectural, language-level, technology-stack, and benchmark-specific constraints. Therefore, manual extraction complements rule-based and model-based methods by injecting expert judgment into experience construction, tool and environment design, and high-level constraint specification for automated issue resolution.

\subsection{Distribution Flow from Knowledge Sources to Extraction Methods}

\begin{figure}[htbp] 
    \centering 
    \includegraphics[width=0.9\linewidth]
    {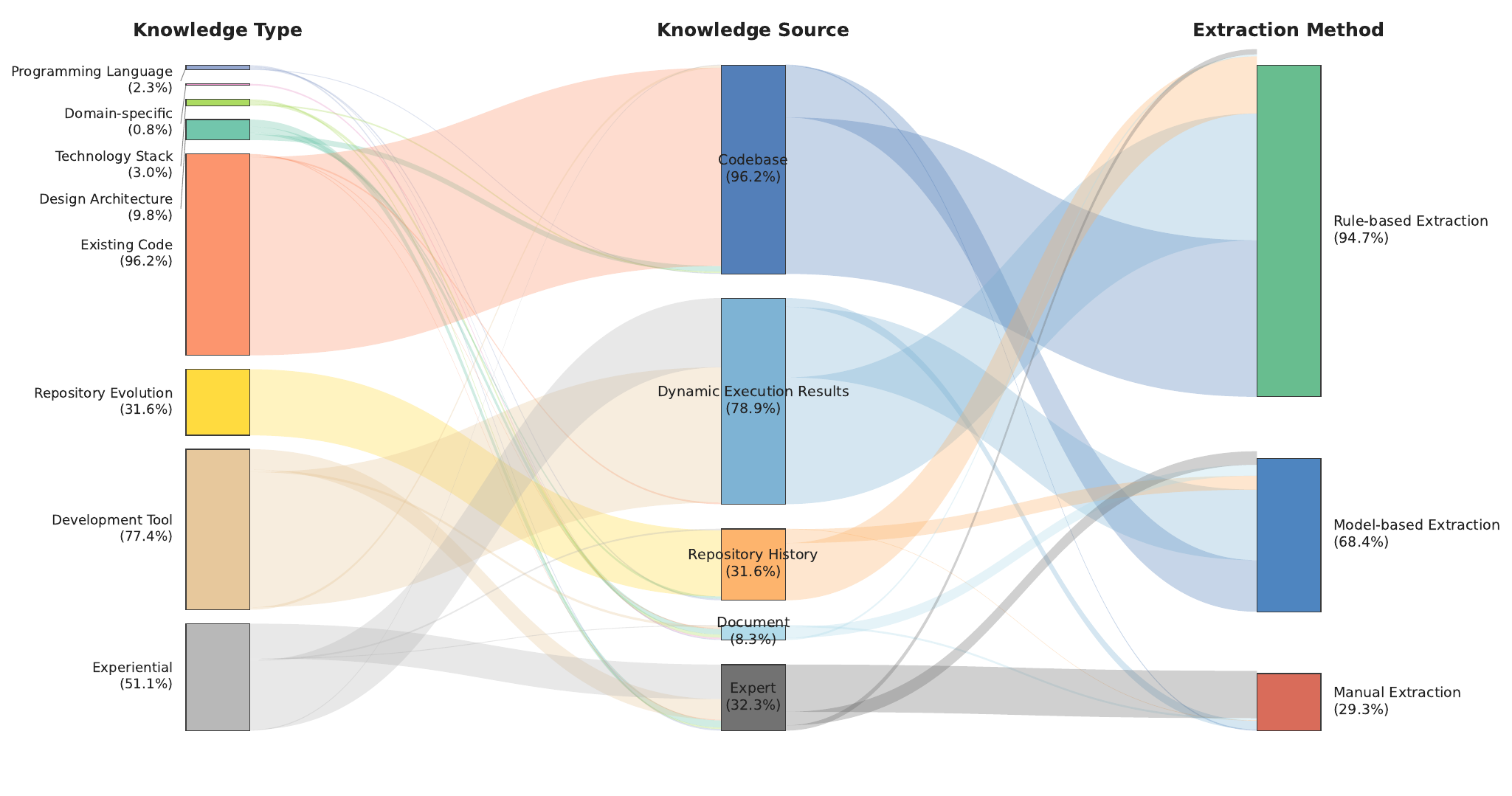} 
    \caption{Sankey-style distribution of knowledge types, knowledge sources, and extraction methods.} 
    \label{fig:source_method} 
\end{figure}

Fig.~\ref{fig:source_method} presents a Sankey-style overview of the relationships among knowledge types, knowledge sources, and extraction methods. Compared with marginal frequency statistics, this visualization further illustrates which source--extraction pathways are more frequently adopted in current studies, which pathways are becoming increasingly visible, and which high-level knowledge sources still have substantial room for further exploration. Overall, the strongest flows are formed around Codebase, Dynamic Execution Results, Existing Code Knowledge, Development Tool Knowledge, and Rule-based Extraction. Existing Code Knowledge accounts for 96.2\% of the analyzed studies and is mainly acquired from the Codebase source, which appears in 96.2\% of the studies. This pattern is expected because repository-level issue resolution first requires agents to inspect the current implementation, identify relevant files, functions, classes, and dependencies, and decide where code changes should be made \cite{jiang2025cosil,xia2025demystifying}. Similarly, Development Tool Knowledge accounts for 77.4\% and is strongly connected with Dynamic Execution Results, which appear in 78.9\% of the studies. This indicates that current systems increasingly rely on test outputs, command-line responses, debugging traces, runtime logs, and tool observations to validate intermediate decisions and refine candidate solutions \cite{aggarwal2025dars,liu2024marscode}. These representative pathways suggest that current knowledge acquisition practices are most frequently built around artifacts that are explicit, executable, and closely tied to the core workflow of locating, modifying, and validating code, rather than demonstrating maturity in a strict methodological sense.

The flow also suggests that different knowledge sources naturally favor different extraction methods. Codebase and Dynamic Execution Results are frequently connected with Rule-based Extraction, which accounts for 94.7\% of the observed extraction methods. This is because source files, repository structures, call relations, execution logs, test outputs, and tool responses often have relatively stable formats or operational semantics, making them suitable for parsers, traversal rules, predefined tool interfaces, and deterministic feedback processing \cite{jiang2025cosil,liu2025context}. By contrast, Model-based Extraction, accounting for 68.4\%, is more suitable for semantically implicit or noisy artifacts, such as issue descriptions, long trajectories, historical issue-resolution records, and cross-modal signals \cite{liu2024codexembed,ouyang2025reasoningbank}. Manual Extraction, accounting for 29.3\%, plays a complementary role when knowledge requires expert judgment, careful environment construction, or qualitative interpretation, such as curating high-quality trajectories, designing agent tools, or defining benchmark-specific constraints \cite{tao2026swe,xie2026arkeval}. This distribution suggests that extraction methods are not interchangeable: rule-based methods are effective for explicit and operational artifacts, model-based methods are better suited for semantic abstraction and experience distillation, and manual methods remain important for high-level constraints, expert workflows, and reliable benchmark design.

Another important observation is that Repository Evolution Knowledge and Experiential Knowledge are becoming increasingly valuable sources of adaptive issue resolution. Repository Evolution Knowledge accounts for 31.6\% and is mainly connected with Repository History, indicating that commits, pull requests, issue-fix links, historical patches, and test changes are being used to understand how repositories evolve and how previous solutions were accepted. Experiential Knowledge accounts for 51.1\%, showing that agent trajectories, failed attempts, reasoning traces, and feedback signals are increasingly treated as reusable knowledge rather than disposable execution records. These two knowledge types are important because they allow agents to move beyond the current repository snapshot: repository history can reveal which code regions are historically related to similar issues, while experiential trajectories can teach agents how to plan, explore, recover from errors, and refine candidate solutions \cite{yang2025kgcompass,mu2025experepair}. This suggests a promising direction in which future systems accumulate reusable knowledge across tasks, turning repository history and agent experience into long-term memory for issue resolution.

In comparison, Document-based sources account for only 8.3\%, but they represent an important expansion space for high-level semantic knowledge. The relatively lower use of documents is partly related to the current research setting: many studies are built on open-source repositories, where documentation may be incomplete, outdated, loosely structured, or disconnected from issue-resolution instances. As a result, current systems more often rely on code and execution feedback, which are easier to obtain and evaluate. However, documents become much more valuable when tasks involve well-maintained specifications, project guidelines, external APIs, dependency migration rules, or enterprise-style development processes. Recent benchmarks and systems have started to make this direction more visible by incorporating specifications, external documents, domain knowledge, project guidelines, or repository-grounded verification criteria \cite{chen2026beyondswe,raghavendra2026agentic}. These examples suggest that as benchmarks begin to provide richer associated documents, Document sources can become a key pathway for extracting design intent, technology-stack conventions, domain requirements, and acceptance criteria.

The flows related to Design Architecture Knowledge, Technology Stack Knowledge, Programming Language Knowledge, and Domain-Specific Knowledge are currently smaller, but they point to the next stage of knowledge acquisition. These knowledge types are often high-level, implicit, and distributed across multiple artifacts, which makes them harder to extract than code snippets or test outputs. Architectural knowledge may be hidden in module organization, historical design decisions, document, and developer conventions; technology-stack knowledge may depend on external libraries, framework versions, and runtime environments; programming-language knowledge may require understanding type systems, ownership rules, or low-resource language constraints; and Domain-Specific Knowledge may require specialized external expertise. Existing studies on architecture-aware issue resolution and low-resource programming ecosystems already illustrate the importance of such knowledge \cite{xie2026arkeval,liu2026architecture}. These cases indicate that future acquisition methods need stronger mechanisms for extracting high-level knowledge from mixed sources, including code, documents, histories, expert annotations, and environment configurations.

Taken together, Fig.~\ref{fig:source_method} shows that knowledge acquisition in automated issue resolution has developed strong and practical pathways for explicit repository artifacts and execution-derived signals, while also revealing several promising directions for future expansion. First, future systems should improve the acquisition of high-quality knowledge sources, especially associated documents, issue discussions, design decisions, review comments, environment specifications, and validated trajectories. Second, knowledge sources should be treated as reusable assets rather than task-specific inputs: repository history, agent trajectories, tool-use records, and expert feedback can be accumulated into project-level memories, skill libraries, or experience banks for future issue resolution \cite{tang2025agent,chen2025swe}. Third, expert involvement can play a larger role in areas where automatic extraction is still difficult, such as design-rationale annotation, workflow construction, abnormal-case handling, benchmark validation, and tool-use strategy design \cite{kon2026swe,zhou2025tom}. Therefore, future progress in RQ2 should not only focus on extracting more information from codebases, but also on building sustainable acquisition pipelines that integrate documents, histories, expert knowledge, and dynamic interaction traces into reusable high-quality knowledge for more context-aware and generalizable issue resolution.

\finding{
The acquisition results show that current automated issue-resolution systems are most mature at acquiring knowledge from sources that are explicit, executable, and closely coupled with the repair workflow. Codebases and dynamic execution results provide relatively stable acquisition pathways because repository structures, code entities, tool outputs, tests, logs, and execution feedback can be directly processed through parsers, static analysis, predefined tool interfaces, or deterministic feedback handling. This explains why rule-based extraction remains a strong foundation for current systems. At the same time, the increasing use of repository history, documents, agent trajectories, and expert guidance suggests that knowledge acquisition is expanding beyond immediate code and runtime evidence. However, these sources often contain implicit semantics, historical rationales, design assumptions, workflow experience, and background constraints, which are harder to acquire systematically through a single extraction method.

This finding suggests that future issue-resolution agents need to move from source-level information extraction toward source-aware knowledge construction. Rather than only retrieving code, collecting logs, or mining isolated historical records, future systems should connect heterogeneous sources into reusable knowledge assets, such as project memories, evolution-aware repair records, experience banks, document-grounded specifications, and expert-informed workflow guidance. Achieving this goal requires hybrid acquisition mechanisms that combine rule-based structural analysis, model-based semantic inference, and human-in-the-loop validation, especially for high-level knowledge such as architectural rationale, technology-stack requirements, programming-language constraints, and domain-specific rules. From this perspective, knowledge acquisition is not merely a preprocessing step for issue resolution, but a core capability for building software agents that can accumulate, verify, and reuse software knowledge across tasks and projects.
}

\section{RQ3: How is knowledge represented for automated issue resolution?}
\label{sec:rq3}

\begin{figure*}[htbp]
\centering
\scriptsize
\definecolor{leafbg}{HTML}{FFFFEC}

\newlength{\KGRepWidth}
\setlength{\KGRepWidth}{8.0cm} 

\newcommand{\KGRepBox}[1]{%
  \begin{varwidth}[t]{\KGRepWidth}
    \normalfont\sffamily\tiny\raggedright
    \textbf{Related Work}\\[-1pt]
    #1%
  \end{varwidth}%
}


\newsavebox{\boxRepGraph}
\sbox{\boxRepGraph}{%
\KGRepBox{%
\textbf{Existing Code Knowledge:} \cite{aggarwal2025dars,chakraborty2025blaze,chang2025bridging,chen2025locagent,chen2025prometheus,jiang2025cosil,li2025swe,liu2024codexgraph,liu2024marscode,liu2025graphlocator,liu2026architecture,luo2026closing,ma2025alibaba,ouyang2024repograph,sohrabizadeh2025nemotron,tang2025synfix,tao2025code,wang2025improving,yang2025kgcompass,yu2025orcaloca,dong2025infcode};
\textbf{Repository Evolution Knowledge:} \cite{yang2025kgcompass};
\textbf{Design Architecture Knowledge:} \cite{liu2026architecture,luo2026closing};
\textbf{Experiential Knowledge:} \cite{zhang2025sealign}.%
}}

\newsavebox{\boxRepStructured}
\sbox{\boxRepStructured}{%
\KGRepBox{%
\textbf{Existing Code Knowledge:} \cite{antoniades2024swe,cao2025siadafix,chen2025locagent,chen2025old,chen2025prometheus,ding2026swe,gautam2024supercoder2,gupta2025sacl,jiang2025cosil,li2025patchpilot,li2025swe,luo2026closing,ma2024lingma,nashid2025issue2test,pabba2025semagent,phan2024hyperagent,raghavendra2026agentic,sohrabizadeh2025nemotron,song2026swe,tao2026swe,tawosi2025meta,vinh2025repeton,wang2025extracting,wu2025git,xia2025demystifying,xiang2026empowering,xiang2026evaluating,xiao2025improving,xie2025swe,xiong2025think,yang2025kimi,dong2025infcode};
\textbf{Experiential Knowledge:} \cite{aggarwal2025dars,antoniades2024swe,chen2024coder,chen2025swe,ding2026swe,gandhi2025agents,kon2026swe,lei2024infant,lei2025infantagent,lin2024llms,lin2025se,ma2024lingma,mu2025experepair,ouyang2025reasoningbank,rastogi2025devstral,reddy2025swerank+,tang2025agent,tang2025boosting,tao2026swe,wang2025aegis,wong2025confucius,wu2025git,yang2025lingxi,yu2025orcaloca,zhang2025darwin,zhou2025tom};
\textbf{Development Tool Knowledge:} \cite{antoniades2024swe,chen2025old,gao2025trae,guo2025swe,kon2026swe,lei2025infantagent,li2025patchpilot,liu2025context,liu2025graphlocator,liu2026dynamic,luo2026closing,ma2025tool,nashid2025issue2test,shum2025swe,sun2025scaling,tang2025boosting,wang2024openhands,wang2025seamlessflow,wei2025toward,xiang2026empowering,xiang2026evaluating,xiao2025improving,yang2025kgcompass,yu2025building,zhang2025one};
\textbf{Repository Evolution Knowledge:} \cite{cao2025siadafix,chakraborty2025blaze,luo2026closing,ma2025tool,soni2026swe,xie2025swe,yang2025kimi,yang2025lingxi};
\textbf{Design Architecture Knowledge:} \cite{li2025swe,liu2026architecture,luo2026closing,raghavendra2026agentic,wu2025git,yang2025lingxi,yu2026does,zhou2025tom};
\textbf{Technology Stack Knowledge:} \cite{pan2024training}.%
}}

\newsavebox{\boxRepUnstructured}
\sbox{\boxRepUnstructured}{%
\KGRepBox{%
\textbf{Existing Code Knowledge:} \cite{aggarwal2025dars,ahmed2025otter,arora2024masai,brown2024large,cao2025siadafix,chakraborty2025blaze,chang2025bridging,chen2024coder,chen2025locagent,chen2025old,chen2025swe,chen2026beyondswe,cheshkov2024exploring,copet2025cwm,da2025agent,dai2025lita,ding2026swe,ehrlich2025codemonkeys,gandhi2025agents,gao2025trae,gautam2024supercoder2,golubev2025training,guo2025swe,gupta2025sacl,huang2025seeing,jain2025r2e,jiang2025cosil,jiang2025putting,kon2026swe,lei2024infant,lei2025infantagent,li2024codetree,li2025infcode,li2025patchpilot,li2025swe,lin2024llms,lin2025se,lindenbauer2025complexity,liu2024codexembed,liu2024marscode,liu2025context,liu2026architecture,liu2026dynamic,luo2026closing,ma2024lingma,ma2024repository,ma2025alibaba,ma2025sorft,ma2025thinking,ma2025tool,mu2025experepair,nashid2025issue2test,ouyang2024repograph,ouyang2025reasoningbank,pan2024training,phan2024hyperagent,pimpale2025forecasting,raghavendra2026agentic,rastogi2025devstral,reddy2025swerank,reddy2025swerank+,rombaut2024watson,ruan2025specrover,shi2025aime,shum2025swe,sohrabizadeh2025nemotron,soni2025coding,soni2026swe,sonwane2025bugpilot,sun2025scaling,suresh2024cornstack,tang2025boosting,tang2025co,tang2025synfix,tao2024magis,tao2025code,tao2026swe,tawosi2025meta,vaghasiya2025corethink,wang2024openhands,wang2025aegis,wang2025improving,wang2025mcts,wang2025practitioner,wang2025seamlessflow,wang2025swe,wei2025swe,wei2025toward,wong2025confucius,xia2025demystifying,xia2025live,xiang2026evaluating,xie2026arkeval,xiong2025think,yang2024swe,yang2025kimi,yang2025lingxi,yang2025swe,yu2025building,yu2025orcaloca,yu2025utboost,yu2026does,yuan2026swe,zainullina2025guided,zeng2025satori,zeng2025skywork,zhang2024autocoderover,zhang2024codev,zhang2024diversity,zhang2025cast,zhang2025darwin,zhang2025one,zhou2025tom,zhu2025training,dong2025infcode};
\textbf{Development Tool Knowledge:} \cite{aggarwal2025dars,ahmed2025otter,arora2024masai,brown2024large,cao2025siadafix,chen2024coder,chen2025prometheus,chen2025swe,chen2026beyondswe,cheshkov2024exploring,copet2025cwm,da2025agent,dai2025lita,ding2026swe,ehrlich2025codemonkeys,gandhi2025agents,gao2025trae,gautam2024supercoder2,golubev2025training,guo2025swe,jain2025r2e,jiang2025putting,kon2026swe,lei2024infant,lei2025infantagent,li2025infcode,li2025patchpilot,li2025swe,lin2024llms,lin2025se,lindenbauer2025complexity,liu2024marscode,liu2025context,liu2026architecture,ma2024lingma,ma2025alibaba,ma2025thinking,mu2025experepair,nashid2025issue2test,ouyang2025reasoningbank,pabba2025semagent,pan2024training,pan2025codecor,phan2024hyperagent,pimpale2025forecasting,raghavendra2026agentic,rastogi2025devstral,ruan2025specrover,shi2025aime,sohrabizadeh2025nemotron,song2026swe,soni2025coding,soni2026swe,sonwane2025bugpilot,sun2025scaling,tang2025co,tang2025synfix,tao2024magis,tao2026swe,vaghasiya2025corethink,vinh2025repeton,wang2025aegis,wang2025practitioner,wang2025seamlessflow,wang2025swe,wei2025toward,wong2025confucius,wu2025git,xia2025demystifying,xia2025live,xiang2026evaluating,xie2025swe,xie2026arkeval,yang2024swe,yang2025kimi,yang2025lingxi,yang2025swe,yu2025building,yu2025orcaloca,yu2025utboost,yuan2026swe,zainullina2025guided,zeng2025satori,zeng2025skywork,zhang2024autocoderover,zhang2025darwin,zhang2025sealign,zhou2025tom,zhu2025training,dong2025infcode};
\textbf{Experiential Knowledge:} \cite{aggarwal2025dars,cao2025siadafix,chen2025locagent,chen2026beyondswe,copet2025cwm,da2025agent,ding2026swe,gao2025trae,golubev2025training,jain2025r2e,kon2026swe,li2025patchpilot,lindenbauer2025complexity,liu2025context,ma2024lingma,ma2025alibaba,ma2025thinking,ma2025tool,mu2025experepair,pan2024training,phan2024hyperagent,pimpale2025forecasting,reddy2025swerank+,rombaut2024watson,shum2025swe,song2026swe,sonwane2025bugpilot,su2025learn,tang2025boosting,tang2025co,wang2024openhands,wang2025practitioner,wang2025swe,wei2025toward,wu2025git,xia2025live,xiao2025improving,xie2025swe,yang2024swe,yang2025kimi,yang2025swe,yu2025building,yuan2026swe,zainullina2025guided,zeng2025satori,zeng2025skywork,zhang2024diversity,zhang2025darwin,zhang2025one,zhang2025sealign,zhou2025tom,zhu2025training};
\textbf{Repository Evolution Knowledge:} \cite{ahmed2025otter,chakraborty2025blaze,chang2025bridging,chen2024coder,copet2025cwm,guo2025swe,jain2025r2e,li2025patchpilot,lindenbauer2025complexity,liu2024codexembed,ma2024lingma,ma2024repository,ma2025sorft,ma2025thinking,reddy2025swerank,reddy2025swerank+,sohrabizadeh2025nemotron,soni2026swe,tang2025co,tao2024magis,tao2025code,tao2026swe,wang2025improving,wang2025mcts,wang2025seamlessflow,wang2025swe,wei2025swe,wei2025toward,xie2026arkeval,yang2025kimi,yang2025lingxi,yang2025swe,yu2025utboost,yu2026does,zainullina2025guided,zeng2025satori,zeng2025skywork};
\textbf{Design Architecture Knowledge:} \cite{cao2025siadafix,chen2026beyondswe,huang2025seeing,li2025swe,nashid2025issue2test,zhu2025training};
\textbf{Technology Stack Knowledge:} \cite{chen2026beyondswe,jain2025r2e,xie2026arkeval};
\textbf{Programming Language Knowledge:} \cite{reddy2025swerank+,xiang2026evaluating,xie2026arkeval};
\textbf{Domain-Specific Knowledge:} \cite{chen2026beyondswe}.%
}}

\begin{forest}
for tree={
    grow'=0,
    anchor=west,
    child anchor=west,
    parent anchor=east,
    font=\sffamily\scriptsize,
    inner sep=2.2pt,
    s sep=2.8pt,
    l sep=1.8mm,
    edge={draw=black!80},
    align=left,
},
where level=0{
    draw=purple!50,
    thick,
    rounded corners=4pt,
    fill=purple!6,
    font=\sffamily\bfseries\scriptsize,
    align=center,
    inner xsep=4pt,   
    inner ysep=2pt,
    fit=band
}{},
where level=1{
    draw=blue!40,
    thick,
    rounded corners=4pt,
    fill=blue!6,
    font=\sffamily\bfseries\scriptsize,
    align=center,
    inner xsep=3pt,
    inner ysep=1.5pt,
    fit=band
}{},
where level=2{
    draw=orange!25,
    rounded corners=3pt,
    fill=leafbg,
    font=\sffamily\tiny,
    align=left,
    inner xsep=3pt, 
}{},
[Knowledge\\Representation\\Formats
    [{Unstructured\\Text}
        [{\usebox{\boxRepUnstructured}}]
    ]
    [{Structured\\Text}
        [{\usebox{\boxRepStructured}}]
    ]
    [{Graph}
        [{\usebox{\boxRepGraph}}]
    ]
]
\end{forest}

\caption{
Knowledge representation formats in automated issue resolution.
Each box lists the corresponding knowledge types and related studies.
}
\label{fig:knowledge_representation_tree}
\end{figure*}
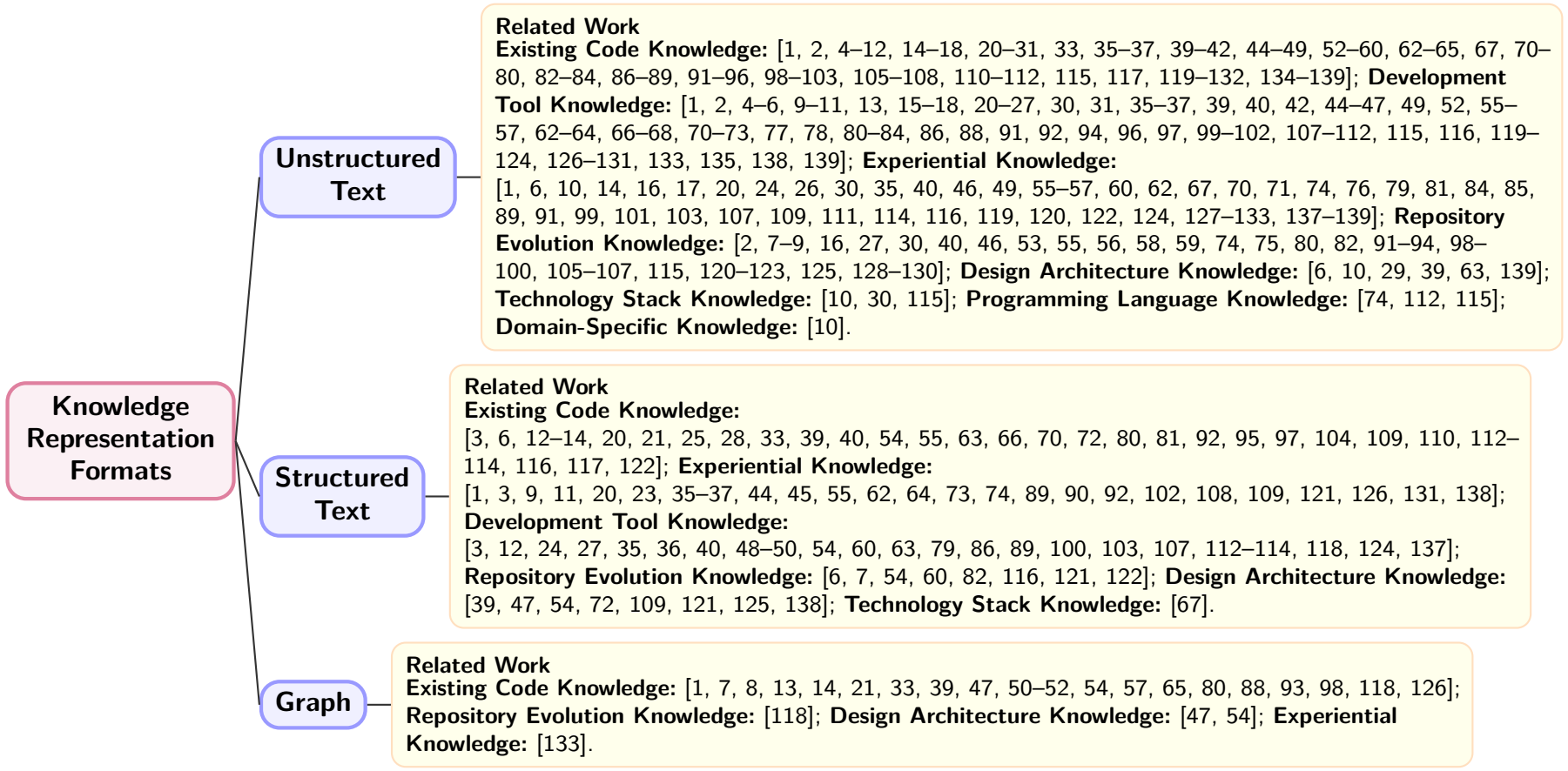

Following the knowledge-representation schema defined in Section~\ref{sec:knowledge_representation_schema}, this section analyzes how acquired knowledge is externalized into formats that can be processed and used by automated issue-resolution systems. The key challenge here is \textbf{representation alignment}: knowledge obtained from codebases, execution feedback, repository histories, documents, or expert guidance must be organized in ways that match the needs of LLM prompting, retrieval, tool interaction, memory reuse, and downstream reasoning. As summarized in Fig.~\ref{fig:knowledge_representation_tree}, existing studies mainly adopt Unstructured Text, Structured Text, and Graph representations. Rather than redefining these formats, this section focuses on how they are instantiated in representative systems, what kinds of knowledge they tend to encode, and how different representation choices affect repository reasoning, context construction, workflow control, and knowledge reuse.

\subsection{Knowledge Representation Formats}

Based on the representation schema introduced in Section~\ref{sec:knowledge_representation_schema}, this subsection examines how knowledge is externalized in existing automated issue-resolution systems. We focus on the empirical roles of Unstructured Text, Structured Text, and Graph in representative studies, including what kinds of knowledge they commonly encode and how they support LLM prompting, retrieval, tool interaction, memory reuse, and repository reasoning. The distribution of these formats across knowledge types is summarized in Fig.~\ref{fig:knowledge_representation_tree}.

\subsubsection{Unstructured Text}

Unstructured Text is the most broadly adopted representation format in current issue-resolution systems. This is mainly because many key artifacts in issue resolution are already textual, including issue reports, code snippets, document fragments, terminal outputs, test logs, tool responses, candidate patches, and agent reasoning traces. Existing studies therefore often preserve these artifacts in their original or lightly processed textual form, making them easy to insert into LLM prompts or agent interaction histories. Across the surveyed literature, Unstructured Text is mainly used in three forms: raw contextual text representation, textual feedback trace representation, and narrative experience representation.

\textbf{Raw Contextual Text Representation.}
Raw contextual text representation is mainly used to preserve task descriptions and repository evidence in forms that can be directly consumed by LLMs. In this form, issue reports, document fragments, code comments, file contents, localized code snippets, and natural-language instructions are retained as textual context for localization and patch generation. This representation is especially common for Existing Code Knowledge, because raw files and code snippets provide direct implementation evidence about where the issue may occur and what code should be modified \cite{aggarwal2025dars,ahmed2025otter,arora2024masai,brown2024large,cao2025siadafix,ouyang2025reasoningbank,liu2024codexembed,ma2025sorft,xie2026arkeval,xiang2026evaluating}. It is also used to expose background-oriented knowledge when programming-language constraints, technology-stack conventions, domain assumptions, or dependency-migration requirements are described in issue statements, benchmark descriptions, documents, or external textual resources \cite{chen2026beyondswe,jain2025r2e,reddy2025swerank+,xiang2026evaluating,xie2026arkeval,zhu2025training}. For example, BeyondSWE includes tasks where dependency-migration requirements or domain-specific assumptions appear as natural-language problem descriptions rather than formally encoded constraints \cite{chen2026beyondswe}. This shows that raw contextual text remains useful when systems need to preserve rich semantic context without imposing additional representation schemas.

\textbf{Textual Feedback Trace Representation.}
Textual feedback trace representation is mainly used to record dynamic observations produced during agent-environment interaction. In many agentic systems, terminal outputs, compilation errors, test logs, runtime exceptions, tool responses, candidate patches, and intermediate observations are passed back to the LLM as textual traces. This form is closely associated with Development Tool Knowledge, because it allows agents to interpret tool-produced evidence and revise subsequent actions accordingly \cite{aggarwal2025dars,ahmed2025otter,arora2024masai,brown2024large,cao2025siadafix,ouyang2025reasoningbank,xiang2026evaluating}. For instance, MASAI exchanges reproduction tests, edit suggestions, possible patches, and test commands as free-form textual artifacts among specialized sub-agents \cite{arora2024masai}, while Large Language Monkeys represents multiple generated solutions or patch attempts as plain textual candidates before using verifier or test feedback for selection \cite{brown2024large}. Compared with raw contextual text, this form is more process-oriented: it captures what the system observes during execution and tool use, thereby supporting failure diagnosis, plan revision, patch refinement, and validation.

\textbf{Narrative Experience Representation.}
Narrative experience representation is mainly used to preserve prior reasoning, expert guidance, and repair experience in natural-language form. Existing systems often summarize successful trajectories, failed attempts, repair plans, critique messages, workflow decisions, or reflective insights as reusable textual guidance for future tasks. This form is frequently used to encode Experiential Knowledge, because process-level lessons are often easier to express as natural-language memories than as fully formalized rules \cite{aggarwal2025dars,cao2025siadafix,chen2025locagent,copet2025cwm,da2025agent,ding2026swe,ouyang2025reasoningbank}. It can also carry Design Architecture Knowledge when expected behaviors, architectural assumptions, project conventions, or workflow rules are expressed through expert-written prompts, project documents, issue descriptions, or visual issue explanations converted into textual reasoning context \cite{cao2025siadafix,chen2026beyondswe,huang2025seeing,li2025swe,nashid2025issue2test,zhu2025training}. This indicates that Unstructured Text is not only a carrier of raw artifacts, but also a flexible medium for preserving abstract reasoning and reusable problem-solving experience.

Overall, Unstructured Text provides strong semantic preservation and prompt compatibility, which explains its wide adoption in LLM-based issue resolution. However, its lack of explicit structure also limits controllability and reuse. Since important knowledge may be buried in long, noisy, or weakly organized textual contexts, systems often need additional retrieval, summarization, ranking, filtering, or prompt engineering mechanisms to make such knowledge useful. Thus, Unstructured Text is most suitable for preserving rich context and interaction traces, but less effective for deterministic parsing, structured comparison, and fine-grained tool integration.

\subsubsection{Structured Text}

Structured Text is adopted when systems need stronger control over knowledge organization while still keeping the representation compatible with LLM prompts. Compared with Unstructured Text, this format makes repository records, tool feedback, procedural memories, and verification criteria more parseable and reusable through explicit fields, templates, lists, records, or hierarchical structures. Existing studies mainly use Structured Text in three forms: structured repository record representation, structured feedback record representation, and structured procedural memory representation.

\textbf{Structured Repository Record Representation.}
Structured repository record representation is mainly used to regularize scattered task and repository artifacts into more stable textual units. Instead of passing raw issue descriptions, code snippets, historical changes, or retrieved contexts directly to the model, systems organize them as file/function metadata, localized code records, training samples, issue-code pairs, or structured retrieval instances. For Existing Code Knowledge, this representation supports localization and patch generation by making relevant files, functions, code chunks, or repository summaries easier to index, compare, and retrieve \cite{luo2026closing,ma2024lingma,nashid2025issue2test,pabba2025semagent}. For Repository Evolution Knowledge, historical bug reports, commits, issue-fix pairs, changesets, or development-process data can also be encoded as structured records, allowing models to learn from previous repository changes rather than treating historical artifacts as isolated text \cite{cao2025siadafix,chakraborty2025blaze,luo2026closing,ma2025tool}. This form is also used when technology-stack information, dependencies, tests, and task specifications are packaged as structured training instances for agent learning \cite{pan2024training}. These studies show that structured repository records improve the reusability of repository-grounded knowledge across retrieval, training, localization, and patch generation.

\textbf{Structured Feedback Record Representation.}
Structured feedback record representation is mainly used to make execution results and tool observations easier for agents to interpret and act upon. Rather than relying on long and noisy terminal outputs, some systems transform execution results, call stacks, variable states, test outcomes, action histories, or debugger observations into explicit reports, state summaries, or action-observation records \cite{liu2026dynamic,luo2026closing,xiang2026empowering,zhang2025one}. This representation is particularly useful for Development Tool Knowledge, because it clarifies what action was performed, what result was observed, what error occurred, and what evidence should guide the next step. Dynamic-analysis-based systems, for example, organize runtime evidence into structured diagnostic reports to support localization and patch refinement \cite{liu2026dynamic,xiang2026empowering}. Repository-navigation agents also structure tool interactions into repeatable records for learning effective search and navigation behaviors \cite{ma2025tool,zhang2025one}. Compared with textual feedback traces, structured feedback records provide stronger support for controlled feedback interpretation, state tracking, and iterative repair.

\textbf{Structured Procedural Memory Representation.}
Structured procedural memory representation is mainly used to organize reusable experience, procedures, constraints, and verification-oriented knowledge. In memory-enhanced or training-based systems, prior demonstrations, reflective insights, reasoning strategies, successful trajectories, and failed attempts are stored as structured memory entries that can be retrieved and composed into future prompts \cite{mu2025experepair,ouyang2025reasoningbank,rastogi2025devstral}. Some systems further abstract recurring problem-solving behaviors into skill-like procedural guidance, such as how to navigate repositories, invoke tools, inspect runtime states, manage long-horizon contexts, or validate patches \cite{mu2025experepair,ouyang2025reasoningbank,ma2025tool,zhang2025one,liu2026dynamic,xiang2026empowering,wu2025git}. In addition, structured text can express repository-specific expectations, contextual verification rubrics, design constraints, or versioned context states, helping agents check whether a patch respects project-level behavior and architectural assumptions \cite{luo2026closing,raghavendra2026agentic,wu2025git}. This shows that Structured Text is not merely a cleaner way to store artifacts; it increasingly functions as an operational interface for memory reuse, workflow guidance, and verification control.

Overall, Structured Text provides a middle-ground representation between flexible natural language and explicit relational structures. It preserves textual compatibility with LLMs while introducing enough organization to improve retrieval precision, workflow orchestration, state tracking, memory reuse, and verification reliability. This makes it especially valuable for agentic issue resolution, where systems must coordinate context, feedback, tool actions, procedural guidance, and correctness criteria across multiple steps.

\subsubsection{Graph}

Graph representations are adopted when issue resolution requires explicit relational reasoning over software entities, historical artifacts, task elements, or agent decisions. Compared with text-based formats, graphs make dependencies, paths, neighborhoods, and structural connections directly accessible. Existing studies mainly use Graph in four forms: code-structure graph representation, repository-aware semantic graph representation, causal and task-oriented graph representation, and architecture-aware planning graph representation.

\textbf{Code-Structure Graph Representation.}
Code-structure graph representation is mainly used to make repository organization and cross-file dependencies explicit. In this form, files, classes, functions, methods, imports, invocations, inheritance relations, call relations, or reference links are modeled as connected code entities. This form is widely used for Existing Code Knowledge because it helps agents move beyond flat textual context and reason over how implementation elements are related across the repository \cite{chen2025locagent,chen2025prometheus,jiang2025cosil,liu2024codexgraph,liu2024marscode,liu2025graphlocator,liu2026architecture,luo2026closing,ouyang2024repograph,wang2025improving,yang2025kgcompass,yu2025orcaloca}. For example, graph-guided localization methods parse repositories into heterogeneous graphs or dynamically construct call graphs so that agents can traverse from an issue description to relevant files, classes, and functions through structural relations \cite{chen2025locagent,jiang2025cosil,ouyang2024repograph,yu2025orcaloca}. This representation is especially useful when relevant implementation logic is scattered across multiple files or modules.

\textbf{Repository-Aware Semantic Graph Representation.}
Repository-aware semantic graph representation is mainly used to connect issue-level semantics, historical artifacts, and current code entities in a unified structure. Instead of representing issue descriptions, pull requests, historical fixes, commits, files, classes, and functions as separate textual records, these systems organize them into connected paths that support path-guided retrieval and repair \cite{yang2025kgcompass}. This form supports both Existing Code Knowledge and Repository Evolution Knowledge, because it links current implementation artifacts with historical issue-fix evidence. Such representations are particularly useful when agents need to reason about how previous issues, pull requests, or code changes relate to the current repository state.

\textbf{Causal and Task-Oriented Graph Representation.}
Causal and task-oriented graph representation is mainly used to organize the reasoning structure of issue resolution. In this form, sub-issues, suspicious code entities, candidate fault locations, intermediate reasoning states, or agent decisions are represented as graph- or tree-shaped structures, where edges encode causal dependencies, propagation paths, decomposition relations, or decision transitions \cite{liu2025graphlocator,zhang2025sealign}. This helps agents move beyond local symptom matching by considering how an observed issue may propagate to hidden root causes or require coordinated edits across multiple locations. In some settings, graph representations can also encode Experiential Knowledge when prior or simulated solving processes are organized as decision graphs for alignment, action optimization, or trajectory-level reasoning \cite{zhang2025sealign}. Compared with code-structure graphs, this form focuses less on repository topology itself and more on how the resolution process unfolds.

\textbf{Architecture-Aware Planning Graph Representation.}
Architecture-aware planning graph representation is mainly used to align high-level functional intent with repository-level implementation structures. In this form, module responsibilities, dependency topology, functional requirements, and implementation units are linked in graph structures to support design-aware planning and impact analysis. This form is particularly relevant to Design Architecture Knowledge because it helps agents understand not only which code entities are connected, but also how changes should respect repository-level design, functional decomposition, and structural constraints \cite{liu2026architecture,luo2026closing}. For example, architecture-aware methods construct repository-level graphs to locate dispersed cross-file modification targets, while RPG-based representations align semantic features with dependency topology and can evolve with repository changes \cite{liu2026architecture,luo2026closing}. Compared with general code-structure graphs, this representation emphasizes architectural consistency and change planning.

Overall, Graph provides the most explicit relational representation among the three dominant formats. It supports structural navigation, neighborhood expansion, path-guided retrieval, multi-hop reasoning, history-aware reasoning, architecture-aware planning, and impact analysis. However, current graph usage is still mainly concentrated on Existing Code Knowledge, especially code structures and dependency relations. Background knowledge, experiential trajectories, architectural intent, and tool feedback are less frequently represented in graph form. This suggests that future work could develop more heterogeneous graph representations that connect code entities with issue semantics, repository evolution, design rationales, tool observations, and agent decisions.

\subsection{Knowledge-Type Composition across Representation Formats}

\begin{figure}[htbp] 
\centering 
\includegraphics[width=0.9\linewidth]
{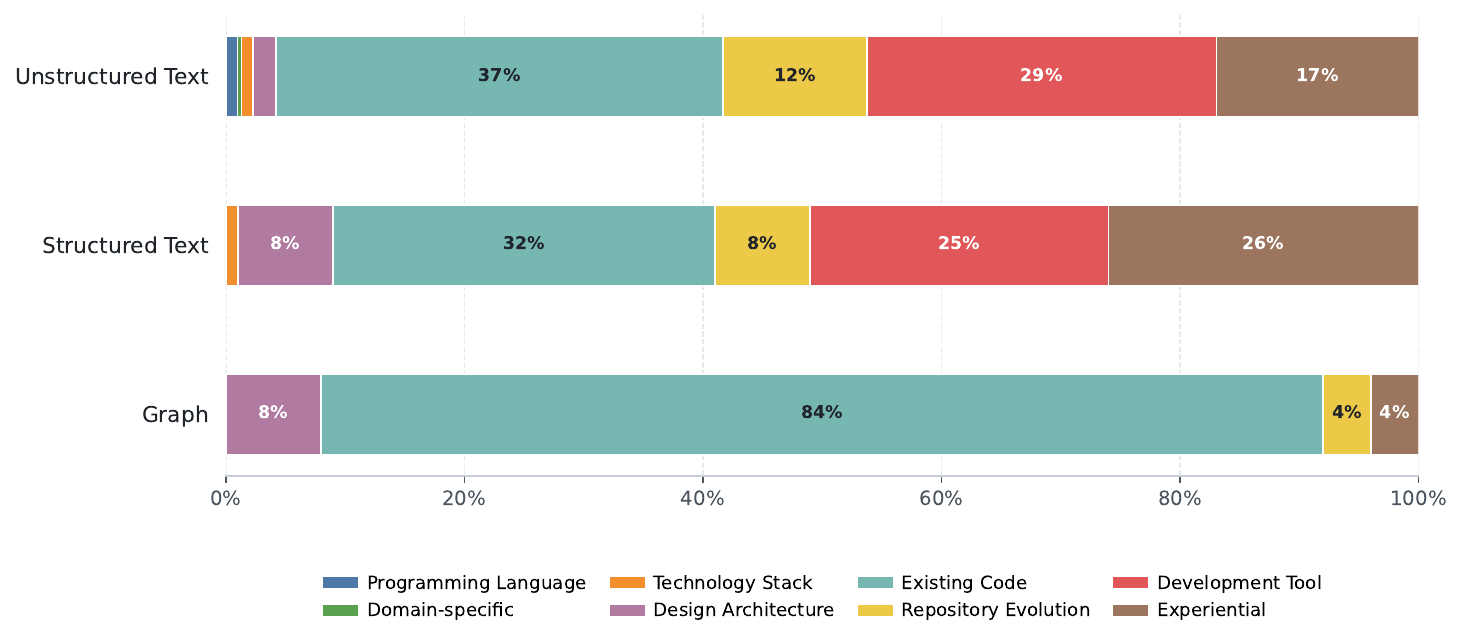} 
\caption{Distribution of knowledge types within each representation format.} 
\label{fig:representation_bar} 
\end{figure}

Fig.~\ref{fig:representation_bar} further quantifies the composition of knowledge types within each representation format. The results show that the three representation formats exhibit clearly different knowledge-type preferences, reflecting their different roles in making knowledge understandable, retrievable, and actionable for LLM-based issue resolution systems. In Unstructured Text, Existing Code Knowledge accounts for the largest proportion at 37\%, followed by Development Tool Knowledge at 29\%, Experiential Knowledge at 17\%, and Repository Evolution Knowledge at 12\%. This distribution suggests that unstructured text is mainly used to preserve rich but loosely organized repository and process-level information, such as raw code snippets, issue descriptions, terminal outputs, tool responses, execution traces, candidate patches, and free-form reasoning trajectories. Its relatively high proportion of Development Tool Knowledge and Experiential Knowledge further indicates that unstructured text is especially suitable for recording interactive and trajectory-based information, where preserving the original context is more important than imposing a strict schema. However, this also reveals a limitation: although unstructured text is highly compatible with LLM prompts, it is less friendly to deterministic retrieval, tool parsing, and fine-grained reuse, because the key knowledge is often embedded in long and noisy natural-language contexts.

Structured Text presents a more balanced composition. Existing Code Knowledge remains the largest category at 32\%, but Experiential Knowledge and Development Tool Knowledge also account for substantial proportions, reaching 26\% and 25\%, respectively. Repository Evolution Knowledge contributes 8\%, while Design Architecture Knowledge also appears as a smaller but visible component. This indicates that structured text is not used merely to store code context, but also to organize reusable experience, tool feedback, workflow states, historical evidence, and task constraints into more controllable textual units. Compared with unstructured text, structured text introduces explicit fields, templates, records, lists, or hierarchical organization, making knowledge easier to retrieve, compare, compose, and reuse across different stages of issue resolution. This is particularly important for LLM-based agents: as issue resolution moves from one-shot prompting toward long-horizon tool use, knowledge representation must become not only human-readable, but also model-friendly and tool-friendly. Structured representations such as action-observation records, skill-like procedural guidance, structured memory, verification rubrics, and repository/task metadata provide a more operational interface between LLM reasoning and external tools \cite{mu2025experepair,ouyang2025reasoningbank,ma2025tool}. Therefore, the relatively high proportions of Experiential Knowledge and Development Tool Knowledge in structured text suggest that current systems are already beginning to transform raw experience and feedback into reusable procedural units.

Graph shows the strongest specialization among the three formats. Existing Code Knowledge dominates this format at 84\%, while Repository Evolution Knowledge and Design Architecture Knowledge occupy much smaller proportions. This confirms that graph representation is primarily adopted to model repository-grounded structural relations, such as file-function hierarchies, call relations, dependency links, inheritance relations, reference edges, and issue-code paths. Compared with text-based representations, graphs make structural relations explicit and therefore better support repository navigation, neighborhood expansion, path-guided retrieval, and multi-hop reasoning. This is especially valuable in large repositories, where relevant code entities are often scattered across multiple files and cannot be effectively captured by linear textual context alone. However, the current graph usage is still highly concentrated on Existing Code Knowledge, suggesting that graph-based representation has not yet been fully extended to background knowledge, architectural constraints, experiential trajectories, or tool-interaction processes. Future work may therefore explore more heterogeneous graph representations that connect code entities with issue semantics, historical changes, architectural intent, tool feedback, and agent decisions \cite{chen2025locagent,yang2025kgcompass,liu2026architecture}.

Overall, the distribution indicates a functional division among representation formats. Unstructured Text emphasizes flexible preservation of rich contextual and interactive information; Structured Text improves controllability by organizing knowledge into reusable and parseable textual units; and Graph strengthens relational reasoning by explicitly modeling dependencies and paths among repository entities. More importantly, these results suggest that knowledge representation in automated issue resolution is gradually shifting from human-friendly textual description toward LLM-friendly, retrieval-friendly, and tool-friendly representation. Future research should further investigate representation formats that can bridge natural-language reasoning and executable agent workflows. For example, DSL-like formats, SDD-like specification structures, and skill-like procedural representations may provide promising extensions beyond the current three dominant formats, because they can encode constraints, procedures, tool-use policies, and verification requirements in forms that are both interpretable to LLMs and easier for tools to parse and execute. Such representations may help future systems move beyond simply placing knowledge into prompts, toward more controllable, reusable, and verifiable knowledge interfaces for autonomous issue resolution.

\finding{
The representation results show that the three dominant formats play complementary roles in making knowledge usable for automated issue resolution. Unstructured Text preserves rich semantic and interaction context and remains naturally compatible with LLM prompting. Structured Text improves controllability by organizing repository records, feedback, memory, verification criteria, and procedural guidance into more explicit and reusable units. Graph makes structural and relational dependencies explicit, supporting repository navigation, localization, history-aware retrieval, and architecture-aware reasoning. These patterns indicate that representation formats are not merely storage choices; they determine how knowledge can be accessed by LLMs, retrieved by tools, composed across workflow stages, and reused across tasks.

This finding further suggests that knowledge representation is becoming a key interface between LLM reasoning and software-agent execution. Future issue-resolution systems should move beyond placing raw artifacts into prompts and instead develop representation layers that make software knowledge more structured, traceable, executable, and verifiable. In this sense, representation-aware design is not only a technical improvement for issue resolution, but also a foundation for building more capable software engineering agents: agents that can coordinate natural-language reasoning, repository structures, tool feedback, historical evidence, architectural constraints, and reusable procedural experience within a unified knowledge framework. Beyond the current taxonomy, hybrid and specification-oriented forms, such as DSL-like specifications, SDD-like structures, and skill-oriented procedural representations, may further support this transition by encoding constraints, workflows, tool-use policies, and verification requirements in forms that are both interpretable to LLMs and operable by tools.
}



\section{RQ4: How is knowledge used by LLMs in issue resolution systems?}
\label{sec:rq4}

After analyzing knowledge acquisition and representation, we further examine how represented knowledge is operationalized by LLM-based issue-resolution systems. Based on the usage schema introduced in Section~\ref{sec:knowledge_usage_schema}, this section focuses on how existing studies make knowledge actionable during problem solving, including how agents construct context, explore repositories, interact with tools, interpret feedback, refine patches, and accumulate reusable capabilities. This perspective shifts the analysis from what knowledge is available to how it participates in the actual reasoning and execution process of automated issue resolution.

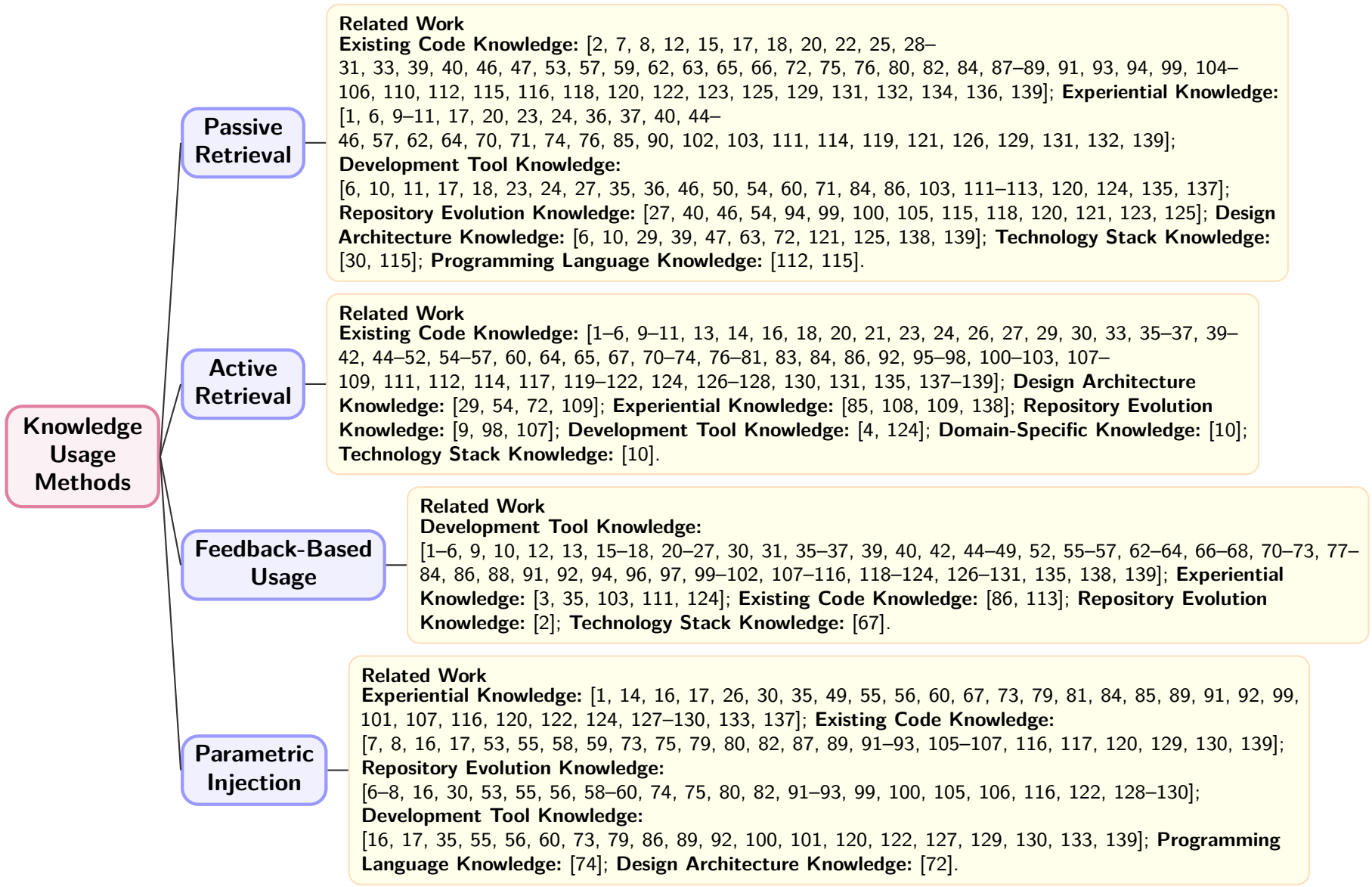
\begin{figure*}[htbp]
\centering
\scriptsize
\definecolor{leafbg}{HTML}{FFFFEC}

\newlength{\KGUsageMWidth}
\setlength{\KGUsageMWidth}{8.2cm} 

\newcommand{\KGUsageBox}[2]{%
  \begin{varwidth}[t]{\KGUsageMWidth}
    \normalfont\sffamily\tiny\raggedright
    \textbf{#1}\\[-1pt]
    #2%
  \end{varwidth}%
}


\newsavebox{\boxUsagePassive}
\sbox{\boxUsagePassive}{%
\KGUsageBox{Related Work}{%
\textbf{Existing Code Knowledge:} \cite{ahmed2025otter,chakraborty2025blaze,chang2025bridging,chen2025old,cheshkov2024exploring,da2025agent,dai2025lita,ding2026swe,ehrlich2025codemonkeys,gautam2024supercoder2,gupta2025sacl,huang2025seeing,jain2025r2e,jiang2025cosil,jiang2025putting,li2025patchpilot,li2025swe,lindenbauer2025complexity,liu2024codexembed,liu2026architecture,ma2025alibaba,ma2025sorft,mu2025experepair,nashid2025issue2test,ouyang2024repograph,pabba2025semagent,raghavendra2026agentic,reddy2025swerank,rombaut2024watson,sohrabizadeh2025nemotron,soni2026swe,sonwane2025bugpilot,suresh2024cornstack,tang2025boosting,tang2025co,tang2025synfix,tao2024magis,tao2025code,wang2025extracting,wang2025mcts,wang2025swe,wei2025swe,xia2025demystifying,xiang2026evaluating,xie2025swe,xie2026arkeval,yang2025kgcompass,yang2025kimi,yang2025swe,yu2025utboost,yu2026does,zeng2025satori,zhang2024codev,zhang2024diversity,zhang2025cast,zhang2025darwin,zhu2025training};
\textbf{Experiential Knowledge:} \cite{aggarwal2025dars,cao2025siadafix,chen2024coder,chen2025swe,chen2026beyondswe,da2025agent,ding2026swe,gandhi2025agents,gao2025trae,lei2024infant,lei2025infantagent,li2025patchpilot,lin2024llms,lin2025se,lindenbauer2025complexity,ma2025alibaba,mu2025experepair,ouyang2025reasoningbank,phan2024hyperagent,pimpale2025forecasting,reddy2025swerank+,rombaut2024watson,su2025learn,tang2025agent,wang2024openhands,wang2025aegis,xia2025live,xiao2025improving,yang2024swe,yang2025lingxi,yu2025orcaloca,zeng2025satori,zhang2024diversity,zhang2025darwin,zhu2025training};
\textbf{Development Tool Knowledge:} \cite{cao2025siadafix,chen2025swe,chen2026beyondswe,da2025agent,dai2025lita,gandhi2025agents,gao2025trae,guo2025swe,kon2026swe,lei2025infantagent,lindenbauer2025complexity,liu2025graphlocator,luo2026closing,ma2025tool,pimpale2025forecasting,sonwane2025bugpilot,sun2025scaling,wang2024openhands,xia2025live,xiang2026empowering,xiang2026evaluating,yang2025swe,yu2025building,zhang2024autocoderover,zhang2025one};
\textbf{Repository Evolution Knowledge:} \cite{guo2025swe,li2025patchpilot,lindenbauer2025complexity,luo2026closing,tao2024magis,wang2025mcts,wang2025seamlessflow,wang2025swe,xie2026arkeval,yang2025kgcompass,yang2025lingxi,yang2025swe,yu2025utboost,yu2026does};
\textbf{Design Architecture Knowledge:} \cite{cao2025siadafix,chen2026beyondswe,huang2025seeing,li2025swe,liu2026architecture,nashid2025issue2test,raghavendra2026agentic,yang2025lingxi,yu2026does,zhou2025tom,zhu2025training};
\textbf{Technology Stack Knowledge:} \cite{jain2025r2e,xie2026arkeval};
\textbf{Programming Language Knowledge:} \cite{xiang2026evaluating,xie2026arkeval}.%
}}

\newsavebox{\boxUsageActive}
\sbox{\boxUsageActive}{%
\KGUsageBox{Related Work}{%
\textbf{Existing Code Knowledge:} \cite{aggarwal2025dars,ahmed2025otter,antoniades2024swe,arora2024masai,brown2024large,cao2025siadafix,chen2024coder,chen2025locagent,chen2025prometheus,chen2025swe,chen2026beyondswe,copet2025cwm,dai2025lita,ding2026swe,gandhi2025agents,gao2025trae,golubev2025training,guo2025swe,huang2025seeing,jain2025r2e,jiang2025cosil,kon2026swe,lei2024infant,lei2025infantagent,li2024codetree,li2025infcode,li2025patchpilot,li2025swe,lin2024llms,lin2025se,lindenbauer2025complexity,liu2024codexgraph,liu2024marscode,liu2025context,liu2025graphlocator,liu2026architecture,liu2026dynamic,luo2026closing,ma2024lingma,ma2025alibaba,ma2025thinking,ma2025tool,ouyang2024repograph,ouyang2025reasoningbank,pan2024training,phan2024hyperagent,pimpale2025forecasting,raghavendra2026agentic,rastogi2025devstral,reddy2025swerank+,rombaut2024watson,ruan2025specrover,shi2025aime,shum2025swe,sohrabizadeh2025nemotron,song2026swe,soni2025coding,sonwane2025bugpilot,sun2025scaling,tao2026swe,tawosi2025meta,vaghasiya2025corethink,vinh2025repeton,wang2024openhands,wang2025aegis,wang2025improving,wang2025practitioner,wang2025seamlessflow,wei2025toward,wong2025confucius,wu2025git,xia2025live,xiang2026evaluating,xiao2025improving,xiong2025think,yang2024swe,yang2025kimi,yang2025lingxi,yang2025swe,yu2025building,yu2025orcaloca,yuan2026swe,zainullina2025guided,zeng2025skywork,zhang2024autocoderover,zhang2025darwin,zhang2025one,zhou2025tom,zhu2025training,dong2025infcode};
\textbf{Design Architecture Knowledge:} \cite{huang2025seeing,luo2026closing,raghavendra2026agentic,wu2025git};
\textbf{Experiential Knowledge:} \cite{su2025learn,wong2025confucius,wu2025git,zhou2025tom};
\textbf{Repository Evolution Knowledge:} \cite{chen2024coder,wang2025improving,wei2025toward};
\textbf{Development Tool Knowledge:} \cite{arora2024masai,yu2025building};
\textbf{Domain-Specific Knowledge:} \cite{chen2026beyondswe};
\textbf{Technology Stack Knowledge:} \cite{chen2026beyondswe}.%
}}

\newsavebox{\boxUsageFeedback}
\sbox{\boxUsageFeedback}{%
\KGUsageBox{Related Work}{%
\textbf{Development Tool Knowledge:} \cite{aggarwal2025dars,ahmed2025otter,antoniades2024swe,arora2024masai,brown2024large,cao2025siadafix,chen2024coder,chen2025old,chen2025prometheus,chen2026beyondswe,cheshkov2024exploring,copet2025cwm,da2025agent,dai2025lita,ding2026swe,ehrlich2025codemonkeys,gandhi2025agents,gao2025trae,gautam2024supercoder2,golubev2025training,guo2025swe,jain2025r2e,jiang2025putting,kon2026swe,lei2024infant,lei2025infantagent,li2025infcode,li2025patchpilot,li2025swe,lin2024llms,lin2025se,lindenbauer2025complexity,liu2024marscode,liu2025context,liu2026architecture,liu2026dynamic,ma2024lingma,ma2025alibaba,ma2025thinking,mu2025experepair,nashid2025issue2test,ouyang2025reasoningbank,pabba2025semagent,pan2024training,pan2025codecor,phan2024hyperagent,pimpale2025forecasting,raghavendra2026agentic,rastogi2025devstral,ruan2025specrover,shi2025aime,shum2025swe,sohrabizadeh2025nemotron,song2026swe,soni2025coding,soni2026swe,sonwane2025bugpilot,sun2025scaling,tang2025co,tang2025synfix,tao2024magis,tao2026swe,vaghasiya2025corethink,vinh2025repeton,wang2025aegis,wang2025practitioner,wang2025seamlessflow,wang2025swe,wei2025toward,wong2025confucius,wu2025git,xia2025demystifying,xia2025live,xiang2026empowering,xiang2026evaluating,xiao2025improving,xie2025swe,xie2026arkeval,yang2024swe,yang2025kgcompass,yang2025kimi,yang2025lingxi,yang2025swe,yu2025building,yu2025orcaloca,yu2025utboost,yuan2026swe,zainullina2025guided,zeng2025satori,zeng2025skywork,zhang2024autocoderover,zhang2025darwin,zhou2025tom,zhu2025training,dong2025infcode};
\textbf{Experiential Knowledge:} \cite{antoniades2024swe,kon2026swe,wang2024openhands,xia2025live,yu2025building};
\textbf{Existing Code Knowledge:} \cite{sun2025scaling,xiang2026empowering};
\textbf{Repository Evolution Knowledge:} \cite{ahmed2025otter};
\textbf{Technology Stack Knowledge:} \cite{pan2024training}.%
}}

\newsavebox{\boxUsageParametric}
\sbox{\boxUsageParametric}{%
\KGUsageBox{Related Work}{%
\textbf{Experiential Knowledge:} \cite{aggarwal2025dars,chen2025locagent,copet2025cwm,da2025agent,golubev2025training,jain2025r2e,kon2026swe,liu2025context,ma2024lingma,ma2025thinking,ma2025tool,pan2024training,rastogi2025devstral,shum2025swe,song2026swe,sonwane2025bugpilot,su2025learn,tang2025boosting,tang2025co,tao2026swe,wang2025practitioner,wang2025swe,wei2025toward,xie2025swe,yang2025kimi,yang2025swe,yu2025building,yuan2026swe,zainullina2025guided,zeng2025satori,zeng2025skywork,zhang2025one,zhang2025sealign};
\textbf{Existing Code Knowledge:} \cite{chakraborty2025blaze,chang2025bridging,copet2025cwm,da2025agent,liu2024codexembed,ma2024lingma,ma2024repository,ma2025sorft,rastogi2025devstral,reddy2025swerank,shum2025swe,sohrabizadeh2025nemotron,soni2026swe,suresh2024cornstack,tang2025boosting,tang2025co,tao2025code,tao2026swe,wang2025mcts,wei2025swe,wei2025toward,xie2025swe,xiong2025think,yang2025swe,zeng2025satori,zeng2025skywork,zhu2025training};
\textbf{Repository Evolution Knowledge:} \cite{cao2025siadafix,chakraborty2025blaze,chang2025bridging,copet2025cwm,jain2025r2e,liu2024codexembed,ma2024lingma,ma2024repository,ma2025sorft,ma2025thinking,ma2025tool,reddy2025swerank,reddy2025swerank+,sohrabizadeh2025nemotron,soni2026swe,tang2025co,tao2025code,tao2026swe,wang2025mcts,wang2025seamlessflow,wang2025swe,wei2025swe,xie2025swe,yang2025kimi,zainullina2025guided,zeng2025satori,zeng2025skywork};
\textbf{Development Tool Knowledge:} \cite{copet2025cwm,da2025agent,kon2026swe,ma2024lingma,ma2025thinking,ma2025tool,rastogi2025devstral,shum2025swe,sun2025scaling,tang2025boosting,tao2026swe,wang2025practitioner,wang2025seamlessflow,yang2025kimi,yang2025swe,yuan2026swe,zeng2025satori,zeng2025skywork,zhang2025sealign,zhu2025training};
\textbf{Programming Language Knowledge:} \cite{reddy2025swerank+};
\textbf{Design Architecture Knowledge:} \cite{raghavendra2026agentic}.%
}}

\begin{forest}
for tree={
    grow'=0,
    anchor=west,
    child anchor=west,
    parent anchor=east,
    font=\sffamily\scriptsize,
    inner sep=2.2pt,
    s sep=2.8pt,
    l sep=1.8mm,
    edge={draw=black!80},
    align=left,
},
where level=0{
    draw=purple!50,
    thick,
    rounded corners=4pt,
    fill=purple!6,
    font=\sffamily\bfseries\scriptsize,
    align=center,
    inner xsep=4pt,
    inner ysep=2pt,
    fit=band
}{},
where level=1{
    draw=blue!40,
    thick,
    rounded corners=4pt,
    fill=blue!6,
    font=\sffamily\bfseries\scriptsize,
    align=center,
    inner xsep=3pt,
    inner ysep=1.5pt,
    fit=band
}{},
where level=2{
    draw=orange!25,
    rounded corners=3pt,
    fill=leafbg,
    font=\sffamily\tiny,
    align=left,
    inner xsep=3pt, 
}{},
[Knowledge\\Usage\\Methods
    [{Passive\\Retrieval}
        [{\usebox{\boxUsagePassive}}]
    ]
    [{Active\\Retrieval}
        [{\usebox{\boxUsageActive}}]
    ]
    [{Feedback-Based\\Usage}
        [{\usebox{\boxUsageFeedback}}]
    ]
    [{Parametric\\Injection}
        [{\usebox{\boxUsageParametric}}]
    ]
]
\end{forest}

\caption{
Knowledge usage methods in automated issue resolution.
Each usage-method box lists the corresponding knowledge types and associated studies.
}
\label{fig:knowledge_usage_method_tree}
\end{figure*}

\subsection{Knowledge Usage Methods}

Fig.~\ref{fig:knowledge_usage_method_tree} presents the distribution of the four usage methods identified in our coding results: Passive Retrieval, Active Retrieval, Feedback-Based Usage, and Parametric Injection. The following analysis compares these methods by examining their representative implementations, the knowledge types they commonly operationalize, and their respective strengths and limitations in supporting automated issue resolution.

\subsubsection{Passive Retrieval}

Passive Retrieval is widely adopted because it provides a controllable way to expose LLMs to task-relevant knowledge before generation. In this usage pattern, systems prepare issue descriptions, repository artifacts, historical evidence, experience memories, or structured constraints in advance and then supply them as external context. Across existing studies, Passive Retrieval mainly appears in three forms: prompt-context injection, similarity-based context retrieval, and structured knowledge retrieval. These forms differ in the type and organization of supplied knowledge, but they share a common role: reducing the model's search burden by constructing an initial knowledge context for issue understanding and patch generation.

\textbf{Prompt-Context Injection.}
Prompt-context injection is mainly used when task-level knowledge, refined issue information, external constraints, or background guidance can be prepared before model inference. In this form, the system designer or pipeline decides what information should be exposed to the LLM, allowing the model to reason over a more complete and focused task context. For example, SIADAFIX optimizes and expands issue descriptions before repair, enabling the model to use refined issue-level knowledge as contextual input \cite{cao2025siadafix}. BeyondSWE includes settings where agents rely on external domain knowledge, official document, or human specifications, making such information part of the task context for issue resolution beyond single-repository bug fixing \cite{chen2026beyondswe}. ArkEval standardizes ArkTS problem statements and evaluates repair under a retrieval-augmented workflow, reflecting the use of language- and technology-specific background information as provided context \cite{xie2026arkeval}. These studies show that prompt-context injection is simple and controllable, but its effectiveness depends on whether the prepared context is complete, concise, and relevant enough for the current issue.

\textbf{Similarity-Based Context Retrieval.}
Similarity-based context retrieval is mainly used to select issue-relevant repository artifacts from large codebases before generation. By using lexical matching, embedding similarity, hierarchical localization, AST-based search, or other predefined retrieval mechanisms, systems can identify relevant files, functions, methods, statements, or code chunks and then supply them to the LLM as external context. This form is especially common for Existing Code Knowledge, because repository-level issue resolution usually requires narrowing a large codebase to a smaller set of likely relevant locations. BLAZE dynamically chunks source code and learns bug-report--code matching patterns for cross-project and cross-language bug localization \cite{chakraborty2025blaze}. CodeXEmbed trains generalist embedding models for multilingual and multi-task code retrieval \cite{liu2024codexembed}. BugCerberus performs hierarchical localization at file, function, and statement levels to provide more precise code context for subsequent issue fixing \cite{chang2025bridging}. AutoCodeRover retrieves repository context through AST-based code search, exposing classes, methods, and related implementation details to the LLM for patch generation \cite{zhang2024autocoderover}. Compared with direct prompt-context injection, this form is more scalable to large repositories, but the retrieval strategy is still largely determined by the system rather than by the model's own exploration decisions.

\textbf{Structured Knowledge Retrieval.}
Structured knowledge retrieval extends passive retrieval from raw code or issue text to more organized knowledge assets. Existing systems retrieve and inject repository history, knowledge-graph paths, experience memories, architectural constraints, or technology-stack knowledge after these resources have been organized by the system. KGCompass retrieves paths connecting issues, pull requests, files, classes, and functions, and then inject the mined contextual paths into the repair prompt \cite{yang2025kgcompass}. Experience-driven methods such as SWE-Exp, EXPEREPAIR, and ReasoningBank retrieve prior repair experiences, episodic demonstrations, semantic reflections, or reasoning memories and use them as contextual guidance for new issues \cite{chen2025swe,mu2025experepair,ouyang2025reasoningbank}. Architecture- and constraint-aware studies further provide design constraints, architecture-related context, or technology-stack knowledge to the model so that generated patches can better conform to project-specific or ecosystem-specific requirements \cite{liu2026architecture,yu2026does,jain2025r2e,xiang2026evaluating,xie2026arkeval}. This form shows that Passive Retrieval is no longer limited to code snippet selection; it can also supply higher-level historical, experiential, relational, and constraint-oriented knowledge.

Overall, Passive Retrieval functions as a knowledge-supplying mechanism. It is effective for building a stable initial context and reducing exploration cost, especially when relevant knowledge can be prepared, ranked, or structured before generation. However, its performance depends heavily on the predefined retrieval pipeline, the quality of selected context, and the system's ability to filter noisy or incomplete information. Therefore, Passive Retrieval provides strong controllability, but limited adaptability when the model needs to discover missing knowledge during the resolution process.

\subsubsection{Active Retrieval}

Active Retrieval becomes important when pre-selected context is insufficient for resolving complex repository-level issues. In this usage pattern, the LLM-based agent participates more directly in knowledge seeking: it decides which files to inspect, which code entities to follow, which memories or histories to consult, and when external resources should be accessed. Existing studies mainly instantiate Active Retrieval in four forms: autonomous codebase exploration, structure-aware active navigation, memory- and history-guided retrieval, and external knowledge-seeking retrieval. These forms show how knowledge use shifts from static context supply toward iterative agent-driven exploration.

\textbf{Autonomous Codebase Exploration.}
Autonomous codebase exploration is mainly used when agents need to search, inspect, and refine repository context during localization and repair. Instead of relying on a fixed set of retrieved files or snippets, agents continuously decide which repository regions should be explored next according to intermediate observations, generated hypotheses, and feedback from previous actions. SWE-Search integrates Monte Carlo Tree Search and iterative refinement to allow agents to explore alternative trajectories, backtrack, and update strategies during repository-level tasks \cite{antoniades2024swe}. MASAI decomposes issue resolution into multiple specialized sub-agents, enabling different agents to gather information from different repository regions according to their assigned objectives \cite{arora2024masai}. Trae Agent formulates repository-level issue resolution as an optimal solution search problem and uses modular agents for generation, pruning, and selection to explore large solution spaces \cite{gao2025trae}. Large Language Monkeys shows that repeated sampling and verification can scale inference-time exploration by searching across diverse candidate solutions rather than relying on a single generation path \cite{brown2024large}. These studies indicate that, in active exploration settings, knowledge acquisition becomes part of the agent's reasoning trajectory rather than a one-time preprocessing step.

\textbf{Structure-Aware Active Navigation.}
Structure-aware active navigation uses repository structures, code graphs, dependency relations, or architectural representations to guide the agent's knowledge-seeking process. This form is especially useful in large repositories, where relevant implementation logic may be distributed across files, modules, classes, and functions. LocAgent parses codebases into heterogeneous graphs and lets agents perform multi-hop reasoning over files, classes, functions, and dependency relations to locate relevant code entities \cite{chen2025locagent}. RPG-Encoder provides a unified repository representation that combines semantic intent and structural dependencies, allowing agents to navigate between high-level architectural intent and low-level implementation logic \cite{luo2026closing}. Agentic Rubrics relies on an expert agent that actively explores the repository to construct context-grounded verification criteria for candidate patches \cite{raghavendra2026agentic}. Compared with open-ended repository exploration, this form makes active retrieval more targeted by using structural knowledge to reduce blind search and support multi-hop reasoning across repository entities.

\textbf{Memory- and History-Guided Retrieval.}
Memory- and history-guided retrieval is mainly used when agents need to consult prior repository evolution, contextual memory, reasoning states, or user-related experience to support current decisions. In this form, the agent does not only search the current codebase, but also retrieves knowledge from past commits, linked issues, summaries, previous trajectories, or persistent memories when such information is relevant. Repository-memory methods retrieve from historical commits, linked issues, and summaries of actively evolving code regions to improve localization \cite{wang2025improving}. CodeR organizes issue resolution through multi-agent collaboration and predefined task graphs, allowing agents to retrieve and exchange task-relevant information in a controlled workflow \cite{chen2024coder}. Git Context Controller structures long-horizon agent memory as a versioned and navigable context hierarchy, enabling agents to branch, merge, recover, and revisit prior reasoning states \cite{wu2025git}. ToM-SWE introduces a dedicated theory-of-mind agent that maintains persistent user-intent memory and provides contextual suggestions to the primary SWE agent \cite{zhou2025tom}. These studies show that Active Retrieval is not limited to code search; it can also support temporally informed and experience-aware reasoning by allowing agents to retrieve what has happened before, what has already been tried, and what contextual assumptions should be preserved.

\textbf{External Knowledge-Seeking Retrieval.}
External knowledge-seeking retrieval expands the knowledge boundary beyond the target repository. This form becomes important when issue resolution depends on external repositories, official document, Domain-Specific Knowledge, technology-stack information, or realistic environment interaction. Such cases often appear in dependency migration, cross-repository maintenance, platform-specific repair, or domain-specific issue resolution. BeyondSWE introduces tasks such as cross-repository issue resolution, domain-specific issue resolution, and dependency-driven migration, and its SearchSWE framework investigates how agents interleave deep search with coding to access external repositories, expert-domain knowledge, and official document \cite{chen2026beyondswe}. Learn-by-interact synthesizes and retrieves interaction experiences from realistic environments to improve agent adaptation \cite{su2025learn}. Confucius Code Agent supports long-context reasoning, persistent note-taking, and modular tool extensions for scalable retrieval and reuse of knowledge in real-world codebases \cite{wong2025confucius}. Compared with repository-internal retrieval, this form allows agents to actively acquire background, ecosystem-level, and experiential knowledge that may not be present in the target repository itself.

Overall, Active Retrieval functions as a knowledge-seeking mechanism. It enables agents to revise retrieval goals, inspect new evidence, recover from incomplete context, and expand the search space when current information is insufficient. This makes it more adaptive than Passive Retrieval, especially for long-horizon and under-specified issue-resolution tasks. However, it also introduces higher cost, longer trajectories, possible error accumulation, and stronger requirements for planning, stopping, and verification. Therefore, the effectiveness of Active Retrieval depends not only on the agent's ability to seek knowledge, but also on whether the system can keep exploration efficient, reliable, and grounded.

\subsubsection{Feedback-Based Usage}

Feedback-Based Usage is most visible in systems that ground issue resolution in executable environments and tool-mediated observations. Instead of relying only on pre-existing repository artifacts or retrieved context, these systems generate new knowledge during the resolution process by running tests, executing code, reading logs, observing runtime states, invoking debugging tools, or receiving guidance and reward signals. Existing studies can be grouped into three forms: test-execution feedback usage, dynamic execution and debugging feedback usage, and guidance and reward feedback usage. These forms respectively use validation outcomes, runtime observations, and higher-level feedback signals to support patch generation, refinement, validation, and agent learning.

\textbf{Test-Execution Feedback Usage.}
Test-execution feedback usage is widely adopted because tests provide executable evidence about whether a candidate patch resolves the issue or introduces regressions. In this form, agents generate, select, run, or reuse tests and then use pass/fail signals, regression outcomes, or failure messages to revise their reasoning and patch decisions. Otter generates tests from issue descriptions and uses their execution results to validate and refine candidate patches \cite{ahmed2025otter}. When Old Meets New leverages regression test suites to prune irrelevant tests and provide issue-specific guidance \cite{pan2024training}. In these systems, knowledge is produced through the interaction between candidate code changes and executable tests. The LLM therefore learns not only from static code context, but also from concrete validation feedback, allowing it to reject incorrect patches, refine repair directions, and improve confidence in the final solution.

\textbf{Dynamic Execution and Debugging Feedback Usage.}
Dynamic execution and debugging feedback usage provides more fine-grained diagnostic evidence about how a program behaves during execution. Compared with test-execution feedback, which mainly indicates whether a behavior passes or fails, runtime traces, debugger observations, exception information, state changes, and semantic execution reports can help agents understand where and why a failure occurs. SWE-Gym instruments Python codebases with unit tests and runtime environments, allowing agents to iteratively adjust actions based on execution outcomes \cite{chen2025old}. DAIRA integrates function-level execution traces and semantic reports to provide precise fault localization for complex defects \cite{xiang2026empowering}. LITA allows LLM agents to observe execution results and adapt their subsequent actions accordingly \cite{dai2025lita,zainullina2025guided,zeng2025satori}. This form highlights the value of dynamic information that cannot be fully obtained from static retrieval: by observing how code actually runs, agents can identify hidden dependencies, runtime failures, incorrect assumptions, and intermediate states that are difficult to infer from repository text alone.

\textbf{Guidance and Reward Feedback Usage.}
Guidance and reward feedback usage uses expert guidance, correction signals, reward models, environment feedback, or tool-assisted interaction signals to regulate agent behavior. This form goes beyond automatic test execution because feedback is used not only to validate a patch, but also to shape the agent's process-level behavior. SWE-Protégé enables small LLMs to selectively query an expert agent and follow guidance to avoid action loops \cite{kon2026swe}. Agent-RLVR and OpenHands provide environment-guided instruction, including plan feedback, interaction signals, and reward-based guidance to improve patch quality and agent learning \cite{wang2024openhands,yu2025building}. Such feedback can help agents decide when to revise a plan, abandon an ineffective action sequence, avoid repeated mistakes, or learn a more reliable repair strategy. This makes Feedback-Based Usage closely connected to adaptive and learning-oriented issue resolution.

Overall, Feedback-Based Usage functions as a dynamic knowledge acquisition and correction mechanism. It grounds issue resolution in observable behavior rather than textual plausibility alone, allowing agents to test assumptions, diagnose failures, refine candidate patches, and learn from environment interaction. However, this usage method also faces practical challenges, including execution cost, incomplete test coverage, noisy feedback, tool failures, and uncertainty about when enough feedback has been collected. Its effectiveness therefore depends on reliable feedback channels and control strategies that prevent agents from being misled by partial or inefficient observations.

\subsubsection{Parametric Injection}

Parametric Injection is mainly used when frequently reused knowledge, behaviors, preferences, or representations need to become more stable model or agent capabilities. Instead of making knowledge available only through prompts, retrieval, or tool interaction at inference time, this usage method incorporates knowledge into model parameters, policy parameters, embedding spaces, reward models, or specialized trainable modules through pre-training, supervised fine-tuning, reinforcement learning, distillation, preference optimization, or representation learning. Existing studies mainly instantiate Parametric Injection in four forms: trajectory-based parametric injection, reward-based parametric injection, representation-based parametric injection, and capability-oriented parametric injection.

\textbf{Trajectory-Based Parametric Injection.}
Trajectory-based parametric injection is mainly used to internalize issue-resolution processes into model behavior. Models learn from historical issue-fixing processes, repository evolution traces, developer workflows, or agent trajectories through supervised fine-tuning, post-training, distillation, or process-oriented training. This form primarily injects Experiential Knowledge and Repository Evolution Knowledge by exposing the model to how software issues are localized, reasoned about, edited, and validated. Existing studies instantiate this form through development-process-centric training, repository-structure-aware training, subtask-oriented fine-tuning, long-reasoning trajectory synthesis, context-management training, and expert-augmented agent training \cite{kon2026swe,liu2025context,ma2024lingma,ma2024repository,ma2025sorft,ma2025thinking,yuan2026swe}. Compared with retrieving individual experiences at inference time, this form aims to internalize more generalizable issue-resolution behaviors, allowing models to reproduce effective localization, planning, editing, and validation patterns across tasks.

\textbf{Reward-Based Parametric Injection.}
Reward-based parametric injection is mainly used to optimize agent behavior through verifiable rewards, environment feedback, tool-use rewards, process rewards, preference signals, or reinforcement learning objectives. This form injects Development Tool Knowledge and Experiential Knowledge by teaching agents how to act in software-engineering environments, when to use tools, how to recover from failures, and how to prefer more effective trajectories under executable or feedback-rich settings \cite{da2025agent,yang2025kimi,zeng2025skywork,zhang2025sealign,ma2025tool,zainullina2025guided}. Unlike trajectory-based injection, which learns primarily from demonstrations or synthesized processes, reward-based injection emphasizes optimization through feedback. It therefore supports the internalization of action-selection policies, tool-use preferences, exploration strategies, and failure-recovery behaviors that are difficult to specify through prompt instructions alone.

\textbf{Representation-Based Parametric Injection.}
Representation-based parametric injection is mainly used to improve knowledge access before generation. Instead of inserting knowledge directly into the prompt, systems encode issue-code matching, multilingual code semantics, ranking preferences, or retrieval-oriented knowledge into embedding models, ranking models, retrievers, or code-aware representation spaces. This form mainly supports Existing Code Knowledge and Programming Language Knowledge by training models to capture semantic relations between issues, code entities, programming languages, and retrieval tasks. CodeXEmbed and SWE-Rank+ illustrate this form by training retrieval or ranking models to capture code and issue-code matching knowledge across multilingual and multi-task code retrieval settings \cite{liu2024codexembed,reddy2025swerank+}. In this form, the injected knowledge influences which repository artifacts are retrieved, ranked, or exposed to the downstream LLM, thereby improving the quality of knowledge access in large or multilingual repositories.

\textbf{Capability-Oriented Parametric Injection.}
Capability-oriented parametric injection is mainly used to transform software-engineering priors into persistent model or module capabilities. This form targets verification criteria, architectural constraints, tool-use skills, coding priors, and reusable procedures that can guide patch assessment, tool operation, repository-level reasoning, and workflow execution. Agentic Rubrics converts repository-grounded verification knowledge into structured criteria that can support scalable patch assessment, while synthetic-environment training and agentless-to-agentic training inject broader coding skills, tool-use behavior, and repository-level reasoning priors into models \cite{raghavendra2026agentic,zhu2025training,yang2025kimi}. This form is important because issue resolution requires not only generating patches, but also judging patch correctness, following repository constraints, operating tools effectively, and reusing learned procedures across related tasks. By internalizing these capability-oriented priors, models can reduce repeated prompt engineering and make agent behavior more stable across issue-resolution workflows.

Overall, Parametric Injection functions as a long-term knowledge internalization mechanism. Its advantage is that knowledge can influence model behavior without repeatedly expanding prompts, querying indexes, or invoking tools at every inference step. This is especially useful for recurring repair patterns, repository-level reasoning skills, multilingual code understanding, tool-use policies, verification preferences, and general software-engineering capabilities. However, once knowledge is absorbed into parameters or trainable modules, it becomes less transparent and less easily editable than prompt- or retrieval-based knowledge, and it may become stale when repositories, dependencies, tools, or development conventions evolve. Therefore, Parametric Injection is most effective when combined with external knowledge access mechanisms, while its unique role lies in turning frequently reused knowledge and behaviors into stable model capabilities.

\subsection{Knowledge-Type Distribution across Usage Methods}

\begin{figure}[htbp]
    \centering
    \includegraphics[width=0.9\linewidth]{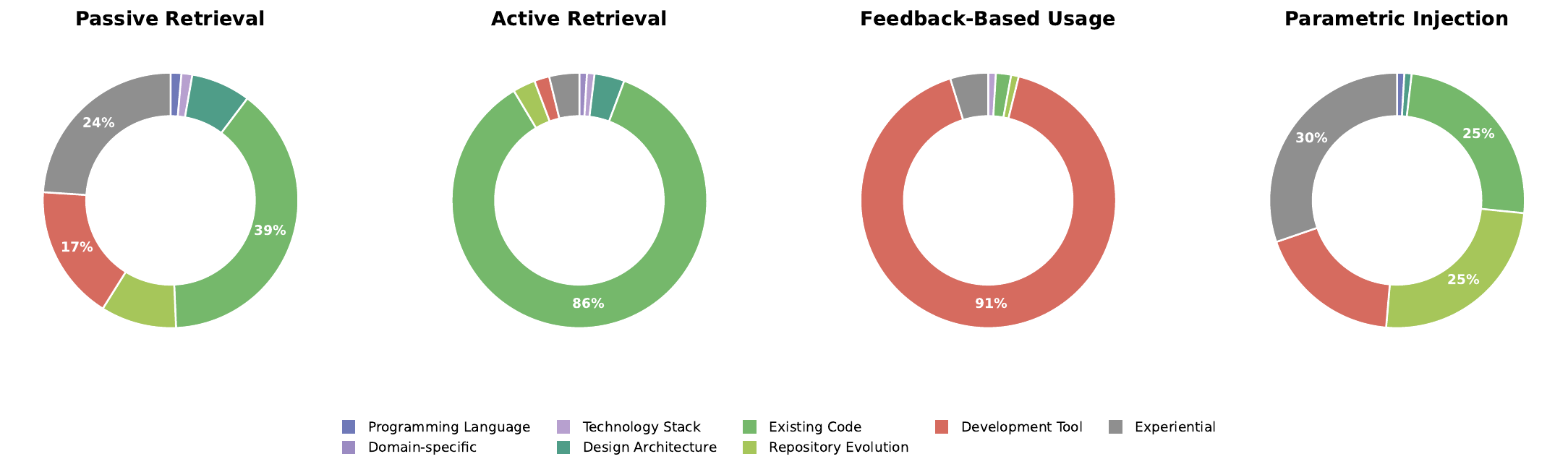}
    \caption{Distribution of knowledge types across knowledge usage methods.}
    \label{fig:usage_method_donut}
\end{figure}

Fig.~\ref{fig:usage_method_donut} further quantifies the distribution of knowledge types across different knowledge usage methods. The percentages indicate the relative share of each knowledge type among the knowledge-use associations observed under a given usage method. They reflect how frequently different knowledge types are associated with each usage method in the reviewed studies.

Passive Retrieval presents a relatively mixed composition in the coded sample: Existing Code Knowledge accounts for the largest share at 39\%, followed by Experiential Knowledge at 24\% and Development Tool Knowledge at 17\%. This suggests that Passive Retrieval is frequently used not only to provide static code snippets, but also to supply preselected repair experiences, tool outputs, validation signals, and other already organized contextual information to the model. In this sense, Passive Retrieval provides a lightweight and controllable way to expose LLMs to task-relevant knowledge when relevant information can be selected in advance. However, this observed usage pattern also indicates a limitation: the usefulness of Passive Retrieval depends heavily on prompt construction, retrieval quality, and the completeness of the preselected context. As issue-resolution tasks become more repository-scale and long-horizon, static or one-shot context provision may become insufficient when the required knowledge cannot be fully anticipated before interaction.

Active Retrieval shows a stronger association with Existing Code Knowledge, which accounts for 86\% of the coded distribution under this usage method. Rather than implying that Active Retrieval is inherently limited to repository exploration, this pattern suggests that current studies most often apply Active Retrieval to situations where LLMs need to inspect large codebases, decide which files or code entities to examine, and iteratively refine repository understanding during localization and repair. Compared with Passive Retrieval, this usage method reflects a more agentic form of knowledge seeking: the LLM can participate in deciding what information is missing, where it may be obtained, and whether the currently available evidence is sufficient for the next action. Such usage is especially relevant for large repositories, multi-file modifications, under-specified issue reports, and tasks where the relevant context is difficult to determine in advance. Recent agentic systems and repository-navigation methods also reflect this tendency toward autonomous knowledge exploration and decision-making \cite{antoniades2024swe,chen2025locagent}.

Feedback-Based Usage exhibits a concentrated association with Development Tool Knowledge, which accounts for 91\% of the coded distribution under this usage method. This result should be interpreted as a knowledge-use preference in the reviewed studies: Feedback-Based Usage is most often observed in settings where agents consume dynamic information produced by execution environments, tests, debuggers, build systems, logs, and other non-search tools. Such feedback can provide important evidence for validation, debugging, and hypothesis revision, especially when issue-resolution failures cannot be diagnosed from static repository context alone. In this sense, Feedback-Based Usage often complements retrieval-based methods by allowing agents to observe whether a candidate action, test, or patch works in the actual environment. However, the distribution should not be read as showing that feedback is the only function of this mechanism or that it causally guarantees better correction; rather, it indicates that current implementations most frequently operationalize feedback through development-tool-related knowledge.

Parametric Injection differs from the other methods by emphasizing long-term knowledge internalization rather than explicit context provision or real-time interaction. In the coded distribution, Experiential Knowledge accounts for the largest share at 30\%, followed by Existing Code Knowledge and Repository Evolution Knowledge, each accounting for 25\%. This suggests that training-based or parameter-internalized methods often absorb reusable repair trajectories, code-understanding capabilities, historical issue-fix patterns, tool-use behaviors, and verification preferences into model parameters, agent policies, retrievers, rankers, reward models, or specialized trainable modules. Compared with prompting or retrieval-based methods, Parametric Injection may reduce repeated context construction and make frequently reused knowledge more persistently available to the system. At the same time, its adoption usually depends on larger datasets, stronger infrastructure, and higher training cost, which may explain why it is more commonly pursued in well-resourced training or reinforcement-learning settings. Existing training- and reinforcement-oriented systems also suggest this direction, where issue-resolution capability is improved not only by inference-time context construction, but also by post-training, reward optimization, and trajectory learning \cite{ma2025tool,yang2025kimi}.

Taken together, these distributions suggest that different usage methods are associated with different knowledge-use preferences in current studies. Passive Retrieval is often used for controllable preselected context provision; Active Retrieval is frequently associated with repository-oriented exploration; Feedback-Based Usage is commonly tied to dynamic tool and execution evidence; and Parametric Injection is often used to internalize reusable experience, repository patterns, and tool-use behaviors. These observations support the view that the four usage methods are complementary rather than mutually substitutable. Future issue-resolution systems may benefit from hybrid usage strategies that combine preselected context, active exploration, dynamic feedback, and long-term knowledge internalization according to task difficulty, cost constraints, efficiency requirements, and the need for generalizable capability. Nevertheless, the figure also shows that Background Knowledge and Design Architecture Knowledge are weakly represented across all usage methods, suggesting that current systems still have substantial room to develop more systematic mechanisms for incorporating domain assumptions, programming-language constraints, technology-stack conventions, and architectural intent into both explicit and internalized knowledge usage pathways.

\finding{
The usage results show that knowledge use in automated issue resolution is moving from static context supply toward a more agentic and adaptive form of knowledge operationalization. Passive Retrieval remains important for low-cost and controllable context construction, while Active Retrieval allows agents to seek missing knowledge during repository exploration. Feedback-Based Usage grounds the resolution process in executable evidence from tests, runtime states, logs, debugging information, and other tool observations. Parametric Injection further internalizes recurring repair patterns, tool-use behaviors, repository-level reasoning skills, and historical experience into more persistent model or agent capabilities. Together, these methods form a complementary usage portfolio in which knowledge is not only supplied to LLMs, but also actively explored, dynamically verified, and gradually transformed into reusable capabilities.

This finding suggests that future issue-resolution systems should be designed around hybrid knowledge-use loops rather than isolated usage mechanisms. Effective software agents need to coordinate external context, agentic exploration, environment feedback, and learned capabilities within the same resolution process. From a knowledge-centric perspective, this means that agents should not only retrieve relevant code, but also decide what knowledge is missing, verify assumptions through tools, preserve useful interaction experience, and update reusable knowledge over time. Such a loop is especially important for incorporating currently underused knowledge, including domain assumptions, programming-language constraints, technology-stack conventions, and architectural intent. By connecting these forms of knowledge with retrieval decisions, feedback interpretation, and capability learning, future systems may become more controllable, verifiable, and semantically grounded in complex repository-level software maintenance.
}

\section{RQ5: At which stages is knowledge applied in issue resolution?}
\label{sec:rq5}

After examining knowledge types, sources, representations, and usage methods, we further analyze how knowledge contributes across the issue-resolution process. RQ5 shifts the focus from the forms and mechanisms of knowledge utilization to its workflow positions, asking where knowledge becomes useful for preparing, guiding, executing, and validating automated issue resolution. This stage-oriented perspective helps connect the preceding knowledge dimensions with the concrete activities performed by issue-resolution agents.

\begin{figure}[htbp]
    \centering
    \includegraphics[width=0.9\linewidth]{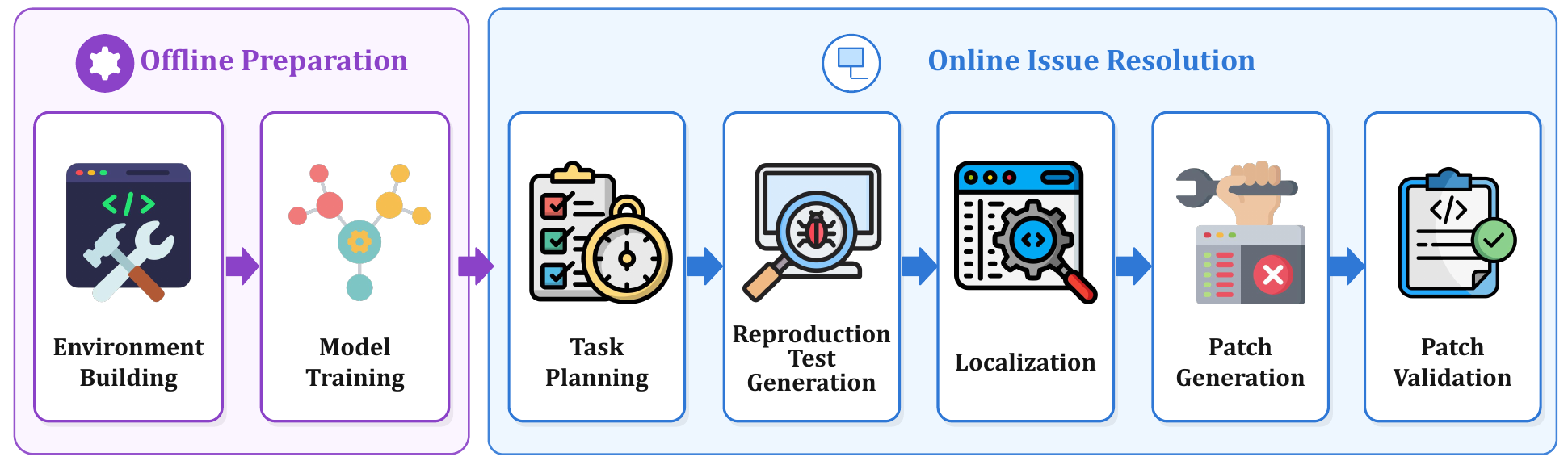}
    \caption{Workflow relationships among knowledge usage stages in automated issue resolution.}
    \label{fig:knowledge_usage_stage}
\end{figure}
\subsection{Knowledge Usage Stages}

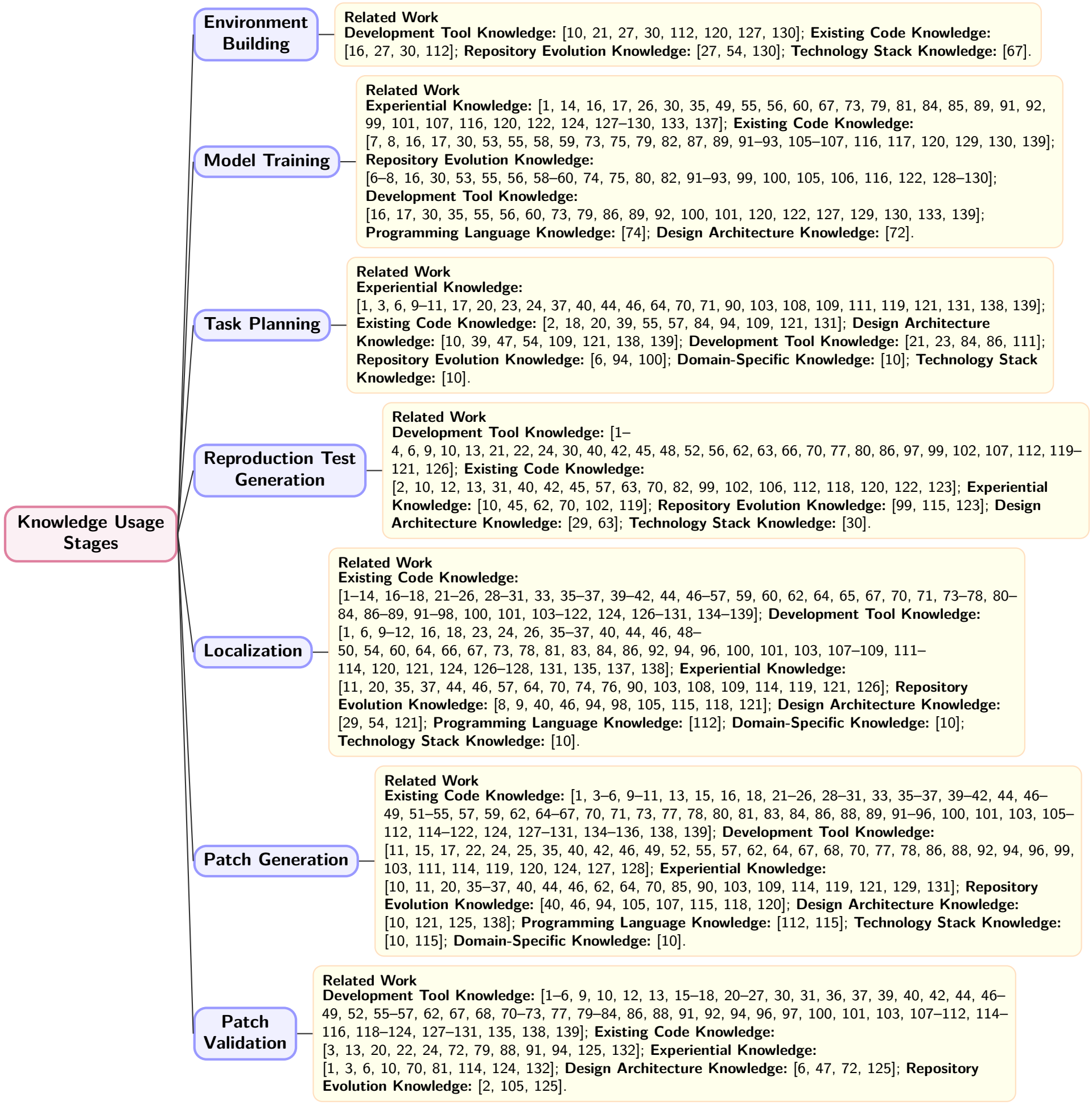
\begin{figure*}[htbp]
\centering
\scriptsize
\definecolor{leafbg}{HTML}{FFFFEC}

\newlength{\KGUsageWidth}
\setlength{\KGUsageWidth}{8.0cm} 

\newcommand{\KGUsageBox}[2]{%
  \begin{varwidth}[t]{\KGUsageWidth}
    \normalfont\sffamily\tiny\raggedright
    \textbf{#1}\\[-1pt]
    #2%
  \end{varwidth}%
}


\newsavebox{\boxUsageEnvBuild}
\sbox{\boxUsageEnvBuild}{%
\KGUsageBox{Related Work}{%
\textbf{Development Tool Knowledge:} \cite{chen2026beyondswe,guo2025swe,jain2025r2e,xiang2026evaluating,yang2025swe,yuan2026swe,zeng2025skywork,dong2025infcode};
\textbf{Existing Code Knowledge:} \cite{copet2025cwm,guo2025swe,jain2025r2e,xiang2026evaluating};
\textbf{Repository Evolution Knowledge:} \cite{guo2025swe,luo2026closing,zeng2025skywork};
\textbf{Technology Stack Knowledge:} \cite{pan2024training}.%
}}

\newsavebox{\boxUsageTraining}
\sbox{\boxUsageTraining}{%
\KGUsageBox{Related Work}{%
\textbf{Experiential Knowledge:} \cite{aggarwal2025dars,chen2025locagent,copet2025cwm,da2025agent,golubev2025training,jain2025r2e,kon2026swe,liu2025context,ma2024lingma,ma2025thinking,ma2025tool,pan2024training,rastogi2025devstral,shum2025swe,song2026swe,sonwane2025bugpilot,su2025learn,tang2025boosting,tang2025co,tao2026swe,wang2025practitioner,wang2025swe,wei2025toward,xie2025swe,yang2025kimi,yang2025swe,yu2025building,yuan2026swe,zainullina2025guided,zeng2025satori,zeng2025skywork,zhang2025one,zhang2025sealign};
\textbf{Existing Code Knowledge:} \cite{chakraborty2025blaze,chang2025bridging,copet2025cwm,da2025agent,jain2025r2e,liu2024codexembed,ma2024lingma,ma2024repository,ma2025sorft,rastogi2025devstral,reddy2025swerank,shum2025swe,soni2026swe,suresh2024cornstack,tang2025boosting,tang2025co,tao2025code,tao2026swe,wang2025mcts,wei2025swe,wei2025toward,xie2025swe,xiong2025think,yang2025swe,zeng2025satori,zeng2025skywork,zhu2025training};
\textbf{Repository Evolution Knowledge:} \cite{cao2025siadafix,chakraborty2025blaze,chang2025bridging,copet2025cwm,jain2025r2e,liu2024codexembed,ma2024lingma,ma2024repository,ma2025sorft,ma2025thinking,ma2025tool,reddy2025swerank,reddy2025swerank+,sohrabizadeh2025nemotron,soni2026swe,tang2025co,tao2025code,tao2026swe,wang2025mcts,wang2025seamlessflow,wang2025swe,wei2025swe,xie2025swe,yang2025kimi,zainullina2025guided,zeng2025satori,zeng2025skywork};
\textbf{Development Tool Knowledge:} \cite{copet2025cwm,da2025agent,jain2025r2e,kon2026swe,ma2024lingma,ma2025thinking,ma2025tool,rastogi2025devstral,shum2025swe,sun2025scaling,tang2025boosting,tao2026swe,wang2025practitioner,wang2025seamlessflow,yang2025kimi,yang2025swe,yuan2026swe,zeng2025satori,zeng2025skywork,zhang2025sealign,zhu2025training};
\textbf{Programming Language Knowledge:} \cite{reddy2025swerank+};
\textbf{Design Architecture Knowledge:} \cite{raghavendra2026agentic}.%
}}

\newsavebox{\boxUsageTaskPlanning}
\sbox{\boxUsageTaskPlanning}{%
\KGUsageBox{Related Work}{%
\textbf{Experiential Knowledge:} \cite{aggarwal2025dars,antoniades2024swe,cao2025siadafix,chen2024coder,chen2025swe,chen2026beyondswe,da2025agent,ding2026swe,gandhi2025agents,gao2025trae,lei2024infant,li2025patchpilot,lin2025se,lindenbauer2025complexity,ouyang2025reasoningbank,phan2024hyperagent,pimpale2025forecasting,tang2025agent,wang2024openhands,wong2025confucius,wu2025git,xia2025live,yang2024swe,yang2025lingxi,zhang2025darwin,zhou2025tom,zhu2025training};
\textbf{Existing Code Knowledge:} \cite{ahmed2025otter,dai2025lita,ding2026swe,li2025swe,ma2024lingma,ma2025alibaba,sonwane2025bugpilot,tao2024magis,wu2025git,yang2025lingxi,zhang2025darwin};
\textbf{Design Architecture Knowledge:} \cite{chen2026beyondswe,li2025swe,liu2026architecture,luo2026closing,wu2025git,yang2025lingxi,zhou2025tom,zhu2025training};
\textbf{Development Tool Knowledge:} \cite{gandhi2025agents,sonwane2025bugpilot,sun2025scaling,xia2025live,dong2025infcode};
\textbf{Repository Evolution Knowledge:} \cite{cao2025siadafix,tao2024magis,wang2025seamlessflow};
\textbf{Domain-Specific Knowledge:} \cite{chen2026beyondswe};
\textbf{Technology Stack Knowledge:} \cite{chen2026beyondswe}.%
}}

\newsavebox{\boxUsageReproduction}
\sbox{\boxUsageReproduction}{%
\KGUsageBox{Related Work}{%
\textbf{Development Tool Knowledge:} \cite{aggarwal2025dars,ahmed2025otter,antoniades2024swe,arora2024masai,cao2025siadafix,chen2024coder,chen2025prometheus,chen2026beyondswe,ehrlich2025codemonkeys,gao2025trae,jain2025r2e,li2025infcode,li2025patchpilot,lin2024llms,liu2024marscode,liu2026dynamic,ma2025thinking,mu2025experepair,nashid2025issue2test,pabba2025semagent,phan2024hyperagent,ruan2025specrover,sohrabizadeh2025nemotron,sun2025scaling,vinh2025repeton,wang2025aegis,wang2025swe,wei2025toward,xiang2026evaluating,yang2024swe,yang2025lingxi,yang2025swe,yu2025orcaloca,dong2025infcode};
\textbf{Existing Code Knowledge:} \cite{ahmed2025otter,chen2025old,chen2025prometheus,chen2026beyondswe,jiang2025putting,li2025infcode,li2025patchpilot,lin2024llms,ma2025alibaba,nashid2025issue2test,phan2024hyperagent,soni2026swe,wang2025aegis,wang2025swe,wei2025swe,xiang2026evaluating,yang2025kgcompass,yang2025kimi,yang2025swe,yu2025utboost};
\textbf{Experiential Knowledge:} \cite{chen2026beyondswe,lin2024llms,mu2025experepair,phan2024hyperagent,wang2025aegis,yang2024swe};
\textbf{Repository Evolution Knowledge:} \cite{wang2025swe,xie2026arkeval,yu2025utboost};
\textbf{Design Architecture Knowledge:} \cite{huang2025seeing,nashid2025issue2test};
\textbf{Technology Stack Knowledge:} \cite{jain2025r2e}.%
}}

\newsavebox{\boxUsageLocalization}
\sbox{\boxUsageLocalization}{%
\KGUsageBox{Related Work}{%
\textbf{Existing Code Knowledge:} \cite{aggarwal2025dars,ahmed2025otter,antoniades2024swe,arora2024masai,brown2024large,cao2025siadafix,chakraborty2025blaze,chang2025bridging,chen2024coder,chen2025locagent,chen2025old,chen2025prometheus,chen2025swe,chen2026beyondswe,copet2025cwm,da2025agent,dai2025lita,ehrlich2025codemonkeys,gandhi2025agents,gao2025trae,gautam2024supercoder2,golubev2025training,gupta2025sacl,huang2025seeing,jain2025r2e,jiang2025cosil,jiang2025putting,kon2026swe,lei2024infant,lei2025infantagent,li2024codetree,li2025infcode,li2025patchpilot,li2025swe,lin2025se,lindenbauer2025complexity,liu2024codexembed,liu2024codexgraph,liu2024marscode,liu2025context,liu2025graphlocator,liu2026architecture,liu2026dynamic,luo2026closing,ma2024lingma,ma2025alibaba,ma2025sorft,ma2025thinking,ma2025tool,mu2025experepair,ouyang2024repograph,ouyang2025reasoningbank,pan2024training,phan2024hyperagent,pimpale2025forecasting,rastogi2025devstral,reddy2025swerank,reddy2025swerank+,rombaut2024watson,ruan2025specrover,shi2025aime,sohrabizadeh2025nemotron,song2026swe,soni2025coding,soni2026swe,sonwane2025bugpilot,sun2025scaling,suresh2024cornstack,tang2025boosting,tang2025co,tang2025synfix,tao2024magis,tao2025code,tao2026swe,tawosi2025meta,vaghasiya2025corethink,vinh2025repeton,wang2024openhands,wang2025extracting,wang2025improving,wang2025mcts,wang2025practitioner,wang2025seamlessflow,wei2025swe,wei2025toward,wong2025confucius,wu2025git,xia2025demystifying,xia2025live,xiang2026empowering,xiang2026evaluating,xiao2025improving,xie2025swe,xie2026arkeval,xiong2025think,yang2024swe,yang2025kgcompass,yang2025kimi,yang2025lingxi,yang2025swe,yu2025building,yu2025orcaloca,yuan2026swe,zainullina2025guided,zeng2025satori,zeng2025skywork,zhang2024autocoderover,zhang2024codev,zhang2025cast,zhang2025darwin,zhang2025one,zhou2025tom,zhu2025training,dong2025infcode};
\textbf{Development Tool Knowledge:} \cite{aggarwal2025dars,cao2025siadafix,chen2024coder,chen2025old,chen2025swe,chen2026beyondswe,copet2025cwm,dai2025lita,gandhi2025agents,gao2025trae,golubev2025training,kon2026swe,lei2024infant,lei2025infantagent,li2025patchpilot,lin2025se,lindenbauer2025complexity,liu2025context,liu2025graphlocator,liu2026dynamic,luo2026closing,ma2025tool,ouyang2025reasoningbank,pabba2025semagent,pan2024training,rastogi2025devstral,shi2025aime,song2026swe,soni2025coding,sonwane2025bugpilot,sun2025scaling,tao2024magis,tao2026swe,vaghasiya2025corethink,wang2024openhands,wang2025practitioner,wang2025seamlessflow,wei2025toward,wong2025confucius,wu2025git,xia2025live,xiang2026empowering,xiang2026evaluating,xiao2025improving,yang2025lingxi,yang2025swe,yu2025building,yu2025orcaloca,yuan2026swe,zainullina2025guided,zhang2024autocoderover,zhang2025darwin,zhang2025one,zhou2025tom};
\textbf{Experiential Knowledge:} \cite{chen2025swe,ding2026swe,kon2026swe,lei2024infant,lin2025se,lindenbauer2025complexity,ma2025alibaba,ouyang2025reasoningbank,phan2024hyperagent,reddy2025swerank+,rombaut2024watson,tang2025agent,wang2024openhands,wong2025confucius,wu2025git,xiao2025improving,yang2024swe,yang2025lingxi,yu2025orcaloca};
\textbf{Repository Evolution Knowledge:} \cite{chang2025bridging,chen2024coder,li2025patchpilot,lindenbauer2025complexity,tao2024magis,wang2025improving,wang2025mcts,xie2026arkeval,yang2025kgcompass,yang2025lingxi};
\textbf{Design Architecture Knowledge:} \cite{huang2025seeing,luo2026closing,yang2025lingxi};
\textbf{Programming Language Knowledge:} \cite{xiang2026evaluating};
\textbf{Domain-Specific Knowledge:} \cite{chen2026beyondswe};
\textbf{Technology Stack Knowledge:} \cite{chen2026beyondswe}.%
}}

\newsavebox{\boxUsagePatchGeneration}
\sbox{\boxUsagePatchGeneration}{%
\KGUsageBox{Related Work}{%
\textbf{Existing Code Knowledge:} \cite{aggarwal2025dars,antoniades2024swe,arora2024masai,brown2024large,cao2025siadafix,chen2024coder,chen2025prometheus,chen2025swe,chen2026beyondswe,cheshkov2024exploring,copet2025cwm,dai2025lita,ehrlich2025codemonkeys,gandhi2025agents,gao2025trae,gautam2024supercoder2,golubev2025training,gupta2025sacl,huang2025seeing,jain2025r2e,jiang2025cosil,jiang2025putting,kon2026swe,lei2024infant,lei2025infantagent,li2024codetree,li2025infcode,li2025patchpilot,li2025swe,lin2025se,lindenbauer2025complexity,liu2024codexembed,liu2024codexgraph,liu2024marscode,liu2025context,liu2026architecture,liu2026dynamic,luo2026closing,ma2024lingma,ma2025alibaba,ma2025sorft,mu2025experepair,ouyang2024repograph,ouyang2025reasoningbank,pabba2025semagent,pan2024training,phan2024hyperagent,pimpale2025forecasting,rastogi2025devstral,ruan2025specrover,shi2025aime,sohrabizadeh2025nemotron,song2026swe,soni2025coding,sonwane2025bugpilot,sun2025scaling,tang2025boosting,tang2025co,tang2025synfix,tao2024magis,tao2025code,tao2026swe,tawosi2025meta,vaghasiya2025corethink,wang2024openhands,wang2025mcts,wang2025practitioner,wang2025seamlessflow,wei2025swe,wei2025toward,wong2025confucius,wu2025git,xia2025demystifying,xia2025live,xiang2026evaluating,xiao2025improving,xie2025swe,xie2026arkeval,xiong2025think,yang2024swe,yang2025kgcompass,yang2025kimi,yang2025lingxi,yang2025swe,yu2025building,yuan2026swe,zainullina2025guided,zeng2025satori,zeng2025skywork,zhang2024autocoderover,zhang2024codev,zhang2025cast,zhang2025darwin,zhou2025tom,zhu2025training,dong2025infcode};
\textbf{Development Tool Knowledge:} \cite{chen2025swe,cheshkov2024exploring,da2025agent,ehrlich2025codemonkeys,gao2025trae,gautam2024supercoder2,kon2026swe,li2025infcode,li2025patchpilot,lindenbauer2025complexity,liu2024marscode,liu2025context,ma2024lingma,ma2025alibaba,mu2025experepair,ouyang2025reasoningbank,pan2024training,pan2025codecor,phan2024hyperagent,ruan2025specrover,shi2025aime,sun2025scaling,tang2025synfix,tao2024magis,tao2026swe,vaghasiya2025corethink,wang2024openhands,wang2025swe,xia2025live,xiao2025improving,yang2024swe,yang2025swe,yu2025building,yuan2026swe,zainullina2025guided};
\textbf{Experiential Knowledge:} \cite{chen2025swe,chen2026beyondswe,ding2026swe,kon2026swe,lei2024infant,lei2025infantagent,li2025patchpilot,lin2025se,lindenbauer2025complexity,mu2025experepair,ouyang2025reasoningbank,phan2024hyperagent,su2025learn,tang2025agent,wang2024openhands,wu2025git,xiao2025improving,yang2024swe,yang2025lingxi,zeng2025satori,zhang2025darwin};
\textbf{Repository Evolution Knowledge:} \cite{li2025patchpilot,lindenbauer2025complexity,tao2024magis,wang2025mcts,wei2025toward,xie2026arkeval,yang2025kgcompass,yang2025swe};
\textbf{Design Architecture Knowledge:} \cite{chen2026beyondswe,yang2025lingxi,yu2026does,zhou2025tom};
\textbf{Programming Language Knowledge:} \cite{xiang2026evaluating,xie2026arkeval};
\textbf{Technology Stack Knowledge:} \cite{chen2026beyondswe,xie2026arkeval};
\textbf{Domain-Specific Knowledge:} \cite{chen2026beyondswe}.%
}}

\newsavebox{\boxUsageValidation}
\sbox{\boxUsageValidation}{%
\KGUsageBox{Related Work}{%
\textbf{Development Tool Knowledge:} \cite{aggarwal2025dars,ahmed2025otter,antoniades2024swe,arora2024masai,brown2024large,cao2025siadafix,chen2024coder,chen2025old,chen2025prometheus,chen2026beyondswe,cheshkov2024exploring,copet2025cwm,da2025agent,dai2025lita,ding2026swe,ehrlich2025codemonkeys,gandhi2025agents,gao2025trae,gautam2024supercoder2,golubev2025training,guo2025swe,jain2025r2e,jiang2025putting,lei2024infant,lei2025infantagent,li2025infcode,li2025patchpilot,li2025swe,lin2025se,lindenbauer2025complexity,liu2024marscode,liu2025context,liu2026architecture,liu2026dynamic,ma2024lingma,ma2025alibaba,ma2025thinking,mu2025experepair,pan2024training,pan2025codecor,phan2024hyperagent,pimpale2025forecasting,raghavendra2026agentic,rastogi2025devstral,ruan2025specrover,shum2025swe,sohrabizadeh2025nemotron,song2026swe,soni2025coding,soni2026swe,sonwane2025bugpilot,sun2025scaling,tang2025co,tang2025synfix,tao2024magis,tao2026swe,vaghasiya2025corethink,vinh2025repeton,wang2024openhands,wang2025practitioner,wang2025seamlessflow,wei2025toward,wong2025confucius,wu2025git,xia2025demystifying,xia2025live,xiang2026evaluating,xiao2025improving,xie2025swe,xie2026arkeval,yang2024swe,yang2025kgcompass,yang2025kimi,yang2025lingxi,yang2025swe,yu2025building,yu2025utboost,yuan2026swe,zainullina2025guided,zeng2025satori,zeng2025skywork,zhang2024autocoderover,zhang2025darwin,zhou2025tom,zhu2025training,dong2025infcode};
\textbf{Existing Code Knowledge:} \cite{antoniades2024swe,chen2025prometheus,ding2026swe,ehrlich2025codemonkeys,gao2025trae,raghavendra2026agentic,shum2025swe,tang2025co,tang2025synfix,tao2024magis,yu2026does,zhang2024diversity};
\textbf{Experiential Knowledge:} \cite{aggarwal2025dars,antoniades2024swe,cao2025siadafix,chen2026beyondswe,phan2024hyperagent,song2026swe,xiao2025improving,yu2025building,zhang2024diversity};
\textbf{Design Architecture Knowledge:} \cite{cao2025siadafix,liu2026architecture,raghavendra2026agentic,yu2026does};
\textbf{Repository Evolution Knowledge:} \cite{ahmed2025otter,wang2025mcts,yu2026does}.%
}}

\begin{forest}
for tree={
    grow'=0,
    anchor=west,
    child anchor=west,
    parent anchor=east,
    font=\sffamily\scriptsize,
    inner sep=2.2pt,
    s sep=2.8pt,
    l sep=1.8mm,
    edge={draw=black!80},
    align=left,
},
where level=0{
    draw=purple!50,
    thick,
    rounded corners=4pt,
    fill=purple!6,
    font=\sffamily\bfseries\scriptsize,
    align=center,
    inner xsep=4pt,
    inner ysep=2pt,
    fit=band
}{},
where level=1{
    draw=blue!40,
    thick,
    rounded corners=4pt,
    fill=blue!6,
    font=\sffamily\bfseries\scriptsize,
    align=center,
    inner xsep=3pt,
    inner ysep=1.5pt,
    fit=band
}{},
where level=2{
    draw=orange!25,
    rounded corners=3pt,
    fill=leafbg,
    font=\sffamily\tiny,
    align=left,
    inner xsep=3pt, 
}{},
[Knowledge Usage\\Stages
    [{Environment\\Building}
        [{\usebox{\boxUsageEnvBuild}}]
    ]
    [{Model Training}
        [{\usebox{\boxUsageTraining}}]
    ]
    [{Task Planning}
        [{\usebox{\boxUsageTaskPlanning}}]
    ]
    [{Reproduction Test\\Generation}
        [{\usebox{\boxUsageReproduction}}]
    ]
    [{Localization}
        [{\usebox{\boxUsageLocalization}}]
    ]
    [{Patch Generation}
        [{\usebox{\boxUsagePatchGeneration}}]
    ]
    [{Patch\\Validation}
        [{\usebox{\boxUsageValidation}}]
    ]
]
\end{forest}

\caption{
Knowledge usage stages in automated issue resolution.
Each usage-stage box lists the corresponding knowledge types and associated studies.
}
\label{fig:knowledge_usage_stage_tree}
\end{figure*}

Following the stage schema introduced in Section~\ref{sec:issue_resolution_stage_schema}, this subsection examines the distribution and roles of knowledge across Environment Building, Model Training, Task Planning, Reproduction Test Generation, Localization, Patch Generation, and Patch Validation. Fig.~\ref{fig:knowledge_usage_stage_tree} summarizes the stage-wise coding results, while Fig.~\ref{fig:knowledge_usage_stage} shows how these stages provide a fine-grained analytical lens for understanding when and for what purposes knowledge is applied within the broader issue-resolution workflow.

\subsubsection{Environment Building}

At the Environment Building stage, knowledge is applied to make repositories executable, reproducible, and verifiable before training, evaluation, or online issue solving. Existing studies show that this stage is less concerned with diagnosing the issue itself and more concerned with constructing a reliable execution substrate for subsequent agent actions. The major concerns include environment operation, repository-grounded setup, historical reproduction, and runtime compatibility.

\textbf{Environment Operation.}
Environment operation mainly involves applying Development Tool Knowledge to prepare executable and controllable environments. Automated issue-resolution systems need to invoke package managers, build tools, shell commands, test runners, containers, sandboxes, and execution backends in a reliable way. SWE-Factory treats evaluation environment construction as a major bottleneck and uses SWE-Builder to iteratively construct executable environments with an environment memory pool \cite{guo2025swe}. R2E-Gym and Skywork-SWE package executable environments or runtime images to support repeated agent interaction and automated unit-test validation \cite{jain2025r2e,zeng2025skywork}. SWE-MiniSandbox further uses lightweight sandboxing and environment pre-caching to reduce the cost of large-scale agent training \cite{yuan2026swe}. These studies indicate that tool knowledge provides the operational basis for executable issue-resolution workflows.

\textbf{Repository-Grounded Setup.}
Repository-grounded setup applies Existing Code Knowledge to adapt environment construction to the concrete repository under repair. Dependency declarations, setup files, build scripts, tests, executable entry points, and project-specific testing conventions all affect whether an environment can be correctly prepared. CWM builds executable repository images by inspecting repository contents and preparing environments capable of running code and tests \cite{copet2025cwm}. SWE-Factory and R2E-Gym also rely on repository-specific configurations, tests, and execution requirements to construct valid task environments \cite{guo2025swe,jain2025r2e}. RustForger further shows that environment construction in Rust repositories must account for project-level build settings, compilation requirements, and dynamic tracing support \cite{xiang2026evaluating}. These examples show that environment construction must be grounded in the actual repository rather than based only on generic setup commands.

\textbf{Historical Reproduction.}
Historical reproduction applies Repository Evolution Knowledge when systems need to recreate a specific issue state or preserve the relationship among an issue, its patch, and its expected test outcomes. This concern is especially important for benchmark construction and training data curation, where fail-to-pass behavior must be verified. SWE-Factory uses issue instances, gold patches, and test outcomes to automate fail-to-pass validation \cite{guo2025swe}. Skywork-SWE scales runtime-validated trajectories from real repositories for data curation and model training \cite{zeng2025skywork}. RPG-Encoder incrementally updates repository representations from commit diffs to maintain consistency with repository evolution \cite{luo2026closing}. These studies show that historical knowledge helps environment construction remain faithful to the correct repository version and issue-fix context.

\textbf{Runtime Compatibility.}
Runtime compatibility applies Technology Stack Knowledge to handle dependencies, package versions, language runtimes, framework conventions, and platform assumptions. SWE-Gym demonstrates this role by constructing task instances with executable runtime environments, software dependencies, and reproducible unit tests for training SWE agents and verifiers \cite{pan2024training}. This concern shows that a repository-grounded environment is not sufficient by itself; the broader technical ecosystem must also be correctly configured for later execution and validation.

Overall, Environment Building transforms heterogeneous knowledge into an executable setting. Development tools make the environment operable, repository artifacts make it task-specific, historical evidence makes it reproducible, and technology-stack knowledge makes it compatible with the required runtime. This stage therefore provides the infrastructure on which later reproduction, localization, patch generation, and validation depend.

\subsubsection{Model Training}

At the Model Training stage, knowledge is applied before online issue solving to improve models, retrievers, verifiers, or agent policies. Instead of using knowledge as temporary inference-time context, this stage converts repeated patterns in repositories, histories, trajectories, feedback, and constraints into reusable capabilities. Existing studies mainly apply knowledge for process behavior learning, repository understanding and editing training, historical supervision, tool-use policy learning, language generalization, and constraint-aware training.

\textbf{Process Behavior Learning.}
Process behavior learning applies Experiential Knowledge to teach models how issue resolution unfolds as a multi-step process. This includes repository exploration, action selection, feedback interpretation, error recovery, and iterative refinement \cite{chen2025locagent,copet2025cwm,song2026swe,aggarwal2025dars}. CWM is representative: it trains on observation-action trajectories from Python execution and agentic Docker environments, allowing the model to learn how actions affect program states and how agents interact with computational environments \cite{copet2025cwm}. These studies show that experience can be transformed from isolated trajectories into reusable behavioral patterns for future agents.

\textbf{Repository Understanding and Editing Training.}
Repository understanding and editing training applies Existing Code Knowledge to help models connect natural-language issues with concrete repository artifacts. This objective involves retrieval-oriented code knowledge, repository-structure knowledge, and edit-oriented code knowledge \cite{liu2024codexembed,ma2024lingma,ma2024repository,ma2025sorft}. CodeXEmbed focuses on multilingual and multi-task code retrieval, while SoRFT decomposes issue resolution into localization and code-editing subtasks \cite{liu2024codexembed,ma2025sorft}. These studies show that code knowledge can be converted into both repository navigation ability and concrete editing ability, which are central to repository-level issue resolution.

\textbf{Historical Supervision.}
Historical supervision applies Repository Evolution Knowledge to train models from prior issue-fix histories, including issue descriptions, pull requests, commits, changed files, and validation outcomes. SWE-Rank and SWE-Rank+ construct issue-code ranking data from public repositories to train retrievers and rerankers for localization, with SWE-Rank+ further extending this setting to multilingual and multi-turn localization \cite{reddy2025swerank,reddy2025swerank+}. Other studies similarly use historical issue-fix signals to support localization and repair training \cite{sohrabizadeh2025nemotron,soni2026swe}. This line of work shows that historical repository evolution provides supervision about how real issues were previously localized, modified, and validated.

\textbf{Tool-Use Policy Learning.}
Tool-use policy learning applies Development Tool Knowledge to train models as interactive software engineering agents rather than static code generators. This objective covers tool invocation, command execution, environment feedback, verifier signals, and workflow-level decision making \cite{copet2025cwm,da2025agent,zhang2025sealign}. Agent-RLVR uses guidance and environment rewards to make reinforcement learning effective in sparse, multi-step agentic environments, while SEAlign aligns models with real-world software engineering workflows through high-quality workflow steps and preference optimization over critical actions \cite{da2025agent,zhang2025sealign}. These studies indicate that model training increasingly targets not only what code to generate, but also how agents should act, observe, revise, and stop.

\textbf{Language Generalization.}
Language generalization applies Programming Language Knowledge when training targets multilingual or language-specific issue-resolution ability. SWE-Rank+ explicitly trains multilingual issue-localization models \cite{reddy2025swerank+}. This concern shows that issue-resolution models need to learn language-dependent code semantics, naming conventions, type constraints, and structural differences, because valid patches must satisfy both repository-level requirements and language-specific correctness rules.

\textbf{Constraint-Aware Training.}
Constraint-aware training applies Design Architecture Knowledge when models or verifiers need to capture repository-level intent, interface constraints, project conventions, and design expectations. Agentic Rubrics is representative: an expert agent explores the repository and generates context-grounded criteria for file changes, specification alignment, integrity, and runtime behavior, turning architectural and codebase-specific constraints into scalable verifier signals \cite{raghavendra2026agentic}. This direction shows that training can go beyond issue-code mapping and tool-use behavior by internalizing constraints that help models judge whether patches are consistent with broader repository expectations.

Overall, Model Training provides a long-term pathway for knowledge internalization. It converts experience into agent behavior, code knowledge into repository understanding and editing ability, historical evidence into supervision, tool knowledge into interactive policies, language knowledge into generalization ability, and architectural knowledge into constraint-aware verification. This stage therefore complements inference-time retrieval and feedback by making selected knowledge patterns persistent in models, retrievers, verifiers, or agent policies.

\subsubsection{Task Planning}

At the Task Planning stage, knowledge is applied to organize the resolution process before or during concrete execution. Existing studies use knowledge to decompose issues, decide exploration order, allocate agent roles, select repair strategies, and determine how tools should be coordinated. The major concerns include experience-guided planning, repository-aware planning, architecture-aware planning, tool-oriented workflow planning, history-guided strategy selection, and background-aware planning.

\textbf{Experience-Guided Planning.}
Experience-guided planning applies Experiential Knowledge to guide task decomposition, reasoning order, and strategy selection. Effective plans often rely on reusable problem-solving strategies, prior reasoning traces, user-interaction histories, or empirical expectations about agent behavior \cite{ouyang2025reasoningbank,phan2024hyperagent,pimpale2025forecasting,zhou2025tom}. ReasoningBank distills generalizable reasoning strategies from successful and failed experiences and retrieves them to guide future agent interactions, while HyperAgent emulates human developer workflows through specialized agents for planning, navigation, editing, and execution \cite{ouyang2025reasoningbank,phan2024hyperagent}. These studies show that experience helps agents avoid treating each issue as an isolated task.

\textbf{Repository-Aware Planning.}
Repository-aware planning applies Existing Code Knowledge to ground plans in concrete repository states, including files, modules, APIs, call relations, and repository summaries \cite{ding2026swe,li2025swe,ma2024lingma,ma2025alibaba}. Alibaba LingmaAgent condenses repository information into a knowledge graph and guides agents to summarize, analyze, and plan with repository-level knowledge before generating patches \cite{ma2025alibaba}. This shows that planning becomes more effective when it is grounded in where relevant functionality is implemented and how repository components are organized.

\textbf{Architecture-Aware Planning.}
Architecture-aware planning applies Design Architecture Knowledge when tasks require reasoning about module responsibilities, system-level intent, architectural dependencies, or design constraints \cite{chen2026beyondswe,li2025swe,luo2026closing,zhou2025tom}. RPG-Encoder provides a semantic-structural repository representation that links high-level intent with low-level dependencies, enabling structure-aware navigation and planning across complex repositories \cite{luo2026closing}. This concern is especially important for multi-file and cross-module issues, where planning must consider not only what code to inspect, but also how changes should preserve system organization.

\textbf{Tool-Oriented Workflow Planning.}
Tool-oriented workflow planning applies Development Tool Knowledge to turn high-level plans into executable agent actions. Agents need to decide when to search, edit, run commands, execute tests, branch from prior trajectories, or intervene in inefficient behaviors \cite{gandhi2025agents,sonwane2025bugpilot,xia2025live}. SWE-PRM provides inference-time feedback to correct trajectory-level inefficiencies, while Live-SWE-agent continuously evolves its own scaffold during runtime \cite{gandhi2025agents,xia2025live}. These studies show that planning is increasingly connected with tool control and workflow adaptation rather than being limited to static task decomposition.

\textbf{History-Guided Strategy Selection.}
History-guided strategy selection applies Repository Evolution Knowledge when agents rely on historical issue responses, repository changes, or prior workflow decisions to select a suitable repair strategy \cite{cao2025siadafix,tao2024magis,wang2025seamlessflow}. SIADAFIX classifies and optimizes issue descriptions, then adaptively selects easy, middle, or hard repair modes to guide bug-fix workflows \cite{cao2025siadafix}. This shows that historical knowledge can support planning not only by providing examples of previous fixes, but also by helping systems choose an appropriate resolution mode.

\textbf{Background-Aware Planning.}
Background-aware planning applies Domain-Specific Knowledge and Technology Stack Knowledge when a task depends on domain semantics, external libraries, dependency migration, or framework-specific constraints. BeyondSWE highlights this need through settings such as DomainFix and DepMigrate, where planning must incorporate specialized domain assumptions or technology-stack constraints beyond the local repository \cite{chen2026beyondswe}. This concern shows that some tasks cannot be planned from repository structure alone and require external semantic or ecosystem-level knowledge.

Overall, Task Planning functions as the organizational layer of automated issue resolution. It connects issue understanding with repository evidence, prior experience, tool workflows, historical patterns, architectural intent, and background constraints, thereby shaping how later localization, patch generation, and validation should proceed.

\subsubsection{Reproduction Test Generation}

At the Reproduction Test Generation stage, knowledge is applied to transform informal issue descriptions into executable failure signals. Existing systems use knowledge to generate, execute, and refine tests or scripts that reproduce the reported issue before or during patch generation. The major concerns include test workflow construction, repository-compatible test context, experience-guided test refinement, fail-to-pass behavior recovery, specification-aware reproduction, and runtime-compatible test execution.

\textbf{Test Workflow Construction.}
Test workflow construction applies Development Tool Knowledge to make reproduction test generation executable and feedback-driven. Agents need to invoke test frameworks, execute scripts, inspect runtime errors, analyze logs, and iteratively repair generated tests based on environmental feedback \cite{ahmed2025otter,gao2025trae,liu2024marscode,liu2026dynamic,ma2025thinking,mu2025experepair,nashid2025issue2test}. Otter uses rule-based analysis and a self-reflective action planner to decide which code to read and which tests to write, while Issue2Test iteratively refines generated tests according to compilation and runtime feedback until the failure aligns with the reported issue \cite{ahmed2025otter,nashid2025issue2test}. These studies show that reproduction testing is increasingly treated as an executable feedback loop rather than static test synthesis.

\textbf{Repository-Compatible Test Context.}
Repository-compatible test context applies Existing Code Knowledge to ensure that generated tests match repository APIs, test structures, configuration files, and execution conventions \cite{ahmed2025otter,nashid2025issue2test,wang2025aegis,wang2025swe,wei2025swe,xiang2026evaluating}. AEGIS condenses retrieved bug-related information into structured context and uses a finite-state-machine-guided process to modify reproduction scripts, reducing irrelevant context and repeated edits \cite{wang2025aegis}. RustForger further shows that reproduction tests in Rust repositories must respect compilation requirements, dynamic tracing, and strict type semantics \cite{xiang2026evaluating}. These studies indicate that reproduction tests must be grounded in the repository's actual implementation and testing conventions.

\textbf{Experience-Guided Test Refinement.}
Experience-guided test refinement applies Experiential Knowledge to reuse prior testing strategies, repair experiences, and developer-like workflows \cite{mu2025experepair,phan2024hyperagent,wang2025aegis}. EXPEREPAIR organizes historical repair experience into episodic and semantic memory, while HyperAgent integrates testing and execution with planning, navigation, and editing \cite{mu2025experepair,phan2024hyperagent}. This concern is important because useful reproduction tests often require agents to decide what behavior should be exposed, how incomplete tests should be refined, and how ineffective test-generation loops can be avoided.

\textbf{Fail-to-Pass Behavior Recovery.}
Fail-to-pass behavior recovery applies Repository Evolution Knowledge when reproduction tests are derived from historical issue-fix pairs, pull requests, regression tests, or repository evolution records \cite{wang2025swe,yu2025utboost}. UTBoost augments SWE-bench evaluation by generating additional issue-related tests to expose patches that pass the original tests but do not truly resolve the issue \cite{yu2025utboost}. This shows that repository history can provide temporal evidence about the expected behavioral change associated with an issue.

\textbf{Specification-Aware Reproduction.}
Specification-aware reproduction applies Design Architecture Knowledge when failures depend on project-level specifications, UI behavior, or high-level design intent rather than isolated function outputs \cite{huang2025seeing,nashid2025issue2test}. Seeing is Fixing illustrates this need in visual software issues, where reproduced code and GUI renderings connect visual symptoms with executable contexts and patch validation \cite{huang2025seeing}. This concern shows that reproduction should align with intended system behavior, interface expectations, or user-visible symptoms.

\textbf{Runtime-Compatible Test Execution.}
Runtime-compatible test execution applies Technology Stack Knowledge when reproduction depends on language ecosystems, package versions, test frameworks, or executable environments. R2E-Gym constructs executable task environments with unit tests and natural-language task descriptions, showing that reliable reproduction and verification require compatible runtime and dependency settings \cite{jain2025r2e}. This concern ensures that generated tests can actually run under the required technical conditions.

Overall, Reproduction Test Generation bridges natural-language issue understanding and executable evidence. It uses tool knowledge for executable feedback loops, code knowledge for repository compatibility, experience for test refinement, history for fail-to-pass expectations, architecture knowledge for high-level correctness, and technology-stack knowledge for runtime compatibility.

\subsubsection{Localization}

At the Localization stage, knowledge is applied to map issue descriptions to concrete repository locations. Existing studies use knowledge to reduce the search space, bridge the semantic gap between natural-language reports and code entities, and prioritize files, functions, classes, or regions that should be inspected or modified. The major concerns include code-entity search, interactive localization exploration, experience-guided search control, history-aware location prioritization, architecture-aware localization, and constraint-aware localization.

\textbf{Code-Entity Search.}
Code-entity search applies Existing Code Knowledge to identify files, functions, classes, code regions, file hierarchies, function definitions, call relations, and implementation details that may be relevant to an issue \cite{chen2025locagent,gupta2025sacl,jiang2025cosil}. LocAgent parses repositories into heterogeneous graphs containing code structures and dependencies to support multi-hop search over files, classes, and functions \cite{chen2025locagent}. CoSIL improves function-level localization by dynamically constructing and traversing repository call graphs during LLM-driven search \cite{jiang2025cosil}. These studies show that code knowledge provides the direct repository-grounded basis for narrowing large repositories into likely fault or edit locations.

\textbf{Interactive Localization Exploration.}
Interactive localization exploration applies Development Tool Knowledge to make localization iterative and evidence-driven. Agents can invoke search tools, language-server interfaces, graph traversal utilities, dynamic analysis, debugging tools, and test-time exploration mechanisms to refine suspicious locations \cite{gao2025trae,liu2025graphlocator,xiang2026empowering,xiang2026evaluating}. GraphLocator uses graph-guided causal reasoning to discover symptom locations and expand causal issue graphs, while dynamic-analysis-based approaches provide runtime evidence such as traces, call stacks, and state information to refine suspicious locations \cite{liu2025graphlocator,xiang2026empowering}. This shows that localization can be updated as new tool and runtime evidence is obtained.

\textbf{Experience-Guided Search Control.}
Experience-guided search control applies Experiential Knowledge to guide localization strategies, manage long-horizon search, avoid repeated ineffective exploration, and reuse successful search behaviors \cite{xiao2025improving,chen2025swe,ding2026swe,kon2026swe}. SWE-Replay recycles prior trajectories and branches from critical intermediate steps, while SWE-Prot\'eg\'e learns when to seek expert guidance during difficult software engineering tasks \cite{ding2026swe,kon2026swe}. These studies show that experience helps agents make localization more strategy-aware rather than exhaustive or purely similarity-driven.

\textbf{History-Aware Location Prioritization.}
History-aware location prioritization applies Repository Evolution Knowledge to prioritize likely fault locations based on past commits, linked issues, pull requests, evolving modules, and historical fix patterns \cite{wang2025improving,wang2025mcts,yang2025kgcompass}. Repository Memory augments localization agents with commit-history-based memory about actively evolving modules and associations between bug types and likely fix locations \cite{wang2025improving}. KGCompass links issues, pull requests, and code entities in a repository-aware knowledge graph to narrow the search space through historical and structural paths \cite{yang2025kgcompass}. This concern shows that temporal evidence can help identify where similar problems were previously fixed.

\textbf{Architecture-Aware Localization.}
Architecture-aware localization applies Design Architecture Knowledge when identifying relevant locations requires understanding system intent, component responsibilities, architectural dependencies, or multimodal design constraints \cite{huang2025seeing,luo2026closing,yang2025lingxi}. RPG-Encoder combines semantic features with dependency topology into a unified representation for structure-aware navigation, while Lingxi uses procedural knowledge about design patterns and coding practices to guide analysis and planning before fixing \cite{luo2026closing,yang2025lingxi}. This concern is important when the correct location is not the most textually similar fragment, but the component whose responsibility best matches the issue.

\textbf{Constraint-Aware Localization.}
Constraint-aware localization applies Programming Language Knowledge, Domain-Specific Knowledge, and Technology Stack Knowledge when localization depends on language semantics, domain assumptions, external APIs, framework behavior, dependency migration, or cross-repository context. RustForger shows that Rust issue resolution requires agents to account for strict type and trait semantics during repository-level debugging and localization \cite{xiang2026evaluating}. BeyondSWE highlights cases where relevant locations depend on domain assumptions, external APIs, dependency migration, or cross-repository context \cite{chen2026beyondswe}. This concern shows that some localization tasks require broader constraints beyond immediate repository structure.

Overall, Localization is the stage where knowledge narrows vague issue descriptions into actionable repository locations. It combines code search, tool-mediated exploration, experiential search control, historical prioritization, architectural reasoning, and background constraints to determine where later patches should be applied.

\subsubsection{Patch Generation}

At the Patch Generation stage, knowledge is applied to synthesize candidate code changes after relevant contexts or repair plans have been identified. Existing studies use knowledge not only to produce code, but also to constrain generated patches so that they are locally correct, repository-consistent, behavior-preserving, and aligned with issue semantics. The major concerns include repository-grounded editing, interactive patch refinement, experience-guided repair, history-aware patch synthesis, architecture-constrained generation, and constraint-compliant code generation.

\textbf{Repository-Grounded Editing.}
Repository-grounded editing applies Existing Code Knowledge to synthesize fixes against concrete implementation contexts, including localized snippets, surrounding functions, cross-file references, coding idioms, and repository-specific APIs \cite{golubev2025training,gupta2025sacl,li2025patchpilot,liu2024codexgraph,liu2024marscode}. PatchPilot organizes patch generation within a reproduction, localization, generation, validation, and refinement workflow, showing that concrete code context must be coupled with downstream refinement to produce stable fixes \cite{li2025patchpilot}. CodexGraph illustrates how graph-based repository interfaces help agents retrieve structure-aware code contexts for repository-level code generation and repair \cite{liu2024codexgraph}. These studies show that patch generation must be grounded in the repository rather than treated as isolated code completion.

\textbf{Interactive Patch Refinement.}
Interactive patch refinement applies Development Tool Knowledge to turn code synthesis into an iterative process of editing, executing, observing, and revising. Agents can edit files, invoke tools, run commands, inspect feedback, and refine patches across multiple attempts \cite{li2025patchpilot,mu2025experepair,ouyang2025reasoningbank,pan2024training}. SWE-Gym provides executable runtime environments, tests, and agent trajectories that enable agents and verifiers to learn from realistic software engineering interactions \cite{pan2024training}. This concern shows that tools increasingly shape patch generation itself rather than only evaluating patches after generation.

\textbf{Experience-Guided Repair.}
Experience-guided repair applies Experiential Knowledge to guide patch generation through successful repair demonstrations, failed attempts, reflective insights, and developer-like workflows \cite{mu2025experepair,ouyang2025reasoningbank,phan2024hyperagent,su2025learn}. EXPEREPAIR retrieves episodic demonstrations and semantic repair insights for context-aware patch generation \cite{mu2025experepair}. ReasoningBank distills generalizable reasoning strategies from successful and failed experiences to guide future agent behavior \cite{ouyang2025reasoningbank}. These studies show that repair decisions are often procedural and strategic, involving choices about how much to modify, which alternatives to avoid, and how to recover from unsuccessful attempts.

\textbf{History-Aware Patch Synthesis.}
History-aware patch synthesis applies Repository Evolution Knowledge to ground candidate fixes in historical issue-patch relations, pull requests, co-changed files, accepted modifications, and repository-aware paths \cite{li2025patchpilot,wang2025mcts,wei2025toward,xie2026arkeval,yang2025kgcompass}. KGCompass links issues, pull requests, and code entities in a repository-aware knowledge graph and uses mined entity paths to provide contextual information for generating precise patches and explanations \cite{yang2025kgcompass}. This concern shows that repository history provides temporal grounding for patch synthesis by revealing how similar problems were fixed before.

\textbf{Architecture-Constrained Generation.}
Architecture-constrained generation applies Design Architecture Knowledge to ensure that generated patches follow system intent, module responsibilities, design patterns, coding practices, and user-oriented requirements \cite{chen2026beyondswe,yang2025lingxi,zhou2025tom}. Lingxi highlights the importance of design patterns and coding practices extracted from historical issue-fixing data \cite{yang2025lingxi}. ToM-SWE suggests that patch generation may need to account for user goals, constraints, and preferences when issue descriptions are ambiguous or stateful \cite{zhou2025tom}. This concern is especially important when a locally plausible patch may violate responsibility boundaries or intended behavior.

\textbf{Constraint-Compliant Code Generation.}
Constraint-compliant code generation applies Programming Language Knowledge, Technology Stack Knowledge, and Domain-Specific Knowledge to ensure that patches obey language-level rules, ecosystem constraints, and specialized semantic requirements. RustForger shows that Rust issue resolution is constrained by ownership, type, and trait semantics \cite{xiang2026evaluating}, while ArkEval evaluates and repairs ArkTS code under the conventions and constraints of the HarmonyOS ecosystem \cite{xie2026arkeval}. BeyondSWE further illustrates dependency-driven migration and domain-specific issue resolution, where correct patches require knowledge beyond the immediate codebase \cite{chen2026beyondswe}. This concern ensures that patches are valid under broader technical and semantic constraints.

Overall, Patch Generation is where knowledge is transformed into concrete code changes. Repository context provides editable grounding, tools enable iterative refinement, experience guides repair strategy, history supplies temporal evidence, architecture constrains system-level consistency, and background knowledge ensures language, ecosystem, and domain compliance.

\subsubsection{Patch Validation}

At the Patch Validation stage, knowledge is applied to determine whether candidate patches correctly address the reported issue without introducing regressions or violating repository-level constraints. Existing studies use knowledge to distinguish correct, incomplete, overfitted, or unsafe patches through execution results, contextual checking, candidate comparison, and refinement signals. The major concerns include execution-based verification, search-based candidate evaluation, repository-grounded validation, experience-guided verification, architecture-aware checking, and history-informed validation.

\textbf{Execution-Based Verification.}
Execution-based verification applies Development Tool Knowledge to validate patches through executable feedback, test generation, test running, sandbox interaction, candidate selection, and iterative refinement \cite{aggarwal2025dars,ahmed2025otter,antoniades2024swe,arora2024masai}. Otter generates issue-reproducing tests that can be used to validate whether a patch truly resolves the issue \cite{ahmed2025otter}. SWE-Search and DARS use search or re-sampling strategies to revisit suboptimal decisions and select more promising solutions based on feedback \cite{antoniades2024swe,aggarwal2025dars}. These studies show that execution-based evidence is central to distinguishing plausible patches from actually valid fixes.

\textbf{Search-Based Candidate Evaluation.}
Search-based candidate evaluation applies Development Tool Knowledge to compare multiple candidate patches under controlled exploration strategies \cite{ahmed2025otter,antoniades2024swe,aggarwal2025dars}. This concern is important when several candidate patches appear plausible. Runtime feedback, verifier signals, or internal scoring mechanisms can help systems select among alternatives rather than relying on a single generation path.

\textbf{Repository-Grounded Validation.}
Repository-grounded validation applies Existing Code Knowledge to anchor correctness checks in concrete repository behavior, including modified files, related APIs, test suites, call relations, and component interactions \cite{ehrlich2025codemonkeys,gao2025trae,raghavendra2026agentic,shum2025swe,tang2025co}. CodeMonkeys jointly generates candidate edits and testing scripts, then selects among multiple candidate edits using model-generated tests and a dedicated selection trajectory \cite{ehrlich2025codemonkeys}. This concern ensures that validation considers both local code changes and repository-wide dependencies.

\textbf{Experience-Guided Verification.}
Experience-guided verification applies Experiential Knowledge to reuse prior verification trajectories, failure patterns, and learned preferences about reliable candidate solutions \cite{phan2024hyperagent,song2026swe,xiao2025improving,yu2025building}. HyperAgent integrates execution and verification into the full developer-like workflow, while SWE-Master incorporates real execution feedback and test-time scaling to improve agent performance through feedback-aware post-training and inference \cite{phan2024hyperagent,song2026swe}. These studies show that validation can benefit from accumulated experience about how patches fail, how feedback should be interpreted, and which refinement strategies are more reliable.

\textbf{Architecture-Aware Checking.}
Architecture-aware checking applies Design Architecture Knowledge when validation must preserve high-level design intent, module responsibilities, interface contracts, architectural dependencies, and project-specific correctness criteria \cite{cao2025siadafix,liu2026architecture,raghavendra2026agentic}. Agentic Rubrics generates context-grounded checklist criteria covering file changes, specification alignment, integrity, and runtime behavior, thereby turning architectural and repository-specific expectations into scalable verification signals \cite{raghavendra2026agentic}. SIADAFIX also reflects this role by combining workflow decisions and iterative repair to judge repair effectiveness and avoid error accumulation \cite{cao2025siadafix}. This concern shows that validation should consider more than test pass rates when project-level constraints are involved.

\textbf{History-Informed Validation.}
History-informed validation applies Repository Evolution Knowledge to align candidate patches with historical issue-fix behavior, gold patches, pull requests, and regression expectations \cite{ahmed2025otter,wang2025mcts}. MCTS-Refined CoT uses strict intermediate validation against verified developer patches to improve the quality of reasoning paths for issue resolution \cite{wang2025mcts}. This concern shows that historical evidence can help assess whether a candidate patch matches expected fail-to-pass behavior and verified developer repair patterns.

Overall, Patch Validation is the stage where knowledge supports correctness judgment. Tool feedback provides executable evidence, repository knowledge grounds checks in project behavior, experience improves verification strategy, architecture knowledge supports high-level consistency checking, and history provides reference points for expected repair behavior. This stage closes the loop of automated issue resolution by determining whether generated patches are acceptable under behavioral, repository-level, and historical constraints.

\subsection{Knowledge-Type Distribution across Usage Stages}

\begin{figure}[htbp]
    \centering
    \includegraphics[width=0.9\linewidth]{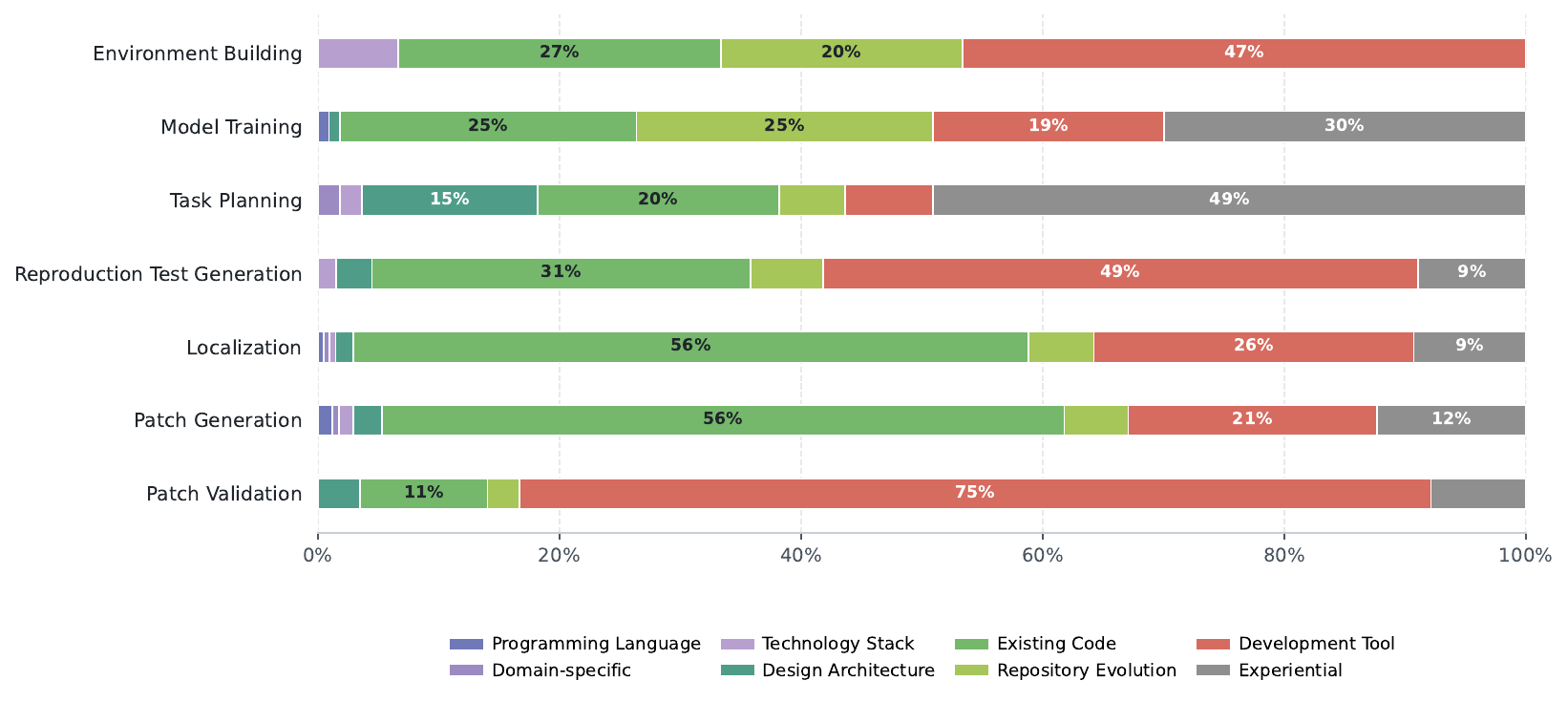}
    \caption{Distribution of knowledge usage across stages.}
    \label{fig:usage_stage_bar}
\end{figure}

Fig.~\ref{fig:usage_stage_bar} further quantifies how different knowledge types are distributed across the seven usage stages of automated issue resolution. The results show a clear stage-specific pattern: different stages do not simply consume more or less knowledge, but rely on different knowledge compositions according to their functional roles. In the Environment Building stage, Development Tool Knowledge accounts for the largest share at 47\%, followed by Existing Code Knowledge and Repository Evolution Knowledge at 27\% and 20\%, respectively. This distribution suggests that, in the reviewed studies, knowledge used for environment construction is strongly associated with tool operation, executable settings, and repository-specific configuration. Before agents can validate patches or interact with a repository in a realistic manner, systems often need to prepare package managers, build systems, test runners, containers, sandboxes, execution interfaces, and related configuration files. Repository-specific code and historical issue states may further help reconstruct faithful execution contexts. Therefore, rather than proving that environment construction is necessarily the first bottleneck, the observed distribution indicates that it is an important preparation stage where tool-related and execution-oriented knowledge is frequently required. Recent efforts on executable environments and scalable sandboxing also reflect the practical importance of supporting this stage for training, evaluation, and online issue solving \cite{guo2025swe,yuan2026swe}.

Model Training and Task Planning exhibit more experience-oriented knowledge usage, but they serve different purposes. In Model Training, Experiential Knowledge accounts for the largest proportion at 30\%, followed by Existing Code Knowledge and Repository Evolution Knowledge, each accounting for 25\%, and Development Tool Knowledge at 19\%. This suggests that training-based methods mainly transform prior trajectories, code-understanding patterns, historical issue-fix data, and tool-interaction behaviors into reusable model capabilities. However, because model training requires large-scale data, executable environments, and substantial infrastructure, it is likely to remain a high-cost optimization path \cite{copet2025cwm,da2025agent,zhang2025sealign}. In contrast, Task Planning shows an even stronger reliance on Experiential Knowledge, which accounts for 49\%, followed by Existing Code Knowledge and Design Architecture Knowledge at 20\% and 15\%, respectively. This indicates that planning is less about directly editing code and more about organizing the resolution process: agents need reusable reasoning strategies, workflow patterns, repository overviews, and architectural constraints to decide what to inspect, how to decompose the task, and which actions should be performed next. Therefore, planning may become increasingly important as issue resolution shifts from fixed pipelines to autonomous agent workflows \cite{ouyang2025reasoningbank,ma2025alibaba,luo2026closing}.

Reproduction Test Generation, Localization, and Patch Generation show how mainstream research is currently concentrated around executable feedback and code-centered repair. In Reproduction Test Generation, Development Tool Knowledge accounts for the largest share at 49\%, while Existing Code Knowledge and Experiential Knowledge account for 31\% and 9\%, respectively. This reflects the fact that requires agents to generate tests, execute scripts, inspect runtime failures, and iteratively refine test cases based on feedback \cite{ahmed2025otter,nashid2025issue2test,yu2025utboost}. Localization and Patch Generation are both dominated by Existing Code Knowledge, which accounts for 56\% in each stage. This confirms that identifying modification targets and producing candidate fixes remain the most repository-grounded stages in current research: systems must reason over concrete files, functions, classes, APIs, and implementation contexts \cite{chen2025locagent,jiang2025cosil,li2025patchpilot,liu2024codexgraph}. However, the secondary distributions also reveal an important transition from static code understanding toward dynamic and interaction-based understanding. Localization uses a substantial amount of Development Tool Knowledge at 26\%, because agents increasingly rely on search tools, graph traversal, dynamic analysis, debugging interfaces, and runtime evidence to narrow the search space \cite{liu2025graphlocator,xiang2026empowering}. Patch Generation also uses Development Tool Knowledge at 21\% and Experiential Knowledge at 12\%, indicating that candidate fixes are increasingly produced through interactive editing, execution-aware refinement, and reuse of prior repair strategies rather than isolated code completion \cite{mu2025experepair,ouyang2025reasoningbank,pan2024training}.

Patch Validation presents the most concentrated knowledge composition. Development Tool Knowledge accounts for 75\%, far exceeding all other knowledge types, while Existing Code Knowledge accounts for 11\%. This indicates that Patch Validation is currently one of the most tool-intensive and feedback-dependent stages. Candidate patches must be tested, compared, selected, or rejected through execution results, generated tests, verifier signals, trajectory selection, or other validation mechanisms. This also explains why validation has become a key optimization target: many patches may appear plausible from static code context but fail at runtime, introduce regressions, or violate repository-specific constraints \cite{ahmed2025otter,antoniades2024swe,aggarwal2025dars,raghavendra2026agentic}. Overall, the distribution reveals a workflow-level transition in knowledge usage. Early preparation and final validation are tool-intensive; model training and planning rely heavily on accumulated experience; reproduction test generation connects issue descriptions with executable failure signals; and localization and patch generation remain centered on Existing Code Knowledge but increasingly depend on dynamic feedback. More importantly, these stage-specific patterns suggest that future systems should not treat knowledge as isolated within each stage. Instead, knowledge should be stored and reused across stages: reproduction tests can support localization and validation, localization trajectories can guide patch generation, validation failures can update planning and repair strategies, and historical issue-fix patterns can inform both training and inference. Cross-project reuse is also promising, especially when the same stage involves reusable knowledge such as environment setup procedures, localization strategies, testing routines, tool-use policies, or validation criteria \cite{wang2025improving,chen2025swe,mu2025experepair}. Therefore, future automated issue resolution systems may benefit from stage-aware knowledge memory, where knowledge accumulated in one stage or project can be transferred, adapted, and reused in similar stages of future issue-resolution tasks.

\finding{
The stage-level results show that knowledge is applied unevenly across the issue-resolution workflow, forming stage-specific knowledge portfolios rather than a uniform knowledge supply. Preparation and verification stages rely strongly on executable and tool-mediated knowledge, while planning and training stages increasingly transform prior trajectories, workflow patterns, and repair strategies into reusable guidance or model capabilities. Localization and patch generation remain strongly grounded in repository knowledge, because agents must connect issue semantics with concrete code locations and generate changes that are consistent with the surrounding implementation. Reproduction test generation further acts as a bridge between issue understanding and executable evidence, converting natural-language reports into failure signals that can support later localization and validation. Overall, these patterns indicate that automated issue resolution is moving from static code-context consumption toward a workflow in which knowledge is prepared, explored, executed, verified, and reused across multiple stages.

This finding suggests that future systems should focus on stage-aware knowledge orchestration. The key challenge is not only to introduce more knowledge sources, but to make knowledge produced at one stage available to other stages in a traceable and reusable way. Reproduction tests can guide localization and validation, localization evidence can constrain patch generation, validation failures can update planning, and accumulated trajectories can improve future training or memory construction. From a broader software-agent perspective, this points to the need for cross-stage knowledge loops that connect environment setup, repository reasoning, tool feedback, patch synthesis, and verification into a coherent process. Such systems should also incorporate higher-level constraints, including architectural intent, programming-language rules, technology-stack conventions, and domain semantics, so that agents can produce fixes that are not only executable and test-passing, but also semantically appropriate, architecturally consistent, and robust in realistic software maintenance scenarios.
}

\section{RQ6: What recent trends and dominant patterns characterize knowledge usage in automated issue resolution?}
\label{sec:rq6}

Building on the previous analyses of knowledge types, acquisition sources, representation formats, usage methods, and application stages, this section synthesizes how knowledge usage in automated issue resolution has evolved and what dominant patterns are emerging in recent systems. Rather than examining each dimension in isolation, we connect them through three complementary views: temporal evolution, representative methodological configurations, and knowledge patterns among high-performing SWE-bench Verified methods. The temporal analysis captures how the adoption of different knowledge types changes over time; the representative-work analysis explains how recent systems combine multiple knowledge types in concrete method designs; and the top-performing-method analysis examines whether similar knowledge portfolios are also visible among systems with strong leaderboard results. This final view is performance-oriented but non-causal, because reported resolved rates are affected by base models, scaffolds, tool settings, inference budgets, and evaluation protocols. Therefore, the goal of this section is not to estimate the independent performance effect of each knowledge type, but to identify how knowledge usage is shifting across the field and what this shift implies for future automated issue resolution systems.

\subsection{Temporal Trends of Knowledge Type Usage}
\label{sec:knowledge_trend}

\begin{figure}
    \centering
    \includegraphics[width=0.7\linewidth]{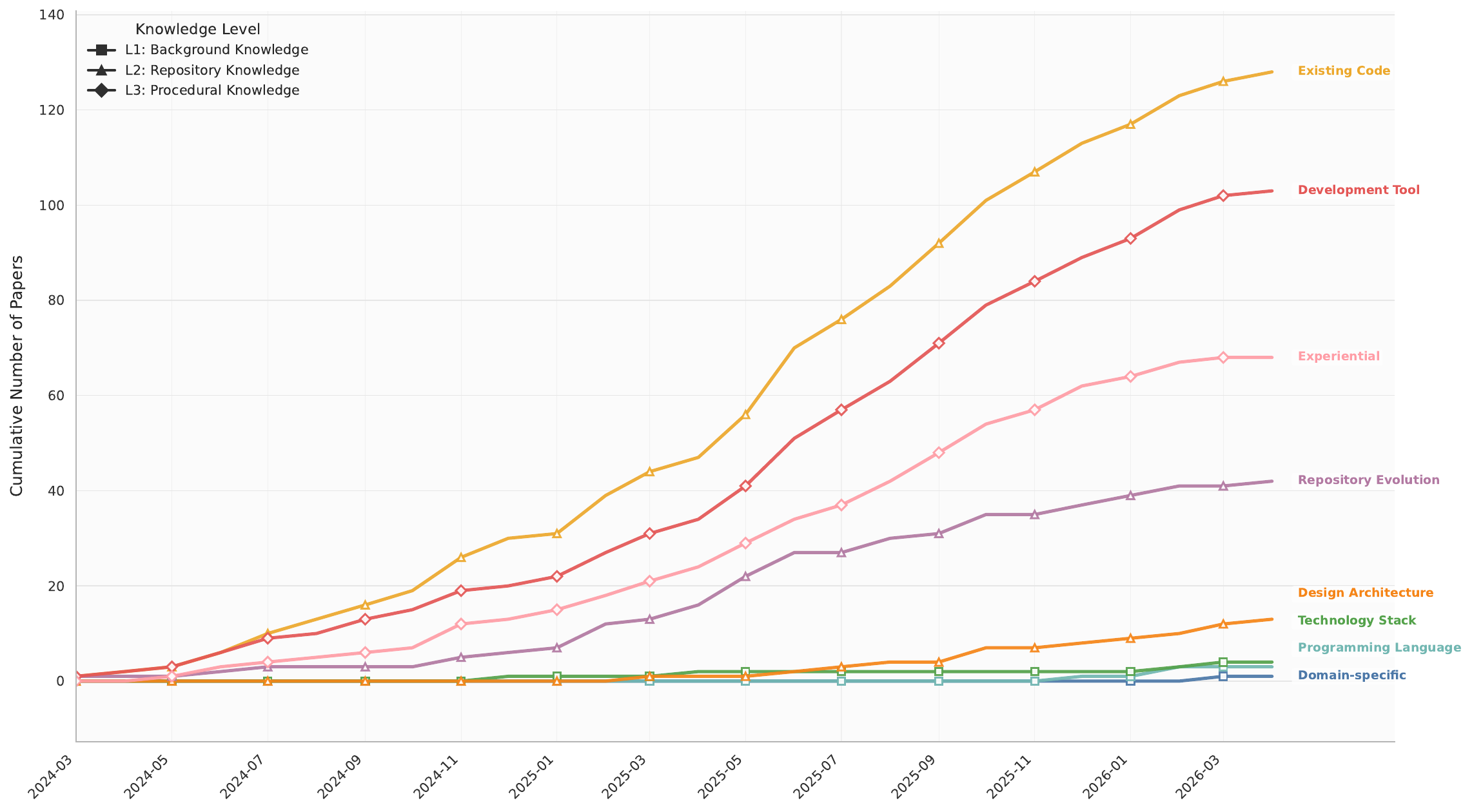}
    \caption{Cumulative adoption trends of different knowledge types over time.}
    \label{fig:knowledge_trend}
\end{figure}

\begin{figure}
    \centering
    \includegraphics[width=0.85\linewidth]{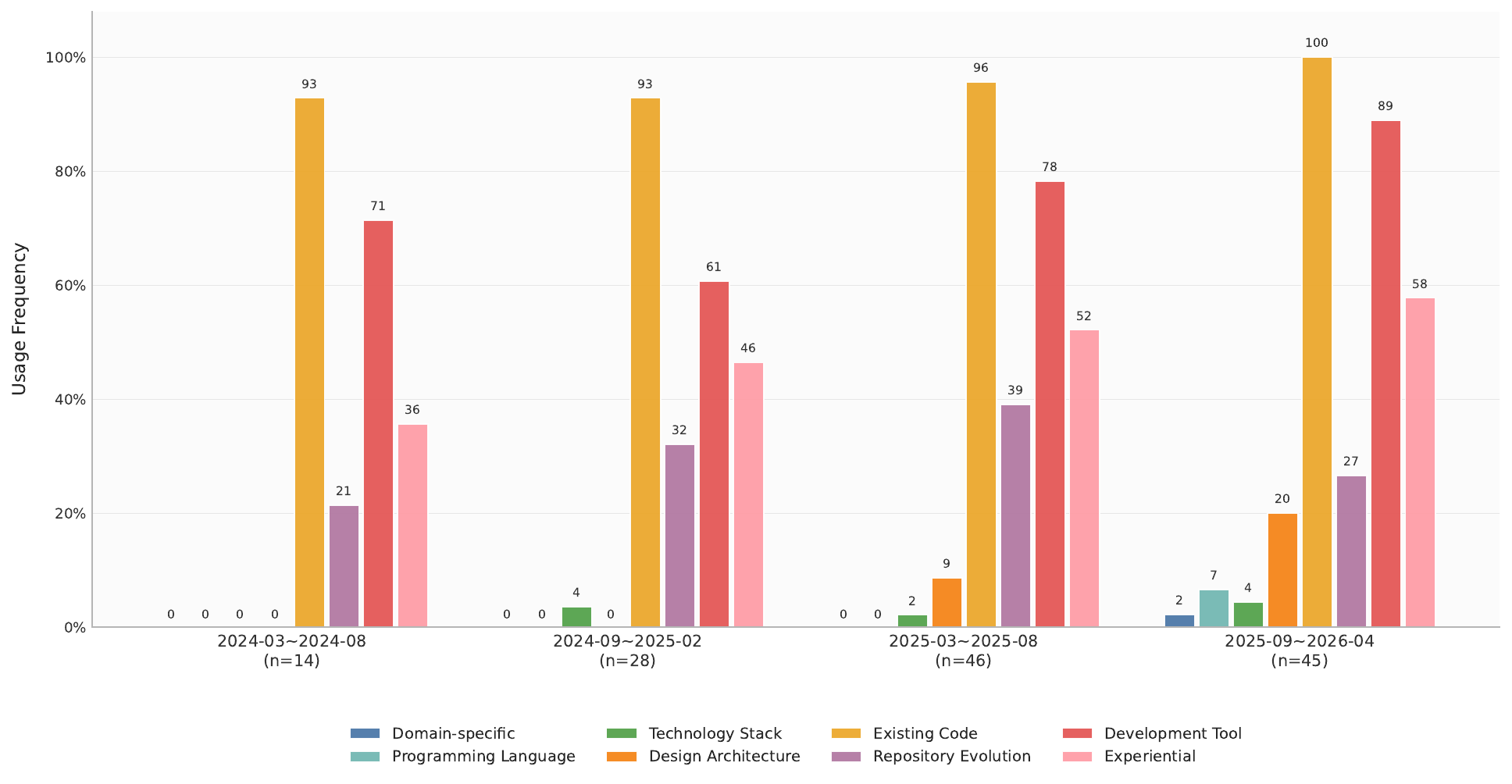}
    \caption{Usage-frequency distribution of knowledge types across consecutive half-year periods.}
    \label{fig:knowledge_frequency_period}
\end{figure}

To understand how knowledge usage has evolved in automated issue resolution, we analyze the temporal patterns of knowledge-type adoption from two complementary perspectives. Fig.~\ref{fig:knowledge_trend} presents the cumulative adoption trends of different knowledge types from March 2024 to April 2026, while Fig.~\ref{fig:knowledge_frequency_period} reports their usage frequencies across four consecutive half-year periods. The former reflects the overall growth trajectory of each knowledge type, whereas the latter reveals how research attention and methodological priorities shift across different stages of field development.

As shown in Fig.~\ref{fig:knowledge_trend}, Existing Code Knowledge grows earlier and faster than other knowledge types, reaching the highest cumulative adoption by the end of the observation period. This confirms that automated issue resolution has remained fundamentally repository-grounded: regardless of whether a system adopts a workflow-based pipeline, an agentic framework, retrieval-augmented generation, or training-based optimization, it must first understand and manipulate the concrete implementation context of the target repository. Development Tool Knowledge also shows rapid growth, especially after early 2025, reflecting the increasing role of tool invocation, execution environments, test runners, shell commands, and repository operations in acquiring task-specific observations. Meanwhile, Experiential Knowledge and Repository Evolution Knowledge steadily increase after 2025, indicating that recent studies increasingly incorporate historical fixes, prior trajectories, intermediate attempts, and feedback loops into the resolution process. These trends suggest that the field is expanding from static repository-context construction toward more interactive, feedback-driven, and process-aware knowledge usage.

Another important pattern in Fig.~\ref{fig:knowledge_trend} is the gradual emergence of higher-level knowledge types. Design Architecture Knowledge, Technology Stack Knowledge, Programming Language Knowledge, and Domain-Specific Knowledge grow more slowly than Existing Code or Development Tool Knowledge, but their recent appearance indicates that automated issue resolution is beginning to move beyond directly observable repository artifacts. This trend is meaningful because high-level knowledge can provide abstractions that are difficult to obtain from code snippets or execution feedback alone, such as architectural intent, framework constraints, language-specific rules, and domain semantics. Therefore, the cumulative trend suggests that future systems may benefit from not only retrieving more code or invoking more tools, but also transforming concrete code, historical changes, and interaction traces into higher-level knowledge that supports design-aware reasoning and cross-repository generalization.

Fig.~\ref{fig:knowledge_frequency_period} further clarifies how research focuses evolve across different periods. In the earliest period, from March 2024 to August 2024, Existing Code Knowledge already appears in 93\% of studies, and Development Tool Knowledge appears in 71\%, while Repository Evolution and Experiential Knowledge account for 21\% and 36\%, respectively. This indicates that the initial stage of automated issue resolution research mainly centered on repository understanding and basic tool-supported execution.

In the second period, from September 2024 to February 2025, Existing Code Knowledge remains high at 93\%, while Repository Evolution and Experiential Knowledge increase to 32\% and 46\%. This suggests that researchers began to enrich repository-centered methods with historical artifacts and experience-like signals, enabling systems to learn from previous changes, repair attempts, and development trajectories.

During the third period, from March 2025 to August 2025, the knowledge portfolio becomes more diversified. Existing Code Knowledge increases to 96\%, Development Tool Knowledge rises to 78\%, Experiential Knowledge reaches 52\%, and Repository Evolution Knowledge reaches 39\%. At the same time, Design Architecture Knowledge begins to appear more visibly, reaching 9\%. This period reflects the rapid development of agentic and feedback-driven issue resolution systems, where agents not only retrieve repository context but also interact with environments, run tests, interpret feedback, and reuse process-level experience.

In the latest period, from September 2025 to April 2026, Existing Code Knowledge reaches 100\%, Development Tool Knowledge increases to 89\%, and Experiential Knowledge rises to 58\%. More notably, Design Architecture Knowledge grows to 20\%, while L1 Background Knowledge also begins to appear more visibly in recent studies, with Programming Language Knowledge, Technology Stack Knowledge, and Domain-Specific Knowledge reaching 7\%, 4\%, and 2\%, respectively. Repository Evolution Knowledge in this period is 27\%. This indicates that the research focus is gradually extending from code-and-tool-centered methods toward richer knowledge compositions involving history, experience, architecture, and background constraints.

Taken together, the two figures show that automated issue resolution is evolving toward a more comprehensive knowledge-use paradigm. Repository-grounded knowledge remains the foundation, procedural knowledge is becoming increasingly central to agentic interaction and validation, and higher-level knowledge is gradually being introduced to support more abstract reasoning. This evolution points to several future research opportunities. First, since Existing Code and Development Tool Knowledge are already widely adopted, future work can further investigate how to abstract reusable design principles, technology-stack constraints, and domain rules from code, execution traces, and tool feedback. Second, the growth of Repository Evolution and Experiential Knowledge suggests opportunities for building systematic memory mechanisms that capture historical repairs, failed attempts, and successful trajectories as reusable guidance. Third, the increasing appearance of Design Architecture and other background knowledge indicates a promising direction for cross-layer knowledge integration, where agents combine low-level code evidence, process-level feedback, and high-level abstractions to support long-horizon planning, architecture-aware repair, and cross-repository transfer.

\subsection{Knowledge Type Usage in Representative Works}
\label{sec:representative_works}

\begin{figure}
    \centering
    \includegraphics[width=0.85\linewidth]{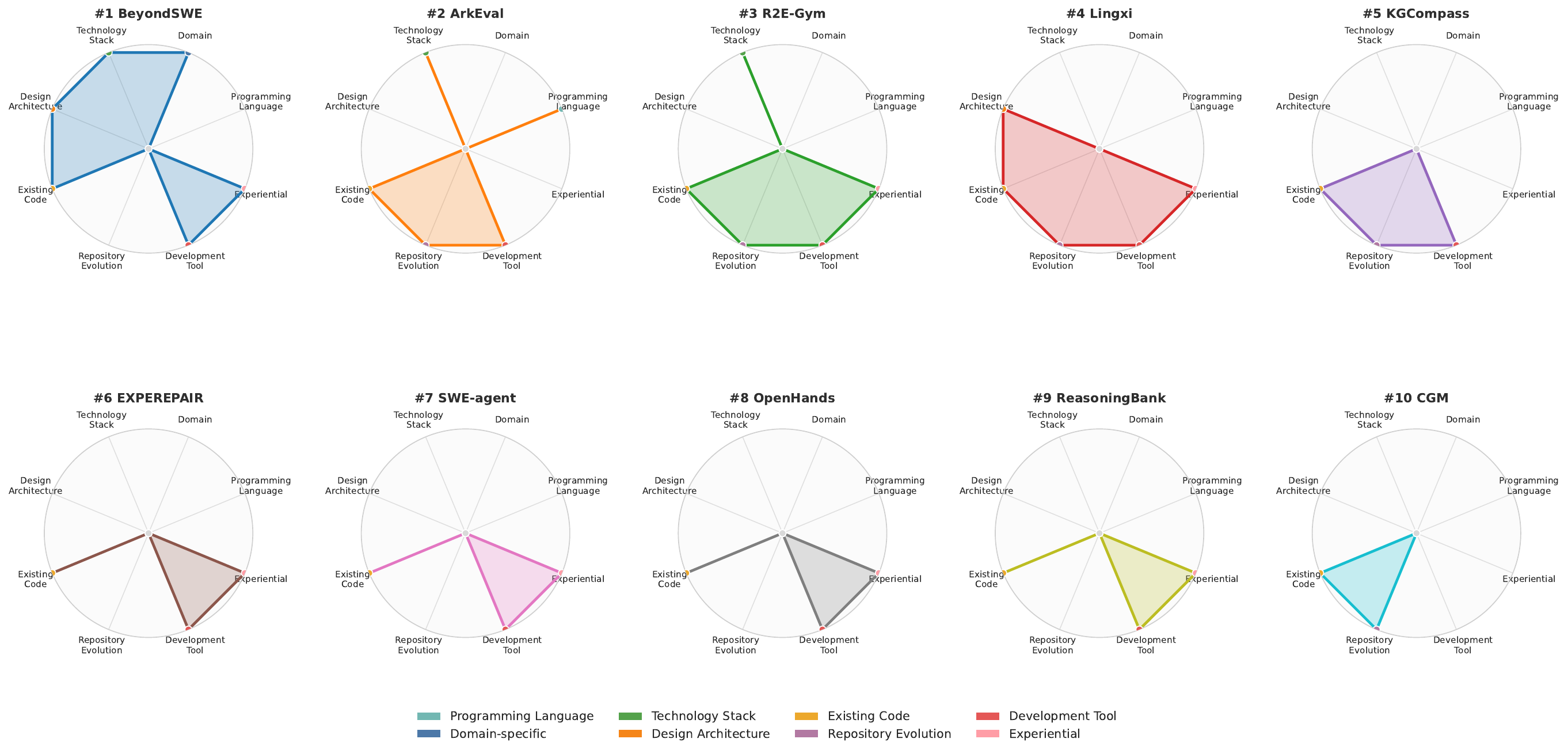}
    \caption{Knowledge-type usage profiles of ten representative automated issue resolution studies.}
    \label{fig:representative_works_radar}
\end{figure}

To complement the macro-level temporal analysis, we examine how ten representative automated issue resolution studies utilize different knowledge types, as illustrated in Fig.~\ref{fig:representative_works_radar}. These studies include BeyondSWE \cite{chen2026beyondswe}, ArkEval \cite{xie2026arkeval}, R2E-Gym \cite{jain2025r2e}, Lingxi \cite{yang2025lingxi}, KGCompass \cite{yang2025kgcompass}, EXPEREPAIR \cite{mu2025experepair}, SWE-agent \cite{yang2024swe}, OpenHands \cite{wang2024openhands}, ReasoningBank \cite{ouyang2025reasoningbank}, and CGM \cite{tao2025code}. These works are selected to cover diverse methodological directions in recent automated issue resolution research, rather than to represent the highest-performing systems according to benchmark scores. They span a range of recent directions, including benchmark construction, executable environment construction, agent-computer interfaces, procedural knowledge scaling, memory-enhanced repair, graph-guided localization, and graph-integrated models. Therefore, they provide a concrete view of how knowledge types are configured in representative systems beyond aggregate frequency statistics.

\textbf{Repository-grounded interaction.}
A common pattern is that representative systems consistently build their core workflow around repository grounding and tool-mediated interaction. SWE-agent \cite{yang2024swe} introduces an agent-computer interface that enables language-model agents to navigate repositories, view and edit files, search code, execute commands, and receive concise environment feedback. OpenHands \cite{wang2024openhands} further provides a general platform where agents interact with sandboxed environments, command lines, browsers, and code-editing tools. R2E-Gym \cite{jain2025r2e} also emphasizes executable environments and hybrid verification, showing that scalable training and test-time scaling depend on reliable execution, testing, and environment construction. In Fig.~\ref{fig:representative_works_radar}, these works show strong coverage of Existing Code and Development Tool Knowledge, indicating that concrete repository artifacts and tool feedback form the operational foundation for automated issue resolution.

\textbf{Procedural and experiential guidance.}
Another prominent pattern is the use of historical and experiential information to guide issue resolution. Lingxi \cite{yang2025lingxi} extracts procedural knowledge from historical issue-fixing data and abstracts the ``how'' and ``why'' behind previous fixes to guide knowledge-driven scaling. EXPEREPAIR \cite{mu2025experepair} organizes historical repair experiences into episodic and semantic memories, enabling the model to retrieve concrete demonstrations and recall abstract repair insights during inference. ReasoningBank \cite{ouyang2025reasoningbank} distills generalizable reasoning strategies from successful and failed agent experiences, allowing agents to reuse prior lessons in future tasks. Their radar profiles cover Experiential Knowledge and, in several cases, Repository Evolution Knowledge, showing that recent systems increasingly treat experience not as raw trajectories, but as reusable guidance for analysis, planning, localization, and patch generation.

\textbf{Structural and architectural reasoning.}
A further pattern is the use of structural representations to support repository-level reasoning. KGCompass \cite{yang2025kgcompass} constructs a repository-aware knowledge graph that links issues, pull requests, files, classes, and functions, and uses graph-mined paths to guide fault localization and patch generation. CGM \cite{tao2025code} integrates code graph structures into the model by combining semantic node representations with structural dependencies through graph-aware attention and an agentless graph RAG framework. These methods highlight the role of structural repository knowledge in organizing scattered code contexts and supporting cross-file, multi-hop reasoning. They also suggest that graph-based representations can serve as an intermediate layer between raw code retrieval and higher-level repository understanding.

\textbf{Background knowledge for expanded task settings.}
Some representative studies show how automated issue resolution is extending beyond conventional single-repository bug fixing. BeyondSWE \cite{chen2026beyondswe} broadens evaluation along both resolution scope and knowledge scope, covering cross-repository issue resolution, domain-specific issue resolution, dependency-driven migration, and document-to-repository generation. ArkEval \cite{xie2026arkeval} focuses on ArkTS repair in the HarmonyOS ecosystem, where agents must handle a low-resource language, platform-specific development constraints, and retrieval-augmented repair workflows. In Fig.~\ref{fig:representative_works_radar}, these works incorporate Domain-specific, Programming Language, and Technology Stack Knowledge more explicitly than most other representative systems. This indicates that as issue resolution expands to new ecosystems and task forms, background knowledge becomes increasingly valuable for interpreting requirements, adapting to language constraints, and handling framework or dependency changes.

\textbf{Insights and future research directions.}
The analysis of these representative works suggests that recent systems are developing several complementary knowledge-use patterns rather than converging toward a single template. Interaction-centered agents emphasize tool feedback and repository exploration; memory-enhanced systems emphasize reusable experience; graph-based systems emphasize structural repository reasoning; executable-environment works emphasize scalable training and verification; and domain- or ecosystem-specific benchmarks introduce background knowledge into issue resolution. These patterns point to several future opportunities. First, tool interaction can become more knowledge-aware, allowing agents to dynamically decide when to search, execute, localize, validate, or revise according to task uncertainty and intermediate feedback. Second, long-term memory mechanisms can be designed to store not only successful patches, but also failed attempts, decision rationales, repair strategies, and stage-specific lessons, enabling agents to transfer experience across similar issues. Third, unified repository knowledge graphs can connect code entities, dependency relations, issue histories, test behaviors, architectural modules, and developer intents, helping agents reason over both local edit locations and global system constraints. Finally, broader integration of programming-language, technology-stack, and domain-specific knowledge can support cross-repository repair, dependency migration, low-resource language repair, and complex long-horizon maintenance tasks. Overall, adaptive cross-layer knowledge integration may become a key direction for building more robust and generalizable automated issue resolution systems.

 \subsection{Knowledge Patterns among High-Performing SWE-bench Verified Methods}
\label{sec:performance_knowledge_verified}

The representative-work analysis above characterizes how different research directions configure knowledge types, but it does not directly show whether methods with strong reported SWE-bench Verified results exhibit common knowledge-use patterns. To complement this perspective, we further examine the top-20 proposed methods with the highest reported SWE-bench Verified resolved rates within our surveyed literature, based on the results reported in the included papers up to our literature collection cutoff in April 2026. Since SWE-bench Verified results and leaderboards may change over time, this analysis should be understood as a descriptive view of high-performing methods covered by our survey corpus rather than a real-time leaderboard ranking. This analysis is performance-oriented but non-causal: since the reported results are obtained with different base models, agent scaffolds, tool settings, inference budgets, and evaluation protocols, we do not attribute resolved rates to individual knowledge types. Instead, we use these high-performing surveyed methods as a descriptive lens to identify which knowledge types are commonly present among systems reporting strong SWE-bench Verified performance.

Table~\ref{tab:top20-performance-knowledge} summarizes the knowledge-type profiles of the top-20 methods. The most evident pattern is that Existing Code Knowledge and Development Tool Knowledge appear in all top-20 systems. This suggests that high-performing SWE-bench Verified methods generally share a repository-operational foundation: agents must ground issue descriptions in the target codebase and then act upon the repository through search, editing, execution, testing, and validation. This pattern is consistent with recent methods that emphasize repository exploration, tool-mediated execution, dynamic analysis, test generation, patch validation, or codebase navigation as core components of issue resolution~\cite{liu2026dynamic,li2025infcode,gao2025trae,chen2025prometheus,xiang2026empowering,sohrabizadeh2025nemotron}. However, since Existing Code Knowledge and Development Tool Knowledge are shared by all top-20 methods, they should not be interpreted as factors that distinguish performance within this group. Rather, they represent the common operational basis on which current high-performing systems are built.

Beyond this shared foundation, Experiential Knowledge is the most frequent additional knowledge type among the top-20 methods, appearing in 12 of the 20 systems. This indicates that many recent high-performing methods are no longer limited to one-shot code retrieval or tool execution at test time. Instead, they increasingly exploit process-level experience, such as previous interaction trajectories, historical fixing procedures, reward signals, post-training traces, replay mechanisms, or reasoning-transfer data~\cite{lin2025se,ding2026swe,yang2025lingxi,shum2025swe,song2026swe,copet2025cwm,tang2025boosting}. This pattern reflects a broader movement from static context use toward experience-informed problem solving, where agents reuse or internalize prior problem-solving processes to guide long-horizon reasoning. Nevertheless, this observation should be understood as a descriptive association rather than causal evidence, since these methods also differ substantially in model capability, inference-time scaling strategy, validation design, and agent architecture.

Repository Evolution Knowledge and Design Architecture Knowledge appear less frequently among the top-20 methods. This limited presence does not imply that historical or architectural knowledge is ineffective. Rather, it suggests that such higher-level knowledge has not yet been widely operationalized in current leaderboard-leading systems. Compared with source code, tool outputs, and execution traces, repository histories, issue-fix trajectories, architectural constraints, module responsibilities, and design rationales are more implicit, distributed, and difficult to extract in a form directly usable by agents. Existing attempts to exploit historical fixes, procedural knowledge, repository-level structure, or codebase graphs indicate the potential value of these knowledge types~\cite{yang2025lingxi,chen2025prometheus,copet2025cwm,sohrabizadeh2025nemotron}, but their selective appearance in the top-performing group suggests that architecture-aware and history-aware issue resolution remains an underexplored direction.

Explicit L1 Background Knowledge, including Programming Language Knowledge, Domain-Specific Knowledge, and Technology Stack Knowledge, is absent from the top-20 methods according to our coding. This observation should be interpreted in relation to the benchmark setting. SWE-bench Verified mainly evaluates general repository-level issue resolution, where performance tends to depend more directly on repository grounding, tool-mediated interaction, patch generation, and validation. In such settings, language, domain, or framework knowledge may be implicitly encoded in the base model or training data rather than explicitly represented as a separate knowledge component. Therefore, the absence of explicit L1 Background Knowledge in Table~\ref{tab:top20-performance-knowledge} should not be read as evidence that these knowledge types are unimportant. As benchmarks expand toward framework-specific bugs, low-resource languages, dependency migration, domain-specific systems, and multimodal software artifacts, explicit background knowledge may become more visible and necessary.

Taken together, the top-20 SWE-bench Verified methods reveal a layered pattern of knowledge use among high-performing issue-resolution systems. Existing Code Knowledge and Development Tool Knowledge form the shared repository-operational foundation, enabling agents to connect issue descriptions with concrete code evidence and to perform executable repair actions. On top of this foundation, Experiential Knowledge has become the most visible additional component, suggesting that recent systems increasingly rely on trajectory reuse, procedural guidance, reward-based feedback, and post-training experience. In contrast, Repository Evolution Knowledge, Design Architecture Knowledge, and explicit L1 Background Knowledge remain less widely represented, pointing to opportunities for future work to better operationalize historical, architectural, and domain-specific knowledge. Overall, this analysis suggests that current progress in automated issue resolution is shifting from static context retrieval toward dynamic, feedback-aware, and experience-oriented knowledge orchestration, while the relationship between knowledge usage and reported performance remains descriptive rather than causal.

\begin{table*}[htbp]
\centering
\small
\caption{Knowledge-type profiles of the top 20 reported methods on SWE-bench Verified.}
\label{tab:top20-performance-knowledge}
\begin{threeparttable}
\setlength{\tabcolsep}{3.0pt}
\renewcommand{\arraystretch}{1.08}

\begin{tabularx}{\textwidth}{
@{}
C{0.55cm}
@{\hspace{0.75em}}
L{2.75cm}
L{3.00cm}
C{1.05cm}
C{0.9cm}
*{8}{C{0.47cm}}
@{}}
\toprule
\multirow{2}{*}{Rank}
& \multirow{2}{*}{Method}
& \multirow{2}{*}{Base Model}
& \multirow{2}{*}{\shortstack{Resolved\\(\%)}}
& \multirow{2}{*}{Citation}
& \multicolumn{8}{c}{Knowledge Types} \\
\cmidrule(lr){6-13}
& & & & 
& PL & DK & TS & DA & EC & RE & DT & EXP \\
\midrule
1  & SE-Agent             & Claude-Sonnet-4             & 80.0 & \cite{lin2025se}                 & \xmark & \xmark & \xmark & \xmark & \cmark & \xmark & \cmark & \cmark \\
2  & DAIRA                & Gemini-3-Flash Preview      & 79.4 & \cite{liu2026dynamic}            & \xmark & \xmark & \xmark & \xmark & \cmark & \xmark & \cmark & \xmark \\
3  & InfCode              & Claude-Sonnet-4.5           & 79.4 & \cite{li2025infcode}             & \xmark & \xmark & \xmark & \xmark & \cmark & \xmark & \cmark & \xmark \\
4  & SWE-Replay           & Gemini-3-Pro                & 75.8 & \cite{ding2026swe}               & \xmark & \xmark & \xmark & \xmark & \cmark & \xmark & \cmark & \cmark \\
5  & Live-SWE-agent       & Claude-Sonnet-3.5           & 75.4 & \cite{xia2025live}               & \xmark & \xmark & \xmark & \xmark & \cmark & \xmark & \cmark & \cmark \\
\addlinespace[0.15em]
6  & Trae Agent           & \nr                         & 75.2 & \cite{gao2025trae}               & \xmark & \xmark & \xmark & \xmark & \cmark & \xmark & \cmark & \cmark \\
7  & Confucius Code Agent & Claude-Sonnet-4             & 74.6 & \cite{wong2025confucius}         & \xmark & \xmark & \xmark & \xmark & \cmark & \xmark & \cmark & \cmark \\
8  & Lingxi               & Claude-Sonnet-4             & 74.6 & \cite{yang2025lingxi}            & \xmark & \xmark & \xmark & \cmark & \cmark & \cmark & \cmark & \cmark \\
9  & SWE-RM               & Qwen3-Coder-Max             & 74.6 & \cite{shum2025swe}               & \xmark & \xmark & \xmark & \xmark & \cmark & \xmark & \cmark & \cmark \\
10 & Prometheus           & GPT-5                       & 74.4 & \cite{chen2025prometheus}        & \xmark & \xmark & \xmark & \xmark & \cmark & \xmark & \cmark & \xmark \\
\addlinespace[0.15em]
11 & SWE-Master           & Qwen-2.5-Coder-32B          & 70.8 & \cite{song2026swe}               & \xmark & \xmark & \xmark & \xmark & \cmark & \xmark & \cmark & \cmark \\
12 & Nemotron-CORTEXA     & Mix                         & 68.4 & \cite{sohrabizadeh2025nemotron}  & \xmark & \xmark & \xmark & \xmark & \cmark & \cmark & \cmark & \xmark \\
13 & Trajectory Reduction & Claude-Sonnet-4             & 66.5 & \cite{xiao2025improving}         & \xmark & \xmark & \xmark & \xmark & \cmark & \xmark & \cmark & \cmark \\
14 & Aime                 & \nr                         & 66.4 & \cite{shi2025aime}               & \xmark & \xmark & \xmark & \xmark & \cmark & \xmark & \cmark & \xmark \\
15 & CodeMonkeys          & Claude-Sonnet-3.5           & 66.2 & \cite{ehrlich2025codemonkeys}    & \xmark & \xmark & \xmark & \xmark & \cmark & \xmark & \cmark & \xmark \\
\addlinespace[0.15em]
16 & InfantAgent-Next     & Claude-Sonnet-3.7           & 66.0 & \cite{lei2025infantagent}        & \xmark & \xmark & \xmark & \xmark & \cmark & \xmark & \cmark & \cmark \\
17 & CWM                  & CWM-32B                     & 65.8 & \cite{copet2025cwm}              & \xmark & \xmark & \xmark & \xmark & \cmark & \cmark & \cmark & \cmark \\
18 & ADI                  & Claude-Sonnet-3.7           & 63.8 & \cite{xiang2026empowering}       & \xmark & \xmark & \xmark & \xmark & \cmark & \xmark & \cmark & \xmark \\
19 & Repairity            & Qwen-2.5-Coder-32B-Instruct & 62.7 & \cite{tang2025boosting}          & \xmark & \xmark & \xmark & \xmark & \cmark & \xmark & \cmark & \cmark \\
20 & Lita                 & Claude-Opus-4               & 62.6 & \cite{dai2025lita}               & \xmark & \xmark & \xmark & \xmark & \cmark & \xmark & \cmark & \xmark \\
\bottomrule
\end{tabularx}

\begin{tablenotes}[flushleft]
\footnotesize
\item \textbf{Notes.} \cmark: used; \xmark: not used; \nr: not reported. 
PL: Programming Language; DK: Domain-specific; TS: Technology Stack; DA: Design Architecture; 
EC: Existing Code; RE: Repository Evolution; DT: Development Tool; EXP: Experiential.
\end{tablenotes}

\end{threeparttable}
\end{table*}

\finding{
Across the temporal trends, representative works, and high-performing SWE-bench Verified methods, a consistent pattern can be observed: automated issue resolution is moving from static, code-centric context provision toward more dynamic, feedback-aware, and experience-oriented knowledge orchestration. Existing Code Knowledge remains the earliest and most consistently adopted knowledge type, while Development Tool Knowledge has become a necessary companion because agents must not only inspect repositories, but also search, edit, execute, test, debug, and validate patches in realistic environments. The increasing use of Experiential Knowledge and Repository Evolution Knowledge further shows that recent systems are beginning to rely on process-level and temporal evidence, such as prior trajectories, failed attempts, reward signals, historical fixes, execution traces, and development histories. However, high-performing systems still share a concentrated repository-operational foundation, and the relatively limited presence of Design Architecture Knowledge and explicit L1 Background Knowledge suggests that architectural intent, domain assumptions, programming-language constraints, and technology-stack conventions have not yet been widely operationalized. These patterns should therefore be interpreted as descriptive trends rather than causal evidence about which knowledge type directly improves leaderboard performance.

This finding points to the next stage of knowledge-centric automated issue resolution: future systems should move beyond retrieving more code or invoking more tools, and instead build cross-layer knowledge infrastructures that connect code artifacts, tool feedback, repository history, architectural intent, and background semantics. Architectural and Domain-Specific Knowledge should be made more operational through structured representations, graph-based reasoning, executable specifications, or agent-usable constraint formats, so that they can participate in retrieval, planning, feedback interpretation, and patch validation. Future benchmarks and evaluations should also better expose cases where such higher-level knowledge is necessary, rather than mainly rewarding repository navigation and test-based validation. At the same time, more rigorous ablation studies and controlled comparison protocols are needed to distinguish descriptive leaderboard associations from the actual contribution of specific knowledge types and knowledge combinations. Overall, the next stage of automated issue resolution is likely to depend not only on stronger base models or larger inference budgets, but also on reusable, traceable, and controllable knowledge systems that support robust reasoning across code, tools, history, architecture, and domain context.
}

\section{Research Opportunities and Road Ahead}

Our knowledge-centric analysis shows that automated issue resolution is moving beyond simply retrieving repository context and generating patches. Existing systems have already made substantial progress in grounding LLMs in codebases, interacting with development tools, using execution feedback, and reusing repair trajectories. However, as issue resolution expands toward feature implementation, dependency migration, low-resource language repair, cross-repository reasoning, and long-horizon software maintenance, the central challenge becomes how to construct, organize, update, reuse, and evaluate software knowledge in a systematic way. Based on the knowledge lifecycle analyzed in this survey, we organize future research opportunities into three temporal levels: near-term directions that can be implemented with current techniques, mid-term directions that require new knowledge representations and memory mechanisms, and long-term challenges that point toward broader knowledge-supported software maintenance. We further discuss industrial private knowledge construction and knowledge-centric evaluation, which are critical for making this research agenda practical and measurable.

\subsection{Near-Term Directions: Repository Knowledge Construction and Cross-Stage Reuse}

A near-term opportunity is to improve the automatic construction and quality control of repository knowledge. Current issue-resolution systems already rely heavily on code artifacts, dependency relations, tests, execution logs, and tool outputs, but such knowledge is often collected in an ad hoc and task-specific manner. Future systems can build more systematic pipelines that extract repository maps, module responsibilities, API usage patterns, dependency constraints, test behaviors, historical change records, and issue-fix links from existing software artifacts. These pipelines should not only retrieve relevant snippets, but also attach provenance, version information, confidence signals, and validity conditions to each knowledge item. For example, a repository summary should indicate which files, commits, documents, or execution results support it, and whether it remains valid under the current repository version. Such quality-aware extraction can reduce noisy context, outdated summaries, and unsupported model-generated assumptions.

Another practical direction is to enable knowledge reuse across different stages of the issue-resolution workflow. In many current systems, intermediate results are consumed locally and then discarded: localization produces suspicious files, patch generation produces candidate edits, and validation produces pass/fail feedback. However, each stage can generate reusable knowledge for later stages. Localization can produce root-cause hypotheses, dependency evidence, alternative candidate locations, and uncertainty estimates. Patch generation can record edit intentions, affected interfaces, and expected behavioral changes. Patch validation can produce failure explanations, regression risks, and constraints that should guide future attempts. By converting these intermediate artifacts into structured and reusable knowledge records, agents can avoid repeatedly rediscovering the same information and can revise earlier hypotheses based on later feedback. This direction is feasible with existing retrieval, summarization, static analysis, and agent-memory techniques, and can directly improve the controllability and efficiency of current issue-resolution systems.

\subsection{Mid-Term Directions: General Knowledge Representations and Self-Evolving Experience Bases}

A mid-term research challenge is to develop more general software knowledge representation schemes for LLM-based agents. Many current systems still represent knowledge as raw text, retrieved code snippets, tool logs, or loosely structured prompts. While these formats are easy to integrate with LLMs, they are often difficult to verify, update, compose, and reuse. Future work can design representation schemas that explicitly encode issue intent, affected components, root-cause hypotheses, architectural constraints, repair rationales, dependency relations, validation evidence, and applicability conditions. Such schemas should remain interpretable to LLMs while also being structured enough for retrieval modules, static analyzers, test systems, and agent controllers to process. In this sense, software knowledge representation should evolve from prompt-level context toward more standardized, traceable, and tool-operable knowledge interfaces.

A related direction is the construction of self-evolving experiential knowledge bases. Recent work has shown the value of repair trajectories, failed attempts, reasoning traces, tool-use records, and validation outcomes. However, many experience memories are still static collections of demonstrations or summaries. Future systems should support continuous experience accumulation, conflict detection, abstraction, and retirement. Successful trajectories can be distilled into reusable repair strategies, while failed attempts can be converted into warning patterns or recovery rules. Repeated tool interactions can reveal effective search and debugging procedures, and validation feedback can be summarized into project-specific repair constraints. The key challenge is not only how to store experience, but how to decide which experiences remain useful, when they become outdated, and how they should be generalized without losing repository-specific grounding. This would allow issue-resolution agents to improve over time through validated experience rather than relying only on one-shot context construction.

\subsection{Long-Term Challenges: Cross-Domain Transfer and Lifecycle-Oriented Software Maintenance}

In the long term, automated issue resolution should support knowledge transfer across repositories, languages, frameworks, and domains. Many software projects share common APIs, architectural patterns, dependency ecosystems, security assumptions, and maintenance practices. If agents can abstract reusable knowledge from one setting and adapt it to another, they may handle tasks such as dependency migration, framework upgrade, cross-repository bug fixing, and domain-specific repair more effectively. However, cross-domain transfer is difficult because useful knowledge must be general enough to transfer, yet specific enough to constrain concrete code changes. For example, a migration rule for one library version may not apply to another; an architectural pattern may have different implementations across repositories; and a domain rule may depend on local business logic or scientific assumptions. Future research therefore needs mechanisms for representing the scope, preconditions, and limitations of transferable software knowledge.

Another long-term challenge is to move from issue-level repair toward knowledge-supported software maintenance across the full lifecycle. Real software maintenance involves requirement understanding, impact analysis, design decision tracking, implementation, testing, deployment, monitoring, and future evolution. Automated issue resolution currently covers only part of this process, usually centered on producing a patch for a given issue. A knowledge-centric view suggests a broader direction: agents should maintain evolving knowledge about system behavior, architecture, dependencies, historical decisions, recurring failures, and validation evidence. Such knowledge can support not only individual patch generation, but also architectural impact analysis, regression prevention, technical-debt management, and future change planning. This does not mean that fully autonomous software maintenance is immediately achievable, but it highlights a long-term research path in which issue-resolution agents become persistent knowledge users and maintainers within evolving software projects.

\subsection{Industrial Perspective: Private Knowledge Systems for Enterprise Repositories}

Industrial deployment introduces an especially important but underexplored direction: building private knowledge systems for enterprise software environments. In real companies, valuable software knowledge is often distributed across private codebases, internal issue trackers, pull requests, code reviews, CI/CD logs, design documents, dependency policies, deployment scripts, monitoring records, and team-specific development conventions. Much of this knowledge is not available in public benchmarks, but it is essential for resolving enterprise-level issues. Therefore, future systems should study how to construct private, project-specific knowledge infrastructures that can support automated issue resolution under realistic engineering constraints.

Such enterprise-oriented knowledge systems should address several practical requirements. They need to continuously synchronize with private repositories and internal development histories, while preserving version consistency and avoiding stale knowledge. They should integrate internal technology-stack knowledge, including framework versions, build systems, deployment environments, API conventions, and organization-specific coding standards. They also need mechanisms for access control, privacy protection, knowledge provenance, and auditability, because industrial code and issue histories often contain sensitive information. More importantly, enterprise knowledge systems should support team-level reuse: historical fixes, review comments, incident reports, and CI failures can be transformed into repository memories, repair-pattern libraries, risk rules, and validation checklists. This direction connects knowledge-centric issue resolution with a core industrial pain point: how to turn fragmented private software artifacts and engineering experience into reliable, reusable, and governable knowledge for software agents.

\subsection{Knowledge-Centric Benchmarks and Evaluation}

Evaluation should also move beyond outcome-only patch correctness toward knowledge-centric assessment. Current benchmarks mainly evaluate whether a generated patch passes tests, which is necessary for measuring end-to-end issue-resolution ability. However, pass rate alone provides limited insight into whether a system retrieved the right evidence, used the correct constraints, understood the root cause, reused prior experience, or updated its knowledge effectively. A system may pass tests through shallow context matching, while another may identify the right evidence and reasoning path but fail because of a small implementation error. Therefore, knowledge capability should be evaluated as an explicit dimension of automated issue resolution.

Future benchmarks can include task-level annotations or hidden evaluation protocols for knowledge requirements, such as relevant code regions, dependency constraints, historical issues, documentation evidence, execution traces, architectural assumptions, or domain rules. These annotations do not need to reveal the solution, but they can support fine-grained analysis of whether agents discover and use appropriate knowledge. Possible metrics include knowledge retrieval precision and recall, evidence coverage, provenance correctness, root-cause localization quality, architectural consistency, domain-rule compliance, validation adequacy, and knowledge reuse rate across stages or tasks. For evolving repositories, evaluation should also consider knowledge freshness, update latency, stale-knowledge rate, and robustness under repository changes. In addition, ablation-based metrics can measure the contribution of different knowledge types by removing code context, history, documents, tool feedback, or experiential memory and observing the performance and behavior changes. Such evaluation would encourage systems that are not only strong patch generators, but also reliable knowledge retrievers, users, updaters, and maintainers.

Overall, the road ahead for automated issue resolution lies in building a more systematic knowledge foundation for software engineering agents. Near-term work can focus on automatic repository knowledge construction, quality assessment, and cross-stage reuse. Mid-term work should develop general knowledge representation schemes and self-evolving experience bases. Long-term research can explore cross-domain software knowledge transfer and lifecycle-oriented software maintenance. At the same time, industrial private knowledge systems and knowledge-centric evaluation are necessary for making these directions practical and measurable. Together, these efforts can help move automated issue resolution from temporary context utilization toward reusable, traceable, and evolvable knowledge infrastructures for more grounded, controllable, and maintainable software agents.
\section{Conclusion}

This paper presents a systematic survey of automated issue resolution from a knowledge-centric perspective. By reviewing 133 studies, we show that issue-resolution agents should not be understood merely as patch generators, but as systems that acquire, represent, operationalize, apply, and update software knowledge throughout repository-level maintenance workflows. The proposed three-layer taxonomy and lifecycle-based analysis provide a structured lens for understanding how current systems rely on background assumptions, repository-specific evidence, and procedural experience to support issue understanding, localization, patch generation, validation, and long-horizon agent behavior. Our findings suggest that future progress will depend not only on stronger base models or larger inference budgets, but also on reusable, traceable, and evolvable software knowledge infrastructures that integrate code artifacts, tool feedback, repository history, architectural intent, technology-stack conventions, domain semantics, and experiential memory. Such a knowledge-centric view can help move automated issue resolution from task-specific patch generation toward more grounded, controllable, and maintainable software engineering agents.

\normalem
\bibliographystyle{ACM-Reference-Format}
\bibliography{ref}

\end{document}